\documentclass[11pt,a4paper]{article}
\pdfoutput=1 

\usepackage{jcappub} 

\usepackage[T1]{fontenc} 

\usepackage{graphicx}
\usepackage{epsfig}
\usepackage{amsmath}
\usepackage{amssymb}
\usepackage{subfigure}
\usepackage{longtable}
\usepackage{graphicx}
\usepackage{arydshln}
\usepackage{textcomp}
\usepackage{color}
\usepackage{url}
\usepackage{relsize}

\def\mission{CORE\ }

\title{\boldmath Exploring Cosmic Origins with \mission\!\!: Extragalactic Sources in Cosmic Microwave Background Maps}

   \author[1]{G. De Zotti,\note{Corresponding author.}}
   \author[2]{J. Gonz\'alez-Nuevo,}
   \author[3]{M. Lopez-Caniego,}
   \author[4]{M. Negrello,}
   \author[5]{J. Greenslade,}
   \author[6]{C. Hern\'andez-Monteagudo,}
   \author[7]{J. Delabrouille,}
   \author[8]{Z.-Y. Cai,}
   \author[25,79]{M. Bonato,}
   \author[9,10]{A. Ach\'ucarro,}
% ...

\affiliation[1]{\scriptsize  INAF-Osservatorio Astronomico di Padova, Vicolo dell'Osservatorio 5, I-35122 Padova, Italy}
\affiliation[2]{\scriptsize  Departamento de F\'isica, Universidad de Oviedo, C. Calvo Sotelo s/n, 33007 Oviedo, Spain
}
\affiliation[3]{\scriptsize  European Space Agency, ESAC, Planck Science Office, Camino bajo del Castillo, s/n, Urbanizaci\'{o}n Villafranca del Castillo, Villanueva de la Ca\~{n}ada, Madrid, Spain}
\affiliation[4]{\scriptsize  School of Physics and Astronomy, Cardiff University, The Parade, Cardiff CF24 3AA, UK}
\affiliation[5]{\scriptsize  Astrophysics Group, Imperial College, Blackett Laboratory, Prince Consort Road, London SW7 2AZ, UK}
\affiliation[6]{\scriptsize  Centro de Estudios de F{\'\i}sica del Cosmos de Arag\'on (CEFCA), Plaza San Juan, 1, planta 2, E-44001, Teruel, Spain}
\affiliation[7]{\scriptsize  APC, AstroParticule et Cosmologie, Universit{\'e} Paris Diderot, CNRS/IN2P3, CEA/lrfu, Observatoire de Paris, Sorbonne Paris Cit{\'e}, 10, rue Alice Domon et L{\'e}onie Duquet, 75205 Paris Cedex 13, France}
\affiliation[8]{\scriptsize  CAS Key Laboratory for Research in Galaxies and Cosmology, Department of Astronomy, University of Science and Technology of China, Hefei, Anhui 230026, China}
\affiliation[9]{\scriptsize  Instituut-Lorentz for Theoretical Physics, Universiteit Leiden, 2333 CA, Leiden, The Netherlands}
\affiliation[10]{\scriptsize  Department of Theoretical Physics, University of the Basque Country UPV/EHU, 48040 Bilbao, Spain}
% ...
\affiliation[25]{\scriptsize  Department of Physics \& Astronomy, Tufts University, 574 Boston Avenue, Medford, MA, USA}
% ...
\affiliation[79]{\scriptsize SISSA, Via Bonomea 265, 34136, Trieste, Italy}
% ...
\emailAdd{gianfranco.dezotti@oapd.inaf.it}

\abstract {We discuss the potential of a next generation space-borne Cosmic Microwave Background (CMB) experiment for studies of extragalactic sources. Our analysis has particular bearing on the definition of the future space project, \mission\!\!, that has been submitted in response to ESA's call for a Medium-size mission opportunity as the successor of the \textit{Planck} satellite. Even though the effective telescope size will be somewhat smaller than that of \textit{Planck}, \mission will have a considerably better angular resolution at its highest frequencies, since, in contrast with \textit{Planck}, it will be diffraction limited at all frequencies. The improved resolution implies a considerable decrease of the source confusion, i.e.~substantially fainter detection limits.  In particular, \mission will detect thousands of strongly lensed high-$z$ galaxies distributed over the full sky. The extreme brightness of these galaxies will make it possible to study them, via follow-up observations, in extraordinary detail. Also, the \mission resolution matches the typical sizes of high-$z$ galaxy proto-clusters much better than the \textit{Planck} resolution, resulting in a much higher detection efficiency; these objects will be  caught in an evolutionary phase beyond the reach of surveys in other wavebands. Furthermore, \mission will provide unique information on the evolution of the star formation in virialized groups and clusters of galaxies up to the highest possible redshifts. Finally, thanks to its very high sensitivity, \mission will detect the polarized emission of thousands of radio sources and, for the first time, of dusty galaxies, at mm and sub-mm wavelengths, respectively. }


\notoc

\begin{document}

\newcommand{\todo}[1]{\textsf{[TODO: #1]}}
\maketitle
\flushbottom

\def\simlt{\mathrel{\rlap{\lower 3pt\hbox{$\sim$}}\raise 2.0pt\hbox{$<$}}}
\def\simgt{\mathrel{\rlap{\lower 3pt\hbox{$\sim$}} \raise
2.0pt\hbox{$>$}}}
\def\lsim{\,\lower2truept\hbox{${<\atop\hbox{\raise4truept\hbox{$\sim$}}}$}\,}
\def\gsim{\,\lower2truept\hbox{${> \atop\hbox{\raise4truept\hbox{$\sim$}}}$}\,}
\def\aap{A\&A}
\def\apj{ApJ}
\def\apjs{ApJS}
\def\apjl{ApJL}
\def\mnras{MNRAS}
\def\aj{AJ}
\def\nat{Nature}
\def\aaps{A\&A Supp.}
\def\aaps{A\&A Supp.}
\def\pra{Phys.Rev.A}         
\def\physrep{Physics Reports}         
\def\prb{Phys.Rev.B}         
\def\prc{Phys.Rev.C}         
\def\prd{Phys.Rev.D}         
\def\prl{Phys.Rev.Lett}      
\def\araa{ARA\&A}       
\def\gca{GeCoA}         
\def\pasp{PASP}              
\def\pasj{PASJ}              
\def\apss{ApSS}
\def\jcap{JCAP}
\def\sovast{Soviet Astronomy}
\def\na{New Astronomy}
\def\aapr{A\&A Rev.}
\def\planss{Planet. Space Sci.}
\def\procspie{Proceedings of the SPIE}
 \vspace{1em}


\section{Introduction}\label{par:intro}

Although not specifically designed for the observation of extragalactic sources, space-borne experiments aimed at investigating the Cosmic Microwave Background (CMB) have the potential to bring breakthroughs also in this field.  An investigation of the impact on studies of extragalactic sources of the project named the Cosmic Origins Explorer plus (COrE$+$), submitted in response to ESA's call for the 4th Medium-size mission (M4) opportunity, was carried out by Ref.~\cite{DeZotti2015}. Various options were considered, with effective telescope sizes of $\simeq 1\,$m, $1.5\,$m  and 2\,m, and a frequency range from 60 to 1200\,GHz. A proposal for ESA's 5th Medium-size mission (M5) is envisaging an instrument (named \mission\!\!) with a baseline telescope size of 1.2\,m and 19 frequency channels, distributed over the 60--600\,GHz frequency range. A decrease or an increase of the telescope size to 1\,m and to 1.5\,m, respectively, were also considered. These options will be referred to as \mission\!\!100 and \mission\!\!150. For the \mission\!\!150 configuration we will also consider the added value of an extension of the frequency range to 800\,GHz. 

Since the analysis by Ref.\cite{DeZotti2015}  was completed, considerable relevant new data have become available and more detailed studies have been carried out, motivating an update for the 5th Medium-size mission (M5) proposal. In particular, most analyses of the \textit{Planck} data have now been published, giving much clearer predictions for the capabilities of next generation CMB experiments.

The plan of the paper is the following. In Sect.~\ref{sec:counts_intens} we present a new assessment of the expected counts of the various classes of extragalactic sources in total intensity. In Sect.~\ref{sect:protocluster} we highlight the \mission potential for detecting galaxy proto-clusters during their early evolutionary phase when they did not yet possess the hot intergalactic medium allowing detection via their X-ray emission and/or the Sunyaev-Zeldovich (SZ) effect. As shown in Sect.~\ref{sect:cluster}, \mission will also provide unique information on the evolution of the star-formation rate (SFR) in virialized clusters. Section~\ref{sect:CIB} deals with the information provided by \mission surveys on the Cosmic Infrared Background (CIB), while the effect of bright sub-mm lines on the power spectra measured in different frequency intervals and the possibility of counts being estimated in lines are considered in Sect.~\ref{sect:lines}. In Sect.~\ref{sect:counts_pol} we discuss counts in polarized flux density and report the results of new simulations aimed at determining the \mission detection limits in polarization, showing that \mission will provide a real breakthrough in this field.  Our main conclusions are summarized in Sect.~\ref{sec:conclusions}.

Throughout this paper we adopt the fiducial $\Lambda$CDM cosmology with the best-fit values of the parameters derived from \textit{Planck} CMB power spectra, in combination with lensing reconstruction and external data: $H_0=67.74\,\hbox{km}\,\hbox{s}^{-1}\,\hbox{Mpc}^{-1}$; $\Omega_\Lambda=0.6911$; and $\Omega_{\rm m}=0.3081$ \cite{Planck_parameters2015}.

This work is part of a series of papers that present the science achievable by the CORE space mission. The performance requirements and the mission concept are described in \cite{Delabrouille2017}. The instrument is described in \cite{deBernardis2017}. Reference \cite{Ashdown2017} explores systematic effects that may represent a threat to the measurement accuracy. Reference \cite{Remazeilles2017} discusses polarised foregrounds and the $B$-mode component separation. The constraints on cosmological parameters and fundamental physics that can be derived from CORE measurements are discussed in \cite{DiValentino2016} while the constraints on inflationary models are discussed in \cite{Finelli2016}. References \cite{Bartlett2017} and \cite{Melin2017} deal large-scale structure and cluster science, respectively, while \cite{Burigana2017} addresses the effect on the CMB of the observer's peculiar motion.

\begin{table}
\centering
\caption{Estimated $4\,\sigma$ \mission detection limits, $S_{\rm d}$ (mJy), for 4 effective telescope sizes. The values of $S_{\rm d}$ were derived from the simulations described in Ref.~\protect\cite{DeZotti2015} and refer to regions of low Galactic emission.  }
\vskip12pt
\begin{tabular}{rrrrr}
\hline
\hline
\multicolumn{1}{l}{$\nu$\,(GHz)}  &  \multicolumn{1}{c}{1\,m} &  \multicolumn{1}{c}{1.2\,m} &  \multicolumn{1}{c}{1.5\,m} &  \multicolumn{1}{c}{2\,m} \\
\hline
       60   &      197.9	 &  147.1  &      94.4 &   	55.3 \\
       70   &      200.1 &	149.5  &     94.8 &  	55.3 \\
       80   &      197.1 &	148.1  &     92.7 &  	53.9 \\
       90   &      190.5 &	144.2  &    89.1 & 	51.6	 \\
       100  &      182.0 &	138.7  &    84.8 &    	49.0 \\
       115  &      169.5 &	130.7  &    78.6 &  	45.2 \\
       130  &      156.7 &	122.2  &    72.5 &  	41.7 \\
       145  &      144.7 &	114.0  &    66.9 &  	38.4 \\
       160  &      131.8 &	105.3  &    61.0 &  	35.1 \\
       175  &      119.2 &	 96.6&   55.2 &   	31.9 \\
       195  &      104.9 &	 86.9&   49.0 &  	28.8 \\
       220  &  	  91.6	 &   78.1&    43.8    &	26.4     \\
       255  &      80.7	 &   70.9&   41.1    &	26.0     \\
       295  &  	  81.0 	 &   73.1&   44.1    &	29.1     \\
       340  &      90.5	 &   83.2&    51.5    &	34.9     \\
       390  &      104.5 &	 97.1&   60.7 &  	41.6 \\
       450  &      121.8 &	113.7  &  71.3 &  	49.3 \\
       520  &  	  140.7	 &  131.5  &    82.5    &	57.6     \\
       600  &      150.5 &	139.8&     90.4 &   	63.5 \\
\hline\hline
\end{tabular}
\label{tab:detlim}
\end{table}

\section{Counts of extragalactic sources in total intensity}\label{sec:counts_intens}

As pointed out by Ref.~\cite{DeZotti2015}, \mission surveys will be confusion limited. This was already the case for the \textit{Planck} High Frequency Instrument (HFI) which however reached the diffraction limit only up to 217\,GHz, while \mission will be diffraction limited over its full frequency range. Since the confusion limit scales roughly as the beam solid angle, i.e.~as the square of the full width at half maximum (FWHM) of the instrument \cite[see figure~3 of Ref.][]{DeZotti2015}, \mission will substantially improve over \textit{Planck}-HFI, even in the case of a somewhat smaller telescope. For example, at 545\,GHz ($550\,\mu$m) the \textit{Planck} beam has an effective $\hbox{FWHM}=4.83'$, while the diffraction limit for its 1.5-m telescope is $1.5'$. Realistic simulations \cite{DeZotti2015} gave, at this frequency, $4\,\sigma$ completeness limits of 86\,mJy for the \mission 1.5-m option and of 142\,mJy for the 1-m option. For comparison the 90\% completeness  limit for the Second \textit{Planck} Catalog of Compact Sources \cite[PCCS\,2;][]{PCCS2_2015} at 545\,GHz is 555\,mJy.

At the lower frequencies covered by the \textit{Planck} Low Frequency Instrument (LFI), whose resolution was also at the diffraction limit, \mission performs better thanks to its lower instrumental noise.  For example at 70\,GHz the PCCS\,2 90\% completeness limit is 501\,mJy, while simulations give, for \mission\!\!, $4\,\sigma$ completeness limits of 95\,mJy  and 200\,mJy for the 1.5-m and 1-m options, respectively \cite{DeZotti2015}.

Reference~\cite{DeZotti2015} found that the $4\,\sigma$ \mission detection limits, $S_{\rm d}$, derived from realistic simulations, are well approximated by the formula:
\begin{equation}\label{eq:Slim}
S_{\rm d}=4[\sigma_{\rm conf}^2+\sigma_{\rm noise}^2+(0.12\,\sigma_{\rm CMB})^2]^{1/2},
\end{equation}
with
\begin{equation}\label{eq:omega_eff}
\sigma_{\rm conf}^2=\sigma_{\rm P, radio}^2+\sigma_{\rm P, dusty}^2+\sigma_{\rm clust, dusty}^2+\sigma_{\rm SZ}^2.
\end{equation}
Here, $\sigma_{\rm conf}$ and $\sigma_{\rm CMB}$ are the rms fluctuations at the instrument resolution due source confusion and to CMB anisotropies, respectively, computed using the formulae in Sect.~2.1 of Ref.~\cite{DeZotti2015}. The source confusion term, due to sources below the detection limit, includes contributions of Poisson fluctuations due to radio sources, $\sigma_{\rm P, radio}$, and to dusty galaxies $\sigma_{\rm P, dusty}$, as well as the contribution of clustering of dusty galaxies, $\sigma_{\rm clust, dusty}$ (the contribution of clustering of radio sources was found to be negligibly small). The contribution of \mission instrumental noise, $\sigma_{\rm noise}$, turns out to be always almost negligible. A tabulation of  $4\,\sigma$ \mission detection limits, $S_{\rm d}$, for the 19 frequency channels and four effective telescope sizes (1\,m, 1.2\,m, 1.5\,m, and 2\,m)  is given in Table~\ref{tab:detlim}.


\subsection{Local dusty galaxies}\label{sect:local_dusty}

The \textit{Planck} surveys offered the first opportunity to accurately determine the luminosity function of dusty galaxies in the very local universe at several (sub-)millimetre wavelengths, using blindly selected samples of low-redshift sources, unaffected by cosmological evolution \cite{Negrello2013}. At 857, 545 and 353\,GHz the local luminosity function could be determined over about three decades in luminosity. Remarkably, \textit{Planck} reached luminosities almost one order of magnitude fainter than \textit{Herschel}'s Spectral and Photometric Imaging REceiver (SPIRE) surveys at the common or nearby frequencies (857 and 545\,GHz, corresponding to wavelengths of 350 and $550\,\mu$m) \cite{Marchetti2016}: \textit{Planck}'s all-sky coverage has proven to be much more effective in detecting local low-luminosity sub-mm galaxies than \textit{Herschel}-SPIRE, in spite of the much deeper flux density levels reached by the latter instrument, on small sky areas.

Estimates of the \mission detection limits \cite{DeZotti2015} show that it will reach, at 500--600\,GHz, flux densities a factor of 5 (1-m option) or 8.6 (1.5-m option) fainter than the PCCS2. \mission  will thus explore a volume larger by a factor of $\simeq 11$ or $\simeq 25$, respectively, thus improving the source statistics by similar factors. As a result, \mission will detect, at 600\,GHz and with $\ge 4\,\sigma$ significance, from $\simeq 14,000$ (1-m option) to $\simeq 30,600$ (1.5-m option) star-forming galaxies out to $z=0.1$ in the ``extragalactic zone'' ($|b|>30^\circ$) defined by Ref.~\cite{PCCS2_2015}.\footnote{Note that table~13 of Ref.~\cite{PCCS2_2015} lists the \textit{total} number of \textit{Planck} detections, not only those above the 90\% completeness limit.}

Within the timescales of the \mission launch, there will be several wide-angle redshift and photometric surveys available that should provide distance information for the majority, if not all, of the galaxies detected by the satellite. Of most interest of course are data sets covering the entire sky. In the near future spectroscopic data for all the Two Micron All Sky Survey (2MASS) galaxies (1 million over most of sky, median $z\sim 0.1$) are expected from the TAIPAN survey \cite{Kuehn2014} and the Low Redshift survey at Calar Alto (LoRCA) project \cite{Comparat2016}. On a longer timescale, the Spectro-Photometer for the History of the Universe, Epoch of Reionization, and Ices Explorer (SPHEREx) mission \cite{Dore2014} aims at providing a low-resolution spectroscopic survey of the entire sky in the near-IR, including millions of galaxies up to $z>1$.

Another avenue towards all-sky 3D catalogues is through combing imaging data-sets and deriving photometric redshifts. This was successfully completed first for 2MASS galaxies \cite[2MPZ;][]{Bilicki2014} and recently also for a combination of WISE and SuperCOSMOS scans of photographic plates \cite{Bilicki2016}. The WISE/SuperCOSMOS catalogue maps the large scale structure (LSS) at $z<0.4$ for 70\% of sky.

The deeper \mission surveys, compared to \textit{Planck}, coupled with redshift information, will make it possible to extend the  determination of the luminosity function of dusty galaxies down to fainter luminosities and to higher redshifts. The information from blind detections on \mission maps can be expanded by exploiting the prior knowledge of galaxy positions from optical/IR surveys to push the detection limits to fainter levels, a technique successfully tested in Ref.~\cite{LopezCaniego2007}. Stacking analyses on positions of known galaxies will allow us to reach still fainter flux densities.

A variety of photometric data will also be available across the electromagnetic spectrum (radio, IRAS, AKARI, WISE, \textit{Euclid}, GALEX, ROSAT, eROSITA) in addition to multi-band optical/near-IR imaging. Combining them with \mission data, it will be possible to carry out an extensive characterization of the properties of galaxies in the nearby Universe \cite[e.g.][]{Malek2014}. \mission data will be crucial in particular to achieve progress towards understanding how dust mass and emission vary with galaxy type and stellar mass. The combination of the \mission data with data in different wavebands will allow us to determine, for each galaxy type and as a function of stellar mass, the distribution of dust temperatures, as well as the dust mass function, the SFR function, the relationship between star formation and nuclear activity, the relative contributions of newborn and evolved stars to dust heating, and more.

The next step will be to relate the properties of galaxies detected by \mission to the underlying dark matter field and to the properties of their host dark matter haloes. Local dusty star-forming galaxies are found to be more clustered and to reside in more massive haloes than galaxies with unobscured star formation \cite{Solarz2015, Pollo2013}. \mission will provide an unprecedented sample to investigate the link between local dusty galaxies of different types and their local environments.


At mm wavelengths we can exploit the synergy with the Stage-IV CMB (CMB-S4) surveys \cite{Abazajian2016}. This next generation ground-based program aims to deploy O(500,000) detectors spanning the 30--300\,GHz frequency range. It will use multiple telescopes and sites to map most of the sky and is expected to be operational between about $2021$ and $2024$. 
Thanks to its higher sensitivity, the CMB-S4 experiment  will extend to longer wavelengths the spectral coverage of dusty galaxies detected by \mission at sub-mm wavelengths. Among other things, this will allow us to test the hints from \textit{Planck} that some fraction of local galaxies have a significant very cold dust component \cite[10--13\,K;][]{RowanRobinsonClements2015} which the mm bands of CMB-S4  will be especially powerful in detecting. 


\begin{figure}
\includegraphics[width=0.48\columnwidth]{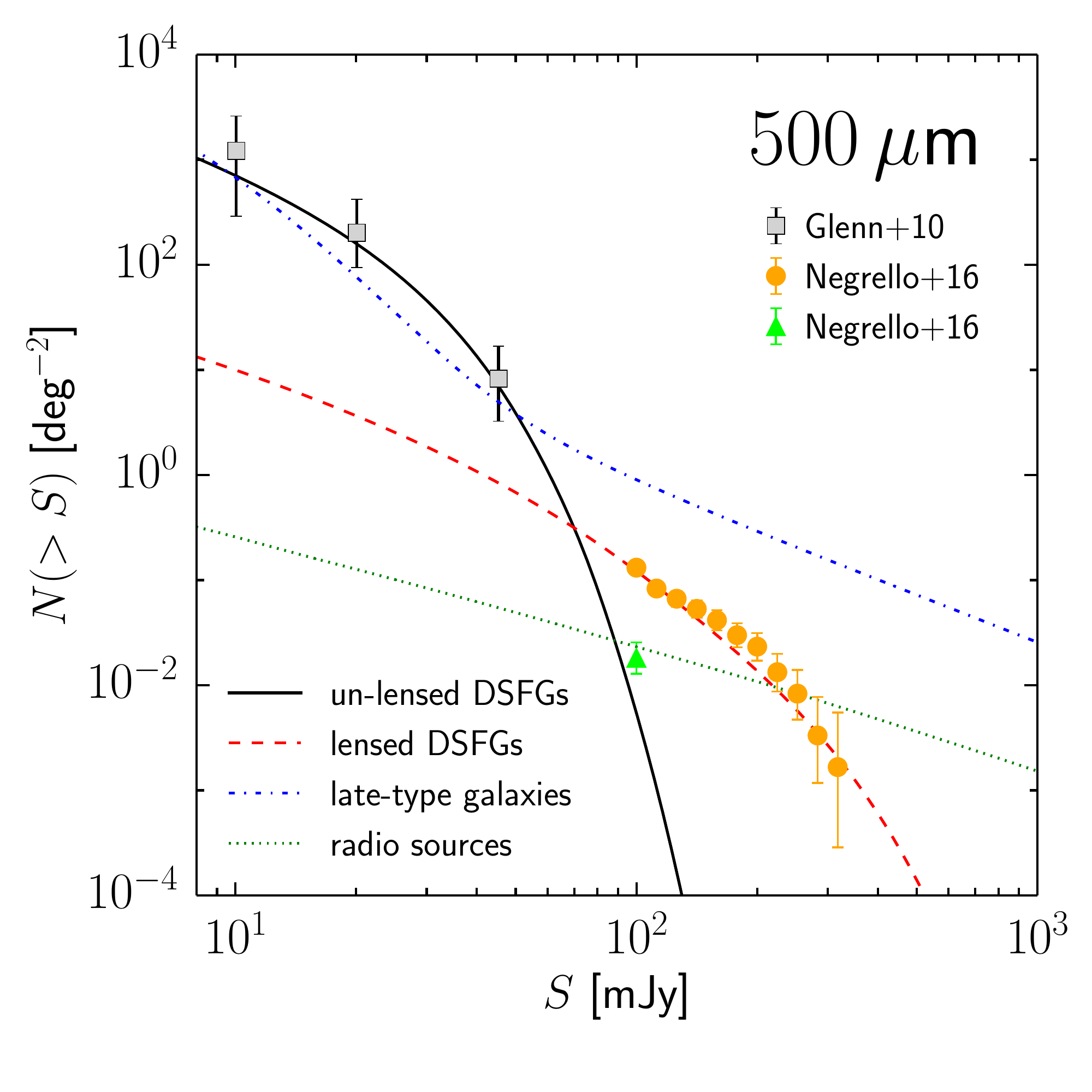}
\includegraphics[width=0.48\columnwidth]{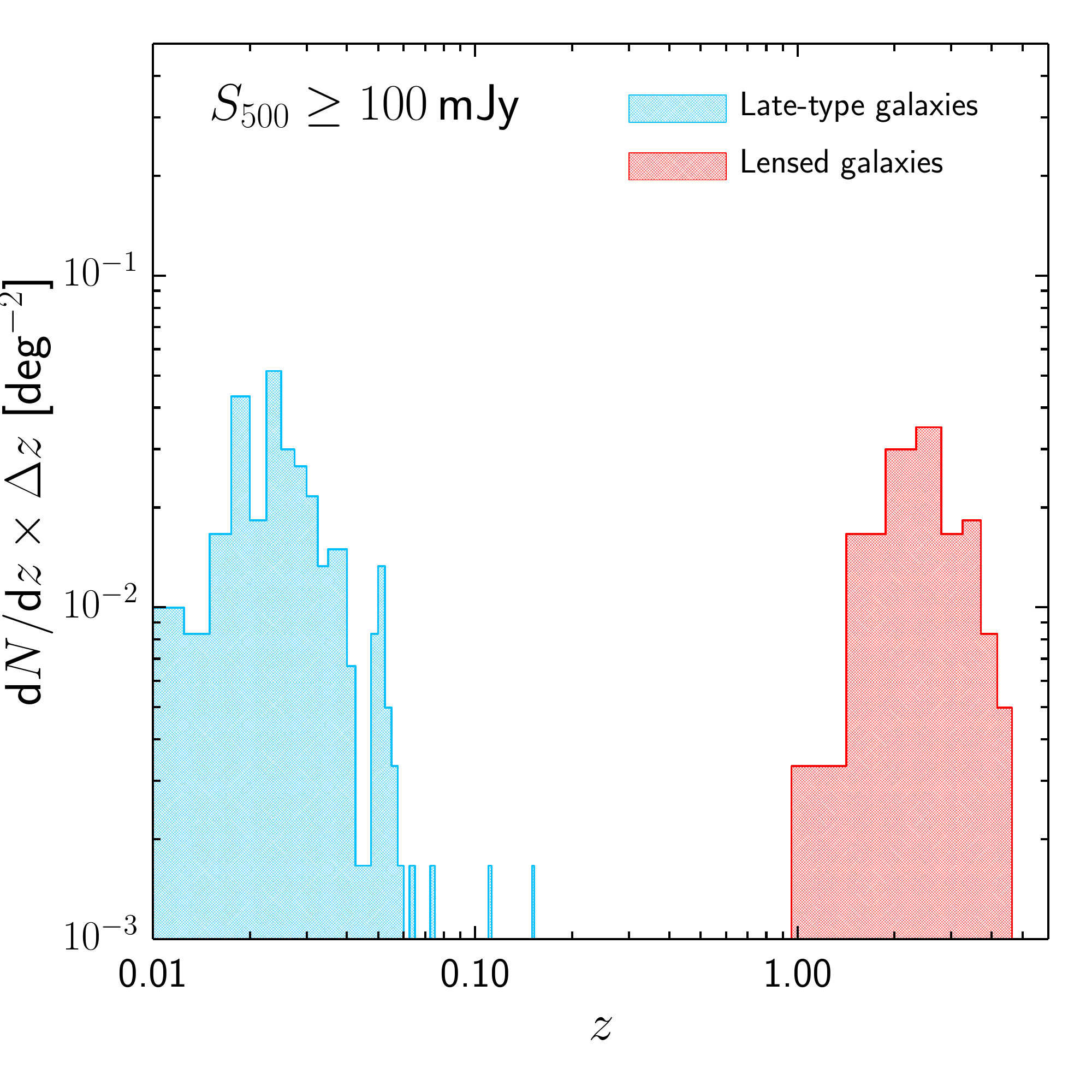}
\caption{\textbf{Left panel.} Counts of candidate gravitationally-lensed dusty star-forming galaxies (DSFGs) from the \textit{Herschel}  Astrophysical Terahertz Large Area Survey (H-ATLAS), over an area of about $600\,\hbox{deg}^2$ \cite[][filled orange circles]{Negrello2016lensed}, compared with the counts of unlensed high-$z$ DSFGs, interpreted as proto-spheroidal galaxies \cite[solid black curve;][]{Glenn2010}, of normal late-type and starburst galaxies (referred to as ``late-type galaxies'' in the inset; dot-dashed blue line) and of radio sources (blazars, dotted green line; the data point, green triangle, is from Ref.~\cite{Negrello2016lensed}). The dashed red line shows the model counts of strongly lensed galaxies for a maximum magnification $\mu_{\rm max}=10$  \cite{Negrello2016lensed}; rare much higher magnifications can be produced by cluster lensing. The counts of normal late-type, starburst and proto-spheroidal galaxies have been computed using the model described in Ref.~\cite{Cai2013}. The effect of lensing has been computed following Ref.~\cite{Lapi2012}. The counts of radio sources are from Ref.~\cite{Tucci2011}. \textbf{Right panel.} Redshift distribution of galaxies brighter than 100\,mJy at $500\,\mu$m, derived from the full H-ATLAS catalogue \cite{Negrello2016lensed}. There is a clear bimodality. On one side we have nearby late-type galaxies, almost all at $z\le 0.06$, and hence easily recognizable in optical/near-infrared catalogues. On the other side we have dust enshrouded, hence optically very faint, gravitationally lensed galaxies at $z\ge 1$ and up to $z>4$.   }
\label{fig:lensed}
\end{figure}

\subsection{High-$z$ dusty galaxies}

At sub-mm wavelengths \mission will reach faint enough flux densities to see the transition from the Euclidean portion of the counts contributed by low-$z$ galaxies to the steeply rising portion caused by cosmological evolution. The precise shape of the counts in this transition region provides critical constraints to evolutionary models \cite{Lapi2011}. There are in fact two main scenarios discussed in the literature for the evolution of galaxies as viewed at far-IR/sub-mm wavelengths. One scenario envisages a direct evolutionary link between local and high-$z$ galaxies, with no new populations. According to the other scenario, a major component of the dusty galaxy population at high-$z$, is made, for example, by massive proto-spheroidal galaxies in the process of forming the bulk of their stars. These galaxies have no low-$z$ counterparts at far-IR/sub-mm wavelengths because they have been essentially passively evolving since $z\simeq 1$--1.5. Clearly the first scenario implies a more gradual steepening of the slope of the counts than the second one.

\mission will provide a complete census of the brightest sub-mm galaxies in the Universe, including those with extreme amplification by gravitational lensing. \textit{Planck} has already offered an exciting foretaste of that: ``\textit{Planck}'s dusty GEMS'' \cite[Gravitationally Enhanced subMillimetre Sources;][]{Canameras2015}. As discussed in Ref.~\cite{Canameras2015}, follow-up CO spectroscopy and multi-frequency photometry of 11 such sources have shown that they are at $z=2.2$--3.6 and have apparent (uncorrected for gravitational amplification) far-IR  luminosities of up to $3\times 10^{14}\,\hbox{L}_\odot$, making them among the brightest sources in the Universe. They are so rare that only very large area surveys can find them.

Ref.~\cite{Nayyeri2016} have found 77 candidate strongly lensed galaxies brighter than 100\,mJy at $500\,\mu$m from \textit{Herschel} surveys covering $372\,\hbox{deg}^2$, corresponding to a surface density of $\simeq 0.2\,\hbox{deg}^{-2}$. A somewhat lower surface density at the same flux density limit has been derived in Ref.~\cite{Negrello2016lensed} from the \textit{Herschel}  Astrophysical Terahertz Large Area Survey (H-ATLAS); they found 79 candidates over an area of $602\,\hbox{deg}^2$. Averaging the two estimates we obtain a surface density of $\simeq 0.16\,\hbox{deg}^{-2}$. The \mission detection limits at $\simeq 500\,\mu$m are $\simeq 150\,$mJy and $\simeq 90\,$mJy for the 1-m and the 1.5-m telescope, respectively. The average surface density of strongly lensed galaxies with $S_{500\mu\rm m}>150\,$mJy is $\simeq 0.0465\,\hbox{deg}^{-2}$ implying that the 1-m option will allow the discovery of about one thousand strongly lensed galaxies in the ``extragalactic zone''. A slight extrapolation to 90\,mJy using the model shown in Fig.~\ref{fig:lensed} leads to an estimate of about 3,000 strongly lensed galaxies detectable by \mission\!\!150 in the ``extragalactic zone''.

At these flux densities the selection of strongly lensed galaxies can be easily done with close to 100\% efficiency \cite{Negrello2007, Negrello2010}. In fact, as illustrated by Fig.~\ref{fig:lensed}, in the relevant flux density range, i.e.~above $S_{500\mu\rm m}\simeq 90\,$mJy, the dominant populations of extragalactic sources are low-$z$ galaxies and high-$z$ strongly lensed galaxies (see the right-hand panel of the figure). The former are easily recognisable in all-sky optical/near-IR surveys, such as the UK Schmidt Telescope and Second Palomar Observatory survey \cite[see the recent SuperCOSMOS all-sky galaxy catalogue;][]{Peacock2016}, the Two-Micron All-Sky Survey \cite[2MASS;][]{Skrutskie2006} and the Wide-field Infrared Survey Explorer \cite[WISE;][]{Wright2010}. In addition there is a small fraction of radio sources, mostly blazars, also easily identifiable by cross-matching with all-sky radio surveys, such as the NRAO VLA Sky Survey \cite[NVSS;][]{Condon1998} plus the  Sydney University Molonglo Sky Survey \cite[SUMSS;][]{Mauch2003}.

There will be a large overlap between the strongly lensed galaxies detected by the CMB-S4 surveys and those detected by \mission\!\!. The two sets of observations will allow a better definition of the SEDs, hence a better photometric redshift estimate, and an improved determination of the source positions, essential for follow-up observation. Due to the shape of the dust emission SED, \mission will be more efficient at detecting objects at $z\simlt 2$, while the longer wavelength observations by the CMB-S4 will be more effective at detecting those at $z\simgt 3$. Note however that the sub-mm \mission measurements will be essential to characterize the emission peak, hence to derive basic quantities like the total IR luminosity, hence the SFR.

These large samples of strongly gravitationally lensed galaxies will be trivially easy targets for ALMA, NOEMA, etc., and the foreground lenses will almost certainly be detectable in, e.g., \textit{Euclid} imaging. Follow-up observations will allow us to determine the total (visible and dark) mass of the lensing galaxies, to investigate their density profiles, and to measure cosmological parameters and especially the Hubble constant using gravitational time delays \cite[e.g.][and references therein]{Eales2015, Meng2015}. The time delay distance measurement are estimated to reach an uncertainty of 5--7\% and are independent of the local distance ladder; they thus provide a crucial test of any potential systematic uncertainties. Although the \textit{Planck} determination of $H_0$ from CMB anisotropies has a much higher precision, it should be noted that time delay distances are completely independent of the properties of the early Universe and may thus allow one to break some of the main degeneracies in the interpretation of CMB data.

While other facilities (e.g., \textit{Euclid}, Gaia, SKA) will also be generating large gravitational lens catalogues on a comparable timescale \cite[see, e.g.,][for a recent review]{Serjeant2016} the critical advantage of \mission and of CMB-S4 will be in extending the sources and lenses to much higher redshifts and in detecting the most extreme amplifications. This has been demonstrated by \textit{Planck}: the magnification factors, $\mu$, of ``\textit{Planck}'s dusty GEMS'' are estimated to be up to 50 \cite{Canameras2015}.

The flux boosting is accompanied by a stretching of images. In general, the image size scales approximately as $\mu$ in one direction (tangentially) while being unchanged in the perpendicular direction (radially). The geometric mean is an overall scaling of $\mu^{1/2}$ as demanded by the conservation of surface brightness. The extreme apparent luminosities and the stretching of images allow us to study the properties and internal structure of these high-$z$ sources in extraordinary detail, much greater than would be possible otherwise. The evolution of dark matter halo substructure and the stellar initial mass function (IMF) can thus be probed to much higher redshifts, providing key tests for galaxy evolution theory.

Another important advantage compared to optical/near-IR searches is that foreground lenses and  background magnified galaxies stand out in different wavebands. The background lensed galaxies \mission will detect are heavily dust obscured, hence are bright at far-IR/(sub-)mm wavelengths but are almost invisible in the optical/near-IR. On the contrary, most foreground lenses are passive spheroidal galaxies, hence bright in the optical/near-IR but almost invisible at far-IR/(sub-)mm wavelengths. Therefore the mutual contamination of their images is small or negligible. This is obviously a major advantage for detailed modelling.

For the unlensed population, the \mission survey will also probe essentially the entire Hubble volume for the most intense hyperluminous starbursts, testing whether there are physical limits to the star-formation rates of galaxies \cite[e.g.~Eddington-limited star formation;][]{vanderWerf2011}.

\begin{figure}
\begin{center}
\includegraphics[width=0.7\columnwidth]{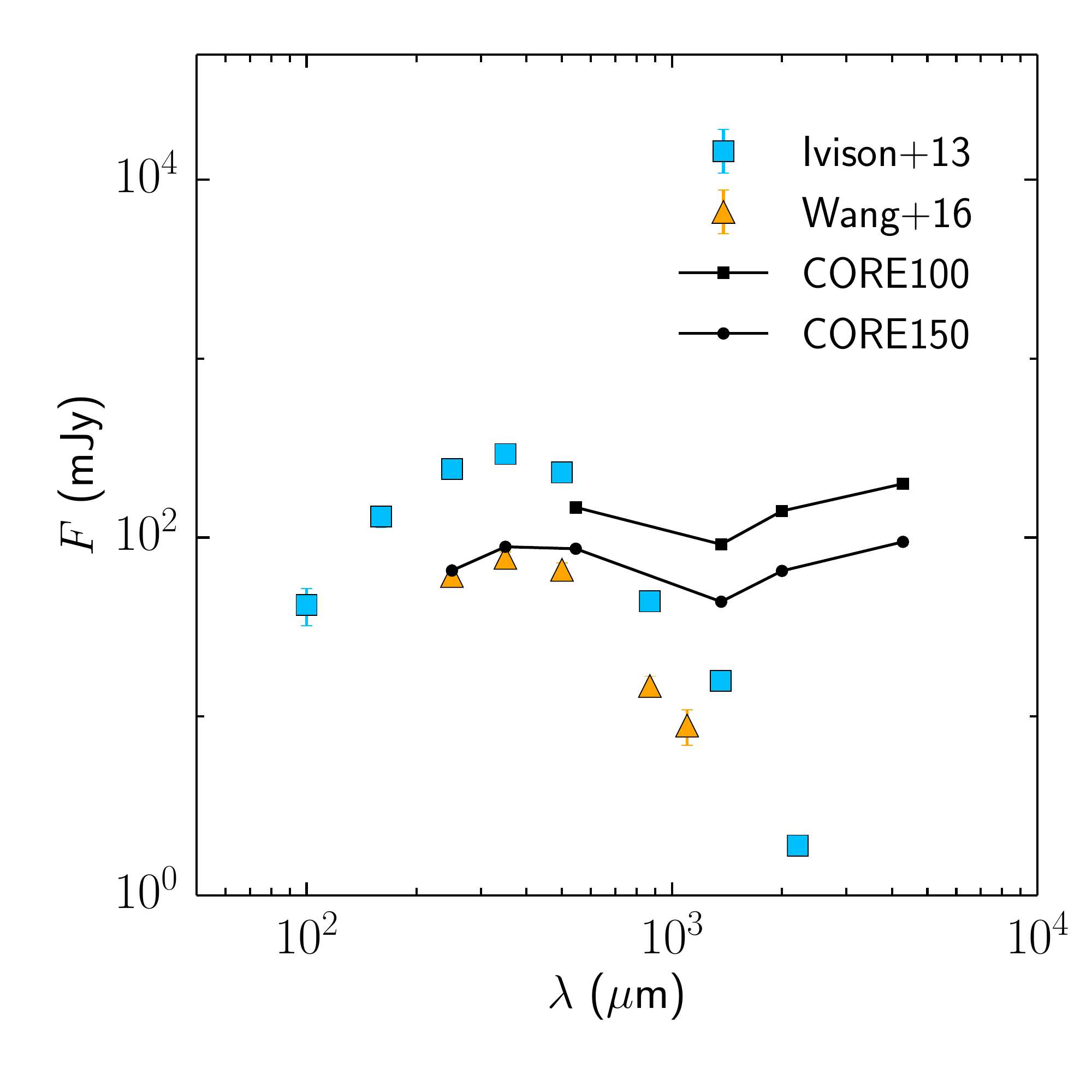}
\caption{Spectral energy distributions (SEDs) of the cores of two clusters of starbursting galaxies. The blue squares show the SEDs of the proto-cluster core at $z=2.41$ \cite{Ivison2013} discovered by means of observations with the Jansky Very Large Array
(JVLA) of a bright H-ATLAS source that did not show indications of strong lensing. The proto-cluster core has a linear size of approximately  $100\,$kpc, i.e.~an angular size of about 12\,arcsec. This object shows extended X-ray emission, suggesting that it is already virialized;  however, it is dominated by star-forming galaxies. The orange triangles show the SED of the proto-cluster core at $z=2.506$ \cite{Wang2016} discovered in the COSMOS field. Its size is about $80\,$kpc. The solid black lines show the \mission detection limits for the 1-m and 1.5-m telescope (from top to bottom).   }
\label{fig:protocluster}
\end{center}
\end{figure}

\begin{figure}
\includegraphics[trim=0cm 0cm 0cm 0cm,clip=true, width=\columnwidth]{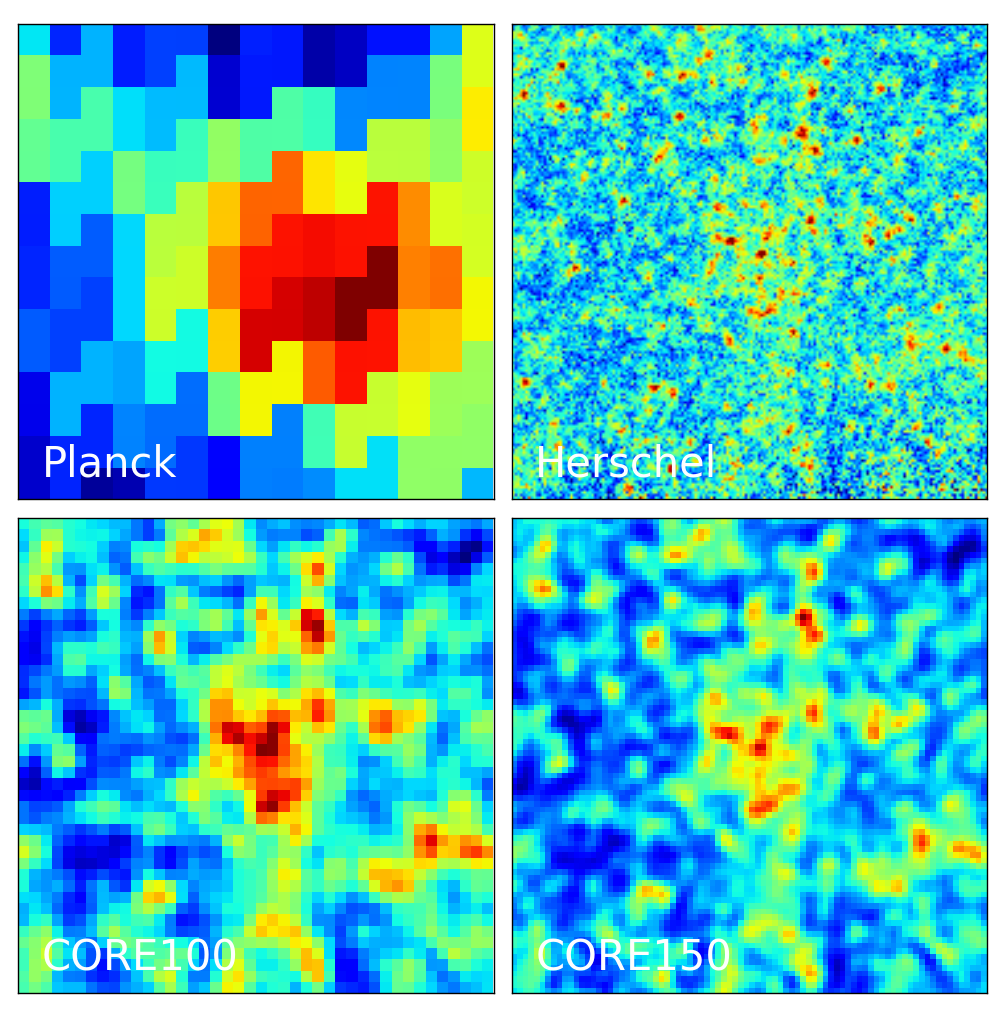}
\caption{\textit{Planck} (upper left panel) and \textit{Herschel} (upper right panel) images at 857 and 600\,GHz, respectively, of a candidate $z = 2.3$ proto-cluster in the Bootes field of the HerMES survey \cite{Clements2014}, compared with the appearance it may take for the diffraction limited beams of \mission\!\!100 and  \mission\!\!150 at 600\,GHz. Pixel sizes for \textit{Planck} and both \mission options are FWHM/2.517. The size of each map is $25'\times 25'$. }
\label{fig:CompareCluster}
\end{figure}

\section{Protoclusters of dusty galaxies}\label{sect:protocluster}


As shown in Ref.~\cite{DeZotti2015}, at sub-mm wavelengths fluctuations in \textit{Planck} maps are dominated by the clustering of dusty galaxies at high $z$. This implies that low-resolution surveys in this spectral range are optimally suited for detecting overdensities of star-forming galaxies that may evolve into present day rich clusters \cite{Negrello2005}.


A blind search on \textit{Planck} maps for candidate high-$z$ proto-clusters was carried out in Ref.~\cite{Planck_high_z2015}, looking for intensity peaks with ``cold'' sub-mm colours, i.e.~with continuum spectra peaking between 353 and 857\,GHz at $5'$ resolution, consistent with redshifts $z>2$ for typical dust emission spectra. Their selection required detections at $\hbox{S/N} > 3$ in all three highest frequency \textit{Planck} bands and applied a flux density threshold of $S_{545\rm GHz} > 500\,$mJy. A total of 2151  high-$z$ source candidates satisfying these criteria were selected in the cleanest 26\% of the sky, corresponding to a surface density of $\simeq 0.2\,\hbox{deg}^{-2}$. Apart from a tiny fraction (around 3\%) of extreme strongly lensed high-$z$ galaxies, these intensity peaks appear to be associated with excess surface densities of high-$z$ dusty star-forming galaxies. This conclusion was substantiated by \textit{Herschel}/SPIRE follow-up of 234 \textit{Planck} ``cold'' targets \cite{PlanckHerschel2015}, finding that about 94\% of them indeed show a significant excess of red 350 and $500\,\mu$m sources in comparison to reference SPIRE fields.

The detected sources have a flux density distribution peaking at $S_{545\rm GHz} \simeq 1\,$Jy and extending up to $S_{545\rm GHz} \simeq 2\,$Jy. If they are indeed proto-clusters at $z>2$, they have extraordinarily high total (8--$1000\,\mu$m) infrared luminosities, $L_{\rm IR}$, peaking around $2\times 10^{14}\,L_\odot$ for a reference dust temperature of 35\,K \cite{Planck_high_z2015}. The corresponding star formation rate (SFR) is $\simeq 3\times  10^4\,\hbox{yr}^{-1}$ \cite{KennicuttEvans2012}.  The associated halo mass, $M_{\rm h}$, obtained from the SFR--$M_{\rm h}$ relation at $z=2$ derived exploiting the abundance matching technique, is $M_{\rm h}\simeq 5\times 10^{14}\,\hbox{M}_\odot$ \cite{Aversa2015}. But the surface density of halos that massive at $z\simeq 2$ is only $\simeq 5\times 10^{-5}\,\hbox{deg}^{-2}$, well below that of \textit{Planck} over-densities.

On the other hand, the halo surface density increases steeply with decreasing $M_{\rm h}$: already at $M_{\rm h}\simeq 10^{14}\,\hbox{M}_\odot$ it reaches $\simeq 0.3\,\hbox{deg}^{-2}$. This suggests that the candidate proto-clusters detected in Ref.~\cite{Planck_high_z2015} may not be individual objects but fluctuations in the number of somewhat smaller, physically unrelated clumps of dusty galaxies along the line of sight. Extensive simulations by Negrello et al. (in preparation) showed that indeed the results of  Ref.~\cite{Planck_high_z2015} can be understood in these terms. Support for this view comes from the fact that the only \textit{Planck} overdensity for which spectroscopic or photometric redshifts of member galaxies have been obtained was found to consist of two structures at $z\simeq 1.7$ and $z\simeq 2$ \cite{FloresCacho2016}. Despite the fact that many of these clumps of emission selected with \textit{Planck} are not single proto-clusters, they are nevertheless signposts of proto-cluster regions along the line of sight.

{An alternative search for candidate high-$z$ proto-clusters has been carried out by Ref.~\cite[][see also Greenslade et al. 2016, in preparation]{Clements2014}, by cross matching the 857\,GHz \textit{Planck} Catalogue of Compact Sources with over 1000 $\hbox{deg}^2$ of \textit{Herschel} survey fields (primarily H-ATLAS \cite{Eales2010} and HerMES \cite{Oliver2012}) to search for evidence of compact emission from cold clumps of high-$z$ galaxies. Selected at 857\,GHz,  these sources are still ``cold'', but ``warmer'' than those selected at 545\,GHz by \textit{Planck} \cite{Planck_high_z2015}, and thus corresponding to a higher dust temperature or  redshifts $z \sim 1-2$ for typical dust SEDs. A visual inspection of the \textit{Herschel} maps at the positions of the \textit{Planck} compact sources for excesses of red sources identifies a total of 59 proto-cluster candidates. As noted in Ref.~\cite{Planck_high_z2015}, few of the \textit{Planck} high-$z$ sources correspond to a PCCS2 source, so these two alternative methods likely probe different populations of proto-clusters.}

{Simulations of the distributions of the sub-mm flux from proto-clusters seen by \textit{Planck}  \cite[][see their figure~7]{Granato2015} suggest that the typical 350-$\mu$m flux density from a proto-cluster is on the order of 100\,mJy, below \textit{Planck}'s sensitivity. The detection of such sources in the PCCS2 however indicates a typical 350\,$\mu$m flux density of  $\simeq 800\,$mJy, 5--10 times greater than predicted from the simulations. Since the typical highest proto-cluster SFR reached in the simulations was $\simeq 1600\,\hbox{M}_\odot\,\hbox{yr}^{-1}$, this implies an observed SFR on the order of  $\simeq 10^4\,\hbox{M}_\odot\,\hbox{yr}^{-1}$, consistent with the observations from blind searches on the \textit{Planck} maps for cooler proto-clusters.}

In any case, the samples of ``cold'' intensity peaks detected by \textit{Planck} (and in much greater abundance by \mission\!\!) are of great interest for the investigation of the evolution of large-scale structure. Follow-up observations of galaxies within them, providing redshift estimates, would be a powerful tool to investigate the early phases of cluster formation, inaccessible by other means. In fact, these overdensities can be detected in the pre-virialization phase, when their intergalactic medium might not have been shock-heated yet, and hence they may not be detectable via X-ray emission or the Sunyaev-Zeldovich (SZ) effect.

The higher resolution of \mission will facilitate a much easier selection of true proto-clusters. The search described in Ref.~\cite{Planck_high_z2015} was carried out adopting an angular resolution of $5'$, corresponding to a physical size of about 2.5\,Mpc at $z=1.5$--2. This resolution is not optimal for detecting proto-clusters. The study by Ref.~\cite{Alberts2014} of 274 clusters with $0.3\le z \le 1.5$ from the \textit{Spitzer} InfraRed Array Camera (IRAC) Shallow Cluster Survey, using \textit{Herschel}/SPIRE 250-$\mu$m imaging, showed that the density of IR-emitting cluster members dominates over the background field level only within $0.5\,$Mpc of the cluster centre, while at $r >0.5$ Mpc the corrected source density of cluster members is only a small enhancement over the field source density.

A linear scale of 0.5\,Mpc corresponds to an angular scale of about $1'$ at redshifts in the range 1.5--2.5, i.e.~to the angular radius of the \mission beam at $600\,$GHz for the 1-m telescope option. The 1.5-m option, with its $\hbox{FWHM}$ of $1.6'$ at 600\,GHz and of $1.2'$ at 800\,GHz, corresponding, in that redshift range, to linear \textit{radii} of $\simeq 395$ or $\simeq 335\,$kpc, respectively, will be much better suited to detect the bright cluster cores of the kind discovered in Ref.~\cite{Ivison2013} at $z=2.41$ and in Ref.~\cite{Wang2016} at $z=2.51$ on scales of $\sim 100$\,kpc (see Fig.~\ref{fig:protocluster}).

{A visual indication of the appearance of such proto-clusters at different resolutions is given in Fig.~\ref{fig:CompareCluster}. Though only a single source is identifiable in the \textit{Planck} maps, both \mission\!\!100 and \mission\!\!150 show that the nature of this source is a clear overdensity of fainter sources, as also seen in the \textit{Herschel} maps. The all-sky nature of \mission would enable us to search for such overdensities over roughly 20 times the area surveyed by \textit{Herschel}.}

The CMB-S4 surveys also have a resolution well suited for proto-cluster detection. However, since most of detectable proto-clusters should be at relatively low redshifts ($z<3$), their signal is much lower at the longer CMB-S4 wavelengths. Thus these objects are considerably easier to detect at the highest \mission frequencies. As in the case of strongly lensed galaxies, the CMB-S4 data will  contribute to determining the SED, hence the photometric redshift (for an assumed dust temperature), the IR luminosity and the SFR of each source.

A complete, unbiased census of these proto-clusters is fundamental to our understanding of the full path of galaxy cluster formation and of the effect of dense environments on galaxy formation and evolution. Such structures, however, are rare and therefore difficult to find. The surface density of proto-clusters serendipitously detectable by \mission (and by CMB-S4) is estimated to be of a few to $\hbox{several}\times 10^{-2}\,\hbox{deg}^{-2}$ \cite[][see also Negrello et al. 2016, in preparation]{Ivison2013}; hence we may expect the detection by \mission of several hundreds to a few thousand violently starbursting proto-cluster cores.

\begin{figure}
\includegraphics[trim=0cm 0cm 2.9cm 15.5cm,clip=true, width=0.48\columnwidth]{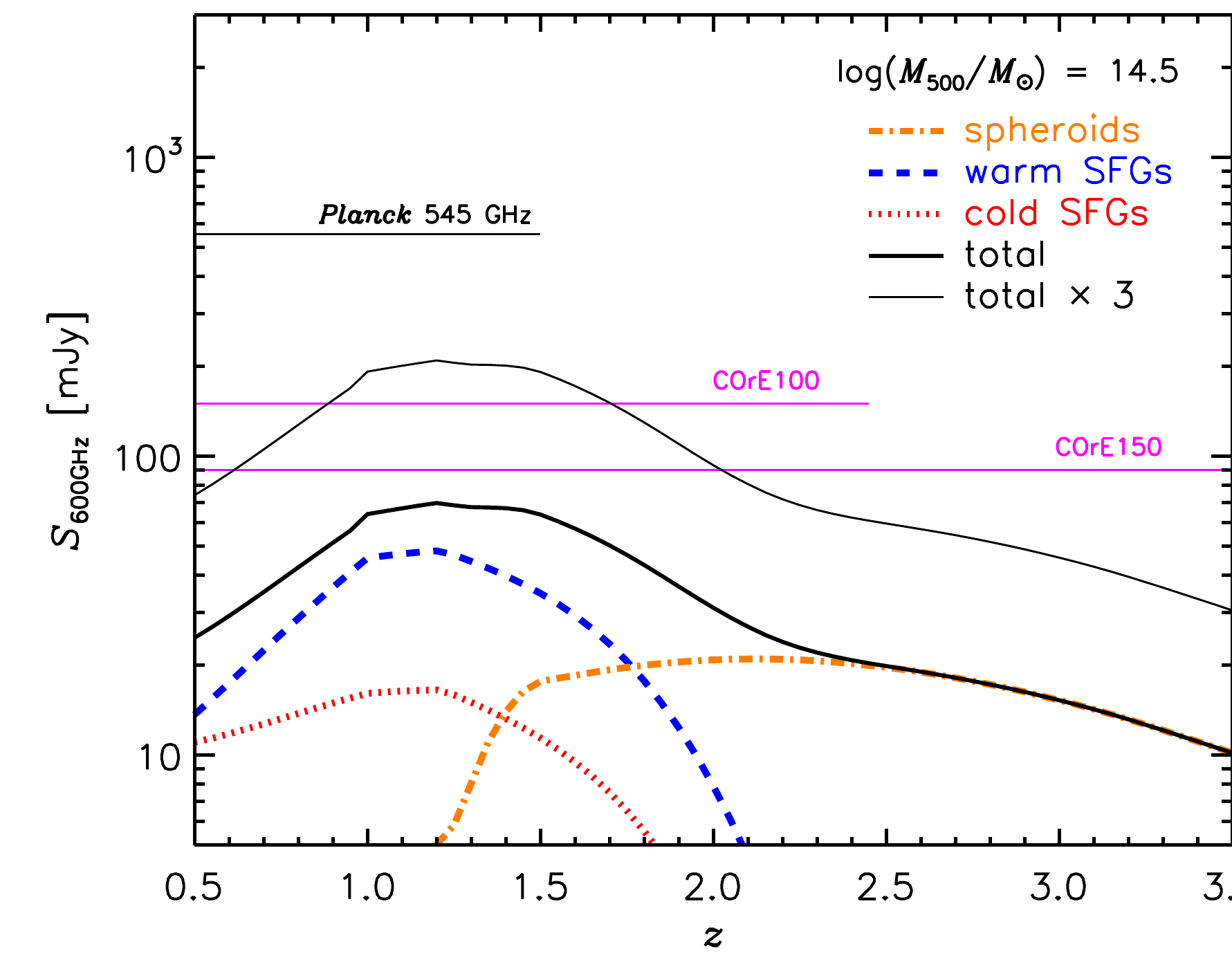}
\includegraphics[trim=0cm 0cm 2.9cm 15.5cm,clip=true, width=0.48\columnwidth]{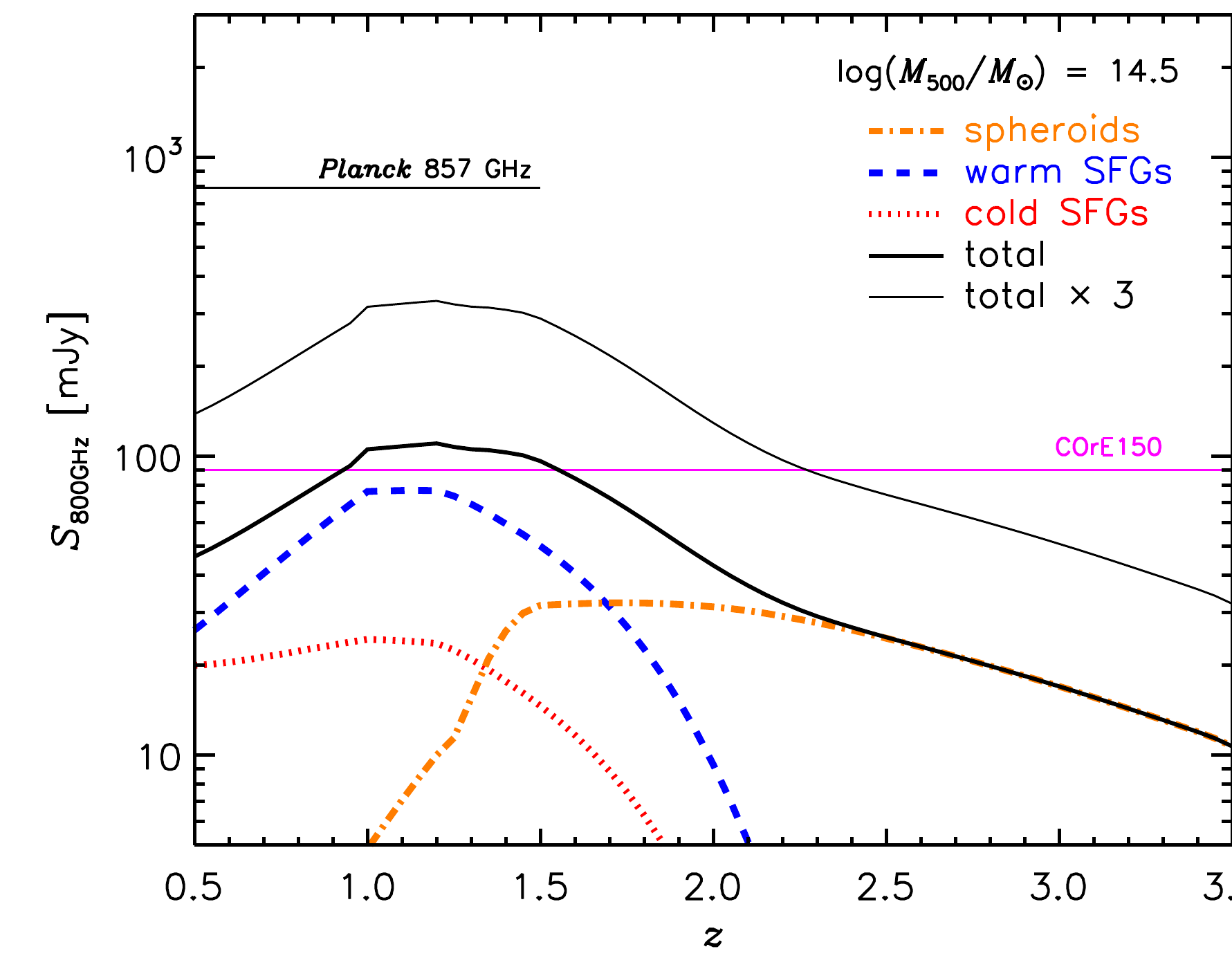}
\caption{Expected flux density at 600 (left panel) and 800 GHz (right panel) due to dust emission of a cluster with $M_{500}=10^{14.5}\,\hbox{M}_\odot$ as a function of the cluster redshift. The cluster luminosity includes contributions from normal late-type and starburst galaxies (warm and cold SFGs, respectively) and from proto-spheroidal galaxies (spheroids), computed using the model by \cite{Cai2013}. The thick curve shows the average flux density, computed using eq.~(\ref{eq:cluster1}). The upper curve is a factor of 3 higher and illustrates the large variance of the cluster IR emission at fixed mass \cite{Alberts2014, Alberts2016}. It must be noted that the \textit{Planck} and the SPT surveys have detected $z>0.5$ clusters with masses of more that $10^{15}\,\hbox{M}_\odot$ (see text). The uppermost horizontal line shows to the PCCS2 90\% completeness level in the `extragalactic zone' for the \textit{Planck} channel nearest in frequency. The two lower horizontal lines in the left panel show the point source detection limits for the 1-m and 1.5-m telescope options. In the right-hand panel only the detection limit for the 1.5-m option is shown.}
\label{fig:IRcluster}
\end{figure}

\section{Evolution of the IR emission of galaxy clusters}\label{sect:cluster}

The serendipitously detected proto-clusters are likely in different evolutionary stages, from early agglomeration towards the collapse to virial equilibrium. There is, however, also clear evidence of strong evolution in the star-formation activity in virialized clusters \citep[][and references therein]{Wagner2016}. The specific SFR increases rapidly from $z\sim 0.2$ to $z\sim 1.3$, mostly driven by the activation of star formation in early-type galaxies at $z\simgt 0.9$. \mission will allow us to investigate the evolution of galaxy populations in such clusters, detected via their X-ray emission \cite{Mehrtens2012, Merloni2012, Bohringer2013} or via the thermal SZ (tSZ) effect \cite{Hasselfield2013, Bleem2015, PlanckCollaboration2015SZcatalog}.

The South Pole Telescope (SPT) tSZ survey sample contains clusters with median mass of $M_{500}=3.5\times 10^{14}\,\hbox{M}_\odot$\footnote{We define $M_{500}$ as the total mass within the radius for which the mean overdensity is 500 times the \textit{mean matter density} of the Universe. Note that frequently $M_{500}$ is defined to correspond to an overdensity 500 times the \textit{critical density}.} and covers the redshift range $0.047 \le z \simlt 1.7$ \cite{Bleem2015}. The \textit{Planck} survey \cite{PlanckCollaboration2015SZcatalog} has been particularly effective at detecting massive ($M_{500}> 5\times 10^{14}\,\hbox{M}_\odot$, and up to $\simeq 2 \times 10^{15}\,\hbox{M}_\odot$) clusters at $z>0.5$ (and up to $z\simeq 1$). Using a stacking approach for a sample of 645 clusters whose SZ signal was detected by \textit{Planck}, Ref.~\cite{PlanckCollaboration2016cluster_dust} found a significant detection of dust emission from 353 to 857 GHz with an average SED shape similar to that of the Milky Way. Further evidence of dust emission from \textit{Planck} SZ clusters was provided by the detection of a cross-correlation between the tSZ effect and the cosmic infrared background \cite[CIB;][]{PlanckCollaboration2015SZ_CIB}.

At $z\lsim 1$  dense cluster cores are preferentially populated by massive, passively evolving, early-type galaxies. Star-forming galaxies are generally found in the cluster outskirts and in the field. Ref.~\cite{Alberts2014} showed that the star-formation activity in cluster core ($r< 0.5\,$Mpc) galaxies from $z = 0.3$ to 1.5 is at the field level at $z \simgt 1.2$ but is suppressed faster at lower redshifts. A similar conclusion was reached by studying a sample of 57 groups and clusters in the range $0<z<1.6$ using the deepest surveys with \textit{Spitzer} MIPS (Multi-Band Imaging Photometer for \textit{Spitzer}) and \textit{Herschel} Photoconductor Array Camera and Spectrometer (PACS) and SPIRE, on blank and cluster fields \cite{Popesso2015}.

The ratio of the mean IR luminosity of galaxies in clusters (within $1\,$Mpc from the cluster centre) to that of galaxies in the field, for $z< 1.2$, was found to be \cite{Alberts2014}
\begin{equation}\label{eq:f}
f(z)={L_{\rm IR, cluster}\over L_{\rm IR, field}}\simeq 5.77 e^{-0.34 t_{\rm Gyr}},
\end{equation}
with large uncertainties; here $t_{\rm Gyr}$ is the cosmic time in Gyr. The cluster IR luminosity then can be written
\begin{equation}\label{eq:cluster1}
L_{\rm IR, cluster, tot}(z) = \Psi_{\rm IR}(z)\,f(z)\,\delta\, V_{\rm comoving} = \Psi_{\rm IR}(z)\,f(z)\,M_{\rm cluster}/\langle \rho_0\rangle,
\end{equation}
where $\Psi_{\rm IR}(z)$ is the IR \textit{comoving} luminosity density,  $\delta$ is the cluster density contrast, $\rho_{\rm cluster}/\langle \rho \rangle$, $\langle \rho\rangle$ is the mean matter density whose present day value is $\langle \rho_0\rangle=2.7755\times 10^{11}h^2 \Omega_{\rm m} \,\hbox{M}_\odot \hbox{Mpc}^{-3}$, $V_{\rm comoving}$ is the cluster volume and $M_{\rm cluster}$ is the cluster mass.

According to figure~17 of Ref.~\cite{Gruppioni2013}, $\Psi_{\rm IR}(z)\propto (1+z)^3$ and for the standard $\Lambda$CDM cosmology we have, approximately, $f(z)\propto (1+z)^{3.8}$ for $z\simlt 1$, so that the cluster IR luminosity at fixed mass evolves very strongly [$L_{\rm IR, cluster, tot}(z)\propto (1+z)^{6.8}$, up to $z\sim 1$].

Ref.~\cite{Alberts2016}  presented a detailed study of star formation in 11 near-IR selected, spectroscopically confirmed, massive ($M\simgt 10^{14}\,\hbox{M}_\odot$) clusters at $1 < z < 1.75$. They found that the star formation in cluster galaxies at $z\simgt 1.4$ is largely consistent with field galaxies at similar epochs, confirming earlier results from the Spitzer/IRAC Shallow Cluster Survey \cite{Brodwin2013}. Although the samples were uniformly selected there are quite strong variations in star formation from cluster to cluster.

We have used these results to make tentative predictions of expected flux densities of clusters as a whole.  
Figure~\ref{fig:IRcluster} shows the expected flux densities due to dust emission from a cluster with mass $M_{500}=10^{14.5}\,\hbox{M}_\odot$ as a function of redshift. The cluster luminosities were obtained using eq.~(\ref{eq:cluster1}), and the contributions of late-type, starburst and proto-spheroidal galaxies were computed using the model from Ref.~\cite{Cai2013}.

As shown in Fig.~\ref{fig:IRcluster} we expect that at 800\,GHz the \mission\!\!150 version will detect the dust emission of the brightest $M_{500}\simeq 10^{14}\,\hbox{M}_\odot$ clusters and of typical $M_{500}\simeq 10^{14.5}\,\hbox{M}_\odot$ clusters at $1\simlt z \simlt 1.5$. More massive clusters will be detected over broader redshift ranges. Stacking will allow us to carry out a statistical investigation of the evolution of the cluster IR emission over a much broader redshift baseline than has been possible so far. Targets for stacking will abound. By the time \mission will fly, eROSITA (extended ROentgen Survey with an Imaging Telescope Array), expected to be launched in 2017, should have provided an all-sky deep X-ray survey detecting $\sim 10^5$ galaxy clusters out to $z>1$ \cite{Merloni2012}. About 38,000 cluster detections out to $z\simgt 2$ are expected, via the tSZ effect, from the baseline \mission mission, and about $50,000$ with \mission\!\!150; combining \mission\!\!150 and CMB-S4 data it will be possible to detect around 200,000 clusters in the sky visible
from the South Pole (Melin et al., in preparation).

For comparison, \textit{Herschel} has allowed the study of the IR emission from clusters up to $z\simeq 1.7$ \cite{Alberts2016}, but the sample comprises only 11 clusters. The \textit{Herschel} data have shown large variations in cluster properties, highlighting the need for evolutionary studies of large, uniform cluster samples over a broad redshift range. \mission will fulfill this need.

The situation is somewhat less favourable at $600\,$GHz, where, however, \mission will nevertheless substantially improve over \textit{Planck}. At this frequency a direct source detection by \mission will be possible only for cluster masses $M_{500}\simgt 10^{14.5}\,\hbox{M}_\odot$. But again, the stacking approach will allow substantial progress in the exploration of the evolution of the IR emission as a function of cluster mass and redshift. This will shed light on the mechanisms which drive the evolution of massive spheroidal galaxies in clusters, from actively star forming to passively evolving in dense environments, and on the role of nuclear activity in this process.

Figure~\ref{fig:IRcluster} shows that, according to the model in Ref.~\cite{Cai2013}, different galaxy populations are expected to dominate the cluster sub-mm emission at different redshifts. At $z \simlt 0.5$ the main contribution comes from late-type galaxies with relatively cold dust temperatures. This is consistent with the dust temperature, $T_{\rm d}\simeq 24\,$K, estimated in Ref.~\cite{PlanckCollaboration2016cluster_dust} for their sample of clusters detected by \textit{Planck}, with an estimated mean redshift  $z=0.26\pm 0.17$. This study also found a slight increase of dust temperature with redshift, comparing clusters at $z$ larger and smaller than 0.25. Such an increase is indeed predicted by the model, due to the increasing contribution of the warmer starburst population that takes over above $z\simeq 0.5$ and out to $z\simeq 1.7$. At still higher redshifts the dominant population is the proto-spheroidal galaxies in the process of forming most of their stars.

Once more, we will take advantage of the synergies with CMB-S4, as pointed out in the previous sections. The characterization of the IR emission in individual clusters, obtained combining \mission and CMB-S4 data, is essential to extract the relativistic correction in the hottest clusters and the contribution from non-thermal electrons to the SZ effect \cite{EnsslinKaiser2000, Colafrancesco2011}. Relativistic corrections would allow us to estimate in an independent way the plasma temperature while non-thermal contributions provide information on outflows from star formation and/or from nuclear activity.



\section{The Cosmic Infrared Background (CIB)}\label{sect:CIB}

Progress in our understanding of the power spectrum of the CIB that will be made possible by CORE has been discussed in Ref.~\cite{DeZotti2015}. Briefly, the  better \mission angular resolution, compared to \textit{Planck}, will allow us to measure, in a uniform way, the CIB power spectrum over an unprecedented range of frequencies and of angular scales (from $\sim 1\,$arcmin to tens of degrees), thus breaking the degeneracy between the Poisson contribution and that of non-linear effects (one-halo term), that complicates the interpretation of \textit{Planck} measurements without resorting to external data. Accurate determinations of the CIB power spectrum at different frequencies provide, on the one hand constraints on the evolution of the cosmic star-formation density and, on the other hand, on halo masses associated with sources of the CIB \cite{PlanckCollaboration2011CIB, PlanckCollaborationXXX2014, Mak2016}.

The exploitation of CIB anisotropies as a probe of star-formation history in the Universe and also as a dark matter tracer  requires a careful removal of foreground radiations, the most important of which, at sub-mm wavelengths, is the Galactic thermal dust emission. The separation of dust and CIB components is challenging since both have frequency spectra scaling approximately as modified black-bodies with similar emissivity indices. An approach successfully applied to \textit{Planck} temperature maps combines the frequency information, to reconstruct the Galactic thermal dust model, with spatial information, taking advantage of different power spectra of the components \cite{PlanckCollaboration2016CIB_Gal_dust}. In fact, the CIB power spectrum scales approximately as $\ell^{-1}$ \cite{PlanckCollaboration2011CIB}, while the Galactic dust power spectrum scales approximately as $\ell^{-2.7}$  or $\ell^{-2.8}$ for $\ell>110$ \cite{PlanckCollaborationXXX2014}. With its much more numerous frequency channels, higher sensitivity and better angular resolution, \mission will substantially improve over \textit{Planck} both in the spectral and spatial characterization, thus enabling a much easier and  more robust separation of the two components. Obviously, a better Galactic thermal dust model will also allow an improved cleaning of CMB maps.

Moreover, the large number of \mission channels will make it possible to investigate in detail the decorrelation of power spectra measured in different frequency bands. A decorrelation is expected because the redshift distribution of CIB sources shifts to higher and higher redshifts with decreasing frequency. Accurate measurements of the CIB cross-spectra for different frequency channels set strong constraints on the frequency dependence of redshift distributions, hence on the evolution of the cosmic SFR. Ref.~\cite{WuDore2016} found that the \mission design is very close to the optimal design for improving the SFR constraints at all redshifts.

As pointed out by Ref.~\cite{Tucci2016}, the angular power spectrum of the CIB is a sensitive probe of primordial non-Gaussianity. It potentially outperforms the forthcoming large-scale galaxy redshift surveys because the CIB traces the large-scale structure over a much larger comoving volume. Again, the key factor is an accurate subtraction of the Galactic thermal dust emission, achievable with the numerous sub-mm channels of \mission\!\!. This will make possible to constrain the ``local type'' primordial non-Gaussianity parameter called $f_{\rm nl}$  down to an amplitude $|f_{\rm nl}|<1$.

Another interesting issue is the possibility of exploiting the multi-frequency measurements of the dipole spectrum to constrain the CIB intensity spectrum \cite{DaneseDeZotti1981, Piat2002, Balashev2015, DeZotti2016}. This arises from the fact that the dipole amplitude is directly proportional to the first derivative of the photon occupation number. High accuracy measurements of the CIB dipole amplitude at sub-mm wavelengths will provide important constraints on the CIB intensity, currently known with a $\simeq 30\%$ uncertainty \citep{Fixsen1998}. Such a large uncertainty constitutes a major limitation to our understanding of the dust-obscured star formation phase of galaxy evolution. The exploitation of this possibility requires a very accurate inter-channel cross-calibration and a careful control of the systematics from the foreground Galaxy subtraction \cite{FixsenKashlinsky2011}. A detailed discussion of this topic will be presented in a companion paper (Burigana et al., in preparation).

\begin{figure}
\centering
\includegraphics[width=9.cm]{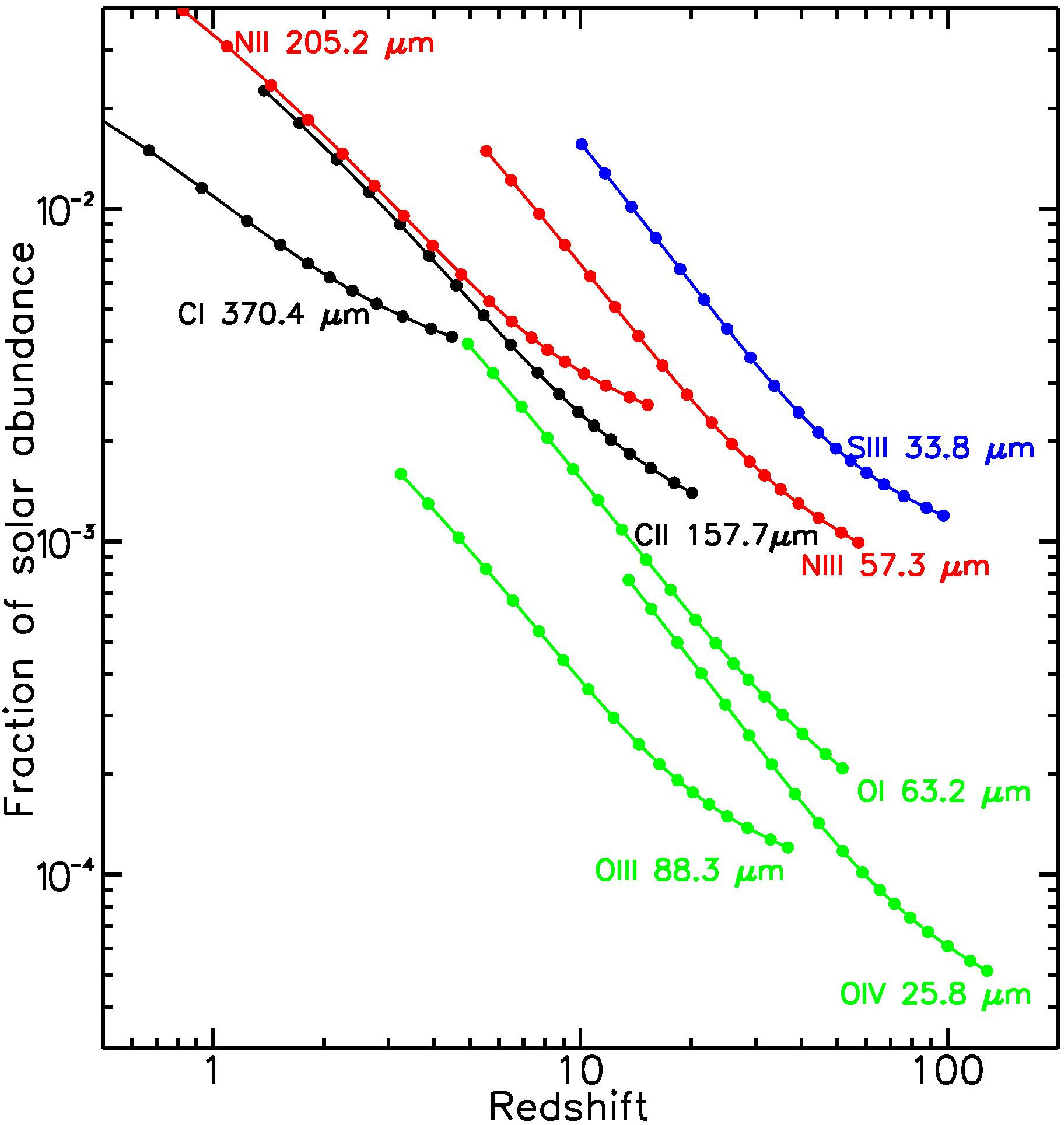}
\caption[fig:metalsI]{Constraints on the abundance of several metal and ion species as a function of redshift, obtained by comparing the power spectra measured in different \mission channels with that measured at 80\,GHz. This assumes an inter-channel calibration accurate at the level of 0.001\,\%, and that the uncertainties in the beam characterization can be neglected.}
\label{fig:metalsI}
\end{figure}

\begin{figure}
\centering
\includegraphics[width=\columnwidth]{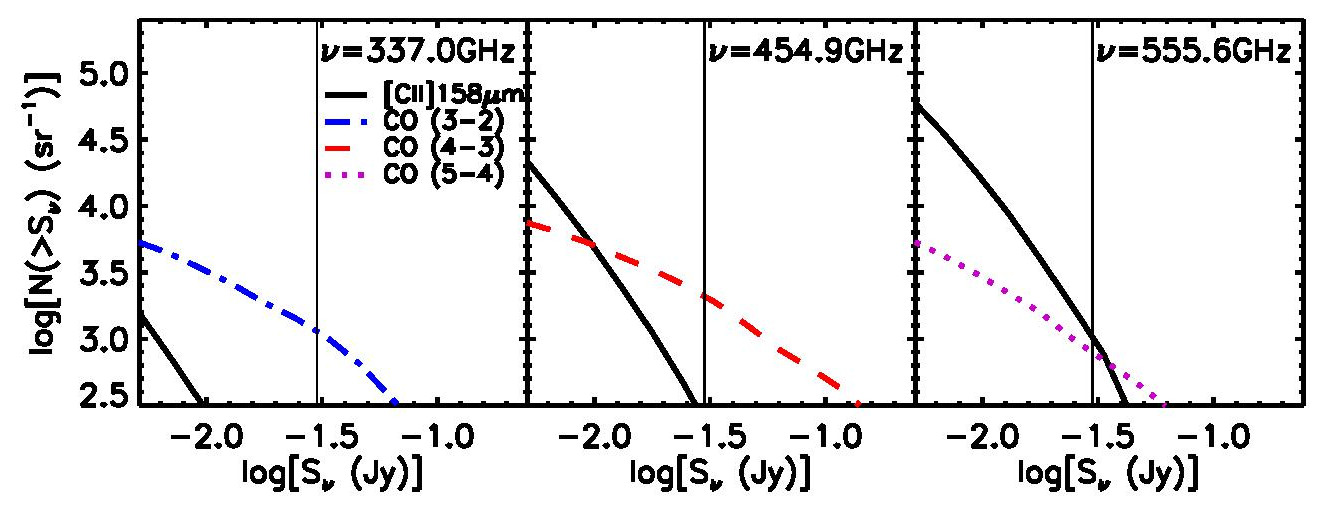}
\caption[fig:lines]{Integral counts of CO and C$+$ lines at three frequencies within the \mission range with a resolution of $R=20$. The vertical solid lines correspond to a detection limit of 30\,mJy. }
\label{fig:lines}
\end{figure}

\section{Resonant scattering on metals and ions at high redshifts}\label{sect:lines}

The multi-frequency \mission observations will not only allow us to separate the usual foreground signals, such as the diffuse Galactic emission and point sources, but might also enable the identification of subtler astrophysical signals. In particular, metals and ions in the intergalactic medium (IGM) leave signatures in the anisotropy pattern of the CMB intensity and polarization.

As shown in Refs.~\cite{Basu04} and \cite{CHMetal07}, in low over-density regions the dominant mechanism coupling the CMB photons with metals and ions is resonant scattering. On large angular scales such scattering blurs the intrinsic anisotropy pattern and induces a frequency dependent variation of the observed angular power spectrum given by
\begin{equation}
\delta C^{ji}_l \equiv C^j_l - C^i_l \approx -2\delta \tau^{ji}_M\, C_l^{\rm intr} =
         -2(\tau^j_M-\tau^i_M) \, C_l^{\rm intr},
\label{eq:dCl1}
\end{equation}
where $C_l^{\rm intr}$ is the intrinsic CMB angular power spectrum and $\delta \tau^{ji}_M$ is the increment of the optical depth
associated with resonant transitions of one or several species corresponding to two different observing frequencies $\nu_i$ and $\nu_j$.


Low frequency channels are affected by resonant lines produced at high redshifts. But at high enough redshifts the abundance of metals and ions is negligible; hence the CMB power spectrum measured at low frequencies is the intrinsic one, so that low frequency channels can be used as a reference, against which to compare observations at higher frequencies that correspond to lower redshift lines. This comparison allows us to set constraints on the abundance of different species of metals and ions at different cosmological epochs \citep[see ][for more details]{Basu04,CHMetal07}.

However, real life is significantly more complicated. Besides foregrounds there are other issues, the most important being inter-channel calibration and beam characterization \citep[][]{CHMetal06}. In the future, the proposed PIXIE experiment \cite{Kogut2016} could provide absolutely calibrated low-resolution maps of the sky to improve the channel inter-calibration and mitigate this problem. If we assume that foregrounds can be projected out down to negligible levels and that we can achieve an inter-channel calibration at the 0.001\% level (approximately $100$ times better than  for the {\it Planck}-HFI cosmological channels), then a comparison of the power spectrum measured by \mission at 80\,GHz with those measured at higher frequencies would yield the constraints (at 3\,$\sigma$ or $> 99.7\,$\% confidence level) on different species shown in Fig.~\ref{fig:metalsI}. These limits are obtained assuming that only one transition contributes to the signal, and that the beams can be characterized with arbitrary precision. For a given species, constraints are tighter at higher redshifts/lower frequencies, where the control of free-free and synchrotron emission is more demanding.

High sensitivity, multi-frequency CMB observations are also sensitive to other signals involving atoms and molecules, such as the collisional/rotational emission of CO and the C$+\,157.7\,\mu$m line \cite[][]{righi08,DeZotti2016,Mashian2016} or the Field-Wouthuysen effect on OI atoms \citep{chmOI07,chmOI08}. The poor spectral resolution of CMB experiments implies a strong dilution of the lines, making them undetectable in individual sources. But an improvement of the spectral resolution, $R=\nu/\Delta\nu$, by a factor of several, at least for some high frequency channels, would make possible the detection of thousands of sources in these lines. Examples are shown in Fig.~\ref{fig:lines} for a modest $R=20$.

The integral counts shown in this figure were worked out exploiting the model in Ref.~\cite{Cai2013} for the cosmological evolution of the IR luminosity function, coupled with relations between line intensities and IR luminosities. For the CO lines we have used the relations  of Ref.~\cite{Greve2014} and for the C$+\,157.7\,\mu$m line that by Ref.~\cite{Bonato2014}. More details on the calculations can be found in Ref.~\cite{DeZotti2016}. As illustrated by Fig.~\ref{fig:lines}, an all-sky survey with $R=20$ is expected to detect line emission brighter than 30\,mJy (a flux density limit achievable by \mission\!\!150) of $\sim 10^4$ galaxies in the ``extragalactic zone''. The C$+\,157.7\,\mu$m line becomes increasingly important with increasing frequency.

\subsection{Radio sources}

The overwhelming majority of extragalactic radio sources detectable by \mission are blazars, i.e. sources whose radio emission is dominated by Doppler-boosted emission from relativistic jets closely aligned with the line of sight \cite{UrryPadovani1995}. These objects, with their extreme properties, are of special interest since they are also strong $\gamma$--ray sources. About 90\% of the firmly identified extragalactic sources and about 94\% of the ``associated'' sources (i.e.~of sources having a counterpart with a $> 80\%$ probability of being the real identification) in the Third \textit{Fermi} Large Area Telescope (LAT) catalogue \cite{Acero2015} are blazars.

Accurate source counts over large flux density intervals provide key constraints on blazar evolutionary models. Because high frequency surveys are still far less extensive than those at low frequencies, evolutionary models for blazar populations, flat spectrum radio quasars (FSRQs) and BL Lacertae sources (BL Lacs), are less developed than those for steep-spectrum radio sources \cite{DeZotti2005, Massardi2010}. For example, while clear evidence for downsizing\footnote{``Downsizing'' refers to the very different evolutionary behaviour of high- and low-luminosity sources, in the sense that the redshift of the peak space density of sources decreases with luminosity \cite{Cowie1996}. } was reported in the case of steep-spectrum sources \cite{Massardi2010, Rigby2015}, the available data are insufficient to test if this is also the case for FSRQs; for BL Lacs the constraints on evolutionary parameters are even weaker. This situation hampers sharp tests of unified models of radio sources.

Furthermore, source counts constrain the compactness of the synchrotron-emitting regions in the jets. Already the counts at mm wavelengths available in 2011 allowed Ref.~\cite{Tucci2011} to rule out simple power-law spectra for these sources and to constrain models for spectral breaks of the synchrotron emission in blazar jets. Furthermore, the assumption of uniform spectral properties for the whole blazar population was found to be, at best, only marginally consistent with the data. A clearly better match with the observed counts was obtained by assuming that BL Lacs have, on average, substantially higher break frequencies than FSRQs, suggesting that the synchrotron emission comes from more compact regions. This led to the prediction of a substantial increase of the BL Lac fraction at still higher frequencies and at bright flux densities. To test this prediction surveys at sub-mm wavelengths, such as those carried out by \mission\!\!, are paramount.

In order to characterize the blazar synchrotron peak it is crucial to push the photometry to sub-mm wavelengths, which will be provided by \mission\!\!. Ref.~\cite{PlanckCollaboration2011radioSED} investigated the global spectral energy distributions (SEDs) of a sample of 104 extragalactic radio sources drawn from the \textit{Planck} Early Release Compact Source Catalogue (ERCSC), combining \textit{Planck} data with simultaneous observations ranging from radio to $\gamma$--rays. They have shown that \textit{Planck} data provide key information on the energy spectrum of relativistic electrons responsible for the synchrotron emission, with implications for the acceleration mechanisms provided by shocks along the relativistic jets.

The further analysis in Ref.~\cite{PlanckCollaboration2016radio_spectra} has confirmed that the flattest high-frequency radio spectral indices are close to zero, indicating that the original energy spectrum of accelerated electrons is much harder than commonly thought, with power-law index around 1.5 instead of the canonical 2.5.  The radio spectra of these sources peak at remarkably high
frequencies, tens of GHz, and there are significant statistical differences between subclasses of Active Galactic Nuclei (AGN) at high frequencies, particularly between highly-polarized quasars and BL Lac objects.

\begin{figure}
\vskip-1cm
\includegraphics[width=\columnwidth, angle=180]{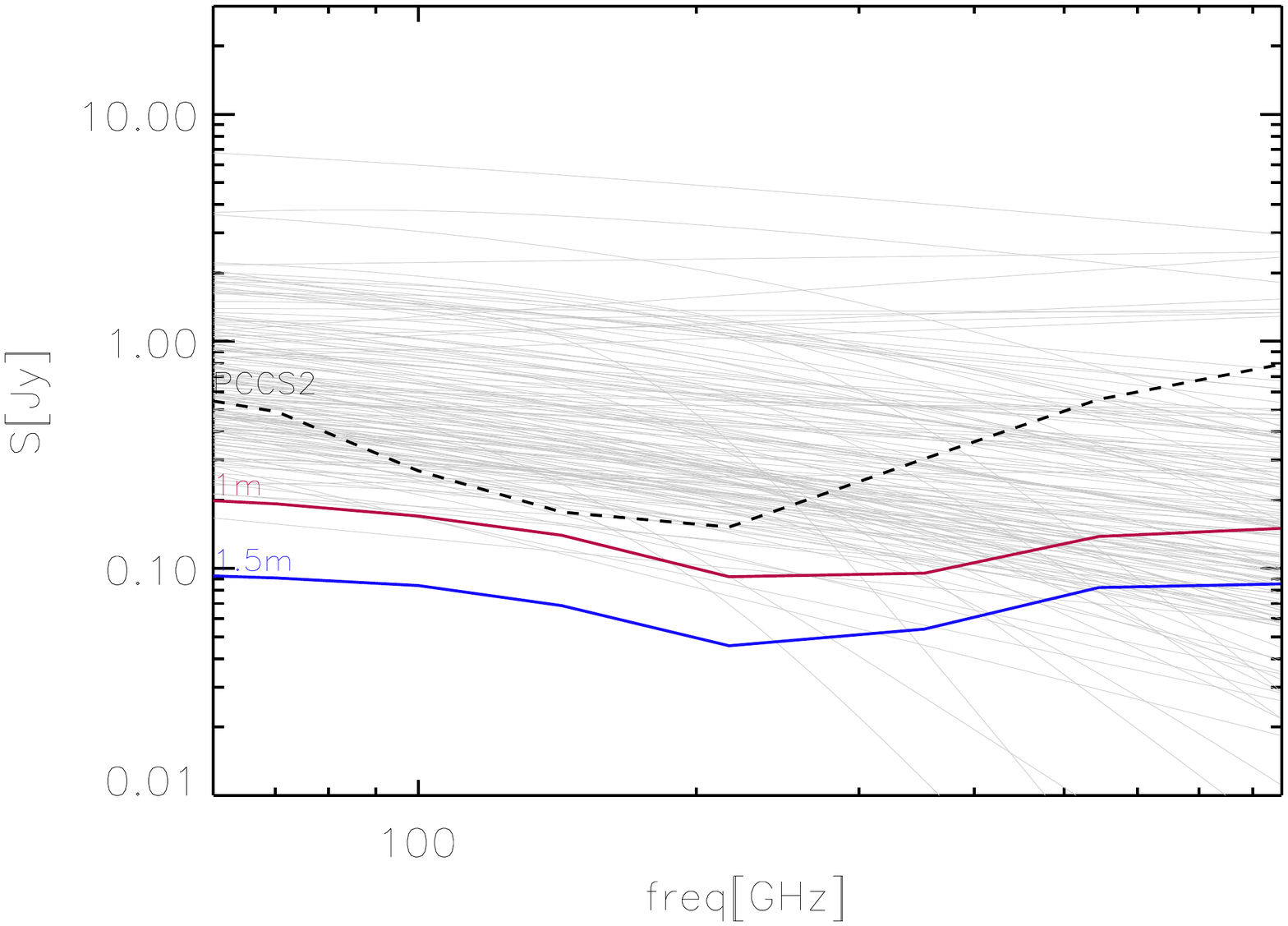}
\vskip-1cm
\caption{Continuum spectra of PACO-bright sources compared with the 90\% completeness limits of the \textit{Planck}/PCCS2 catalogue (dashed black line) and with the $4\,\sigma$ detection limits for the \mission 1-m (\mission\!\!100) and 1.5-m (\mission\!\!150) telescope sizes (solid red and blue lines, respectively). Note that, particularly at the highest frequencies, the fraction of PACO-bright sources whose fitted spectra are above the \textit{Planck}/PCCS2 90\% completeness limits is substantially larger than the number of actual \textit{Planck}/PCCS2 detections, listed in Table~\ref{tab:PACO}; this may suggest that the fits over-predict the flux densities at frequencies beyond the highest frequency detections or that the PCCS2 completeness level is somewhat overestimated.}
\label{fig:PACO}
\end{figure}

An interesting open question is the geometry of the emitting region. The data analysed in Ref.~\cite{Massardi2016}  are consistent with a single synchrotron emission zone, but their analysis was limited to frequencies up to 217\,GHz. However, simultaneous SED data, including \textit{Planck} photometry, for some particularly bright sources suggest double emission zones peaking at different frequencies with a transition occurring at sub-mm/far-IR wavelengths \cite[][and references therein]{Cutini2014}. As illustrated by Fig.~\ref{fig:PACO} and by Table~\ref{tab:PACO}, the deeper sub-mm surveys by \mission will increase by about two orders of magnitude the number of sources detected up to 600\, GHz. The improved constraints on the spectral energy distributions at sub-mm wavelengths entail correspondingly better constraints on the geometry of the emitting regions.

Key information on the flow of the plasma within the relativistic jets is provided by variability. Signatures of evolving shocks in the strongest radio flares were seen in a study of \textit{Planck} sources \cite{PlanckCollaboration2016radio_spectra}, although much of the high frequency variability may be better approximated by achromatic variations. These results are compatible with the standard shocked jet model, but other interpretations are also possible. \mission will shed light on this by providing much larger samples and a broader spectral coverage.

Definite conclusions on all the above issues are currently hampered by the limited statistics. To illustrate the improvement made possible by \mission with respect to \textit{Planck} we have considered the complete \textit{Planck}-ATCA Co-eval Observations (PACO) ``bright'' sample \cite{Massardi2016}. The sample comprises the 189 sources with AT20G flux densities $S_{20\rm GHz} > 500\,$mJy, only a few of which were detected by \textit{Planck} beyond 217\,GHz (see Table~\ref{tab:PACO}). Figure~\ref{fig:PACO} shows the continuum spectra of the subset of 145 sources for which we could obtain good fits with smooth functions that can be used to extrapolate the spectra to higher frequencies.\footnote{We have updated the analysis of Ref.~\cite{Massardi2016} using the PCCS2 data. The number of sources with smooth spectra decreased from 147 to 145, with no significant change to the earlier conclusions.} Using such fits we find that the fraction of these sources detectable by \mission is $\ge 95\%$ for \mission\!\!100, and $\simeq 100\%$ for \mission\!\!150, up to 353\,GHz (cf. Fig.~\ref{fig:PACO} and Table~\ref{tab:PACO})

\begin{table*}
\centering
\caption{Number of \textit{Planck}/PCCS2 counterparts to sources in the PACO-bright sample, compared with the number of expected \mission detections for the 1-m and 1.5-m telescope options. }
\vskip12pt
\scriptsize
\begin{tabular}{lrrrrrr}
\hline
\hline
Frequency [GHz]  & \multicolumn{1}{c}{70}            &   \multicolumn{1}{c}{100}        &   \multicolumn{1}{c}{143}        &    \multicolumn{1}{c}{217}         &    \multicolumn{1}{c}{353}       &  \multicolumn{1}{c}{545}   \\ \hline 
PCCS2           & 103 (71\%)     &   132 (91\%)  &   135 (93\%) &     91 (63\%) &      3 (2\%) &    1 (0.7\%)   \\ \hline 
\mission\!\!100  & 143 (99\%)     &   142 (98\%)  &   141 (97\%) &    141 (97\%) &    137 (95\%) &  91 (63\%)  \\ \hline 
\mission\!\!150 & 145 (100\%)   & 145  (100\%)  &  145 (100\%) &  145 (100\%)  &    142 (98\%) &  120 (83\%) \\  
\hline
\hline
\end{tabular}
\label{tab:PACO}
\end{table*}

Together with \mission\!\!, CMB-S4 will extend by a factor of 30 the frequency range over which accurate number counts of radio sources down to tens to hundred mJy will be determined. At the moment, counts down to at least these levels are available only at $\nu \le 20\,$GHz \cite[see Ref.][for a review]{DeZotti2010}, and between 95 and 220\,GHz, where the \textit{Planck} counts are supplemented at fainter flux densities by those from the South Pole Telescope \cite[SPT;][]{Mocanu2013} and from the Atacama Cosmology Telescope \cite[ACT;][]{Marsden2014}. At other frequencies $\nu\ge 30\,$GHz the best current information on radio source counts comes from \textit{Planck} \cite{PlanckCollaboration2011stat_prop} and is limited to flux densities not much below 1\,Jy.

The characterization of the blazar emission at both mm and sub-mm wavelengths is an essential ingredient for understanding the physics of these sources \cite{PlanckCollaboration2011radioSED, Giommi2012, LeonTavares2012}. There is an obvious synergy, in this respect, between CMB-S4 and \mission surveys. The synergy extends also to the study of variability properties. Multi-epoch (sub-)millimetre observations can serendipitously detect variable objects and flares in BL Lacs and FSRQs, particularly in regions with high scan coverage, as seen by \textit{Planck} in the blazar S$5\,1803+78$ \cite{Rachen2016}. Planck/HFI carried out five full surveys of the sky, while \mission is expected to make six to ten full surveys. In combination with CMB-S4 observations, \mission will enable a multifrequency study of variability on scales from months to several years.

\begin{table*}
\centering
\begin{small}
\caption{Simulations of extragalactic sources in polarization: technical specifications}\label{tab:simula}
\vskip 12pt
\begin{tabular}{rrrrrrrrrrr}
\hline
\hline
     &  \multicolumn{4}{c}{\mission\!\!100 (1-m telescope)} &\ & \multicolumn{4}{c}{\mission\!\!150 (1.5-m telescope)} \\
   \cline{2-5}\cline{7-10}  \vspace{-8pt}\\

\multicolumn{1}{c}{Freq.} & \multicolumn{1}{c}{FWHM} & \multicolumn{1}{c}{Noise} & \multicolumn{1}{c}{Pix. size} & \multicolumn{1}{c}{Pixel no.} &  & \multicolumn{1}{c}{FWHM} & \multicolumn{1}{c}{Noise} & \multicolumn{1}{c}{Pix. size} & \multicolumn{1}{c}{Pixel no.}\\
\multicolumn{1}{c}{[GHz]} & \multicolumn{1}{c}{[arcmin]} & \multicolumn{1}{c}{[mJy]} &  \multicolumn{1}{c}{[arcsec]} & & & \multicolumn{1}{c}{[arcmin]} & \multicolumn{1}{c}{[mJy]} &  \multicolumn{1}{c}{[arcsec]} \\ \hline\hline
60  & 21.0  &  3.26 & 175 &   $512\times 512$ & & 14.0 & 1.54 & 110 &  $512\times 512$ \\
100 & 12.6  &  1.88 & 105 & $1024\times 1024$ & &  8.4 & 0.52 &  70 & $1024\times 1024$ \\
145 &  8.7  &  0.98 &  70 & $1024\times 1024$ & &  5.8 & 0.40 &  50 & $1024\times 1024$ \\
220 &  5.7  &  1.18 &  45 & $1024\times 1024$ & &  3.8 & 0.52 &  30 & $2048\times 2048$ \\
340 &  3.7  &  3.10 &  30 & $2048\times 2048$ & &  2.5 & 1.06 &  20 & $2048\times 2048$ \\
450 &  2.8  &  5.30 &  20 & $2048\times 2048$ & &  1.9 & 1.66 &  10 & $2048\times 2048$ \\
600 &  2.1  &  6.70 &  20 & $2048\times 2048$ & &  1.4 & 2.84 &  10 & $2048\times 2048$ \\
800 &  ...  &  ...  & ... & ...               & &  1.1 & 4.04 &  10 & $2048\times 2048$ \\
\hline
\hline
\end{tabular}
\end{small}
\end{table*}

\begin{figure}
\includegraphics[width=\columnwidth]{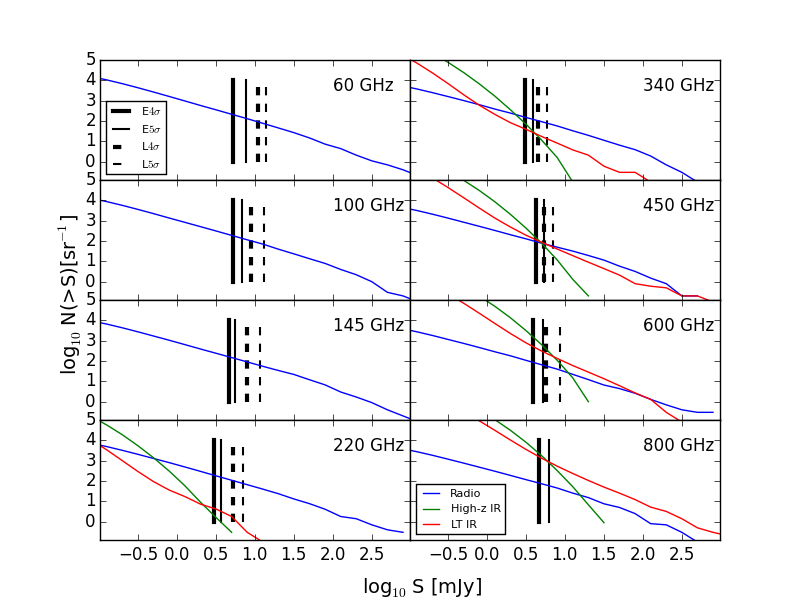}
\caption{Comparison of the estimated source number counts in polarization for a selection of \mission channels and different source populations: radio sources (solid blue line); and two populations of dusty galaxies (proto-spheroids and late-type, spiral and starburst, galaxies). Proto-spheroids, labelled ``High-z IR'' (solid green line) dominate at faint flux densities while late-types  (LT IR, solid red lines) dominate at the brighter flux densities. The vertical lines show the $4\sigma$ and  $5\sigma$ detection limits obtained from the simulations for the 1-m ({dashed}) and 1.5-m ({solid}) telescope (see Tables \ref{tab:light} and \ref{tab:extended} for more details).}
\label{fig:both}
\end{figure}

\begin{figure}
\begin{center}
\includegraphics[width=0.7\columnwidth]{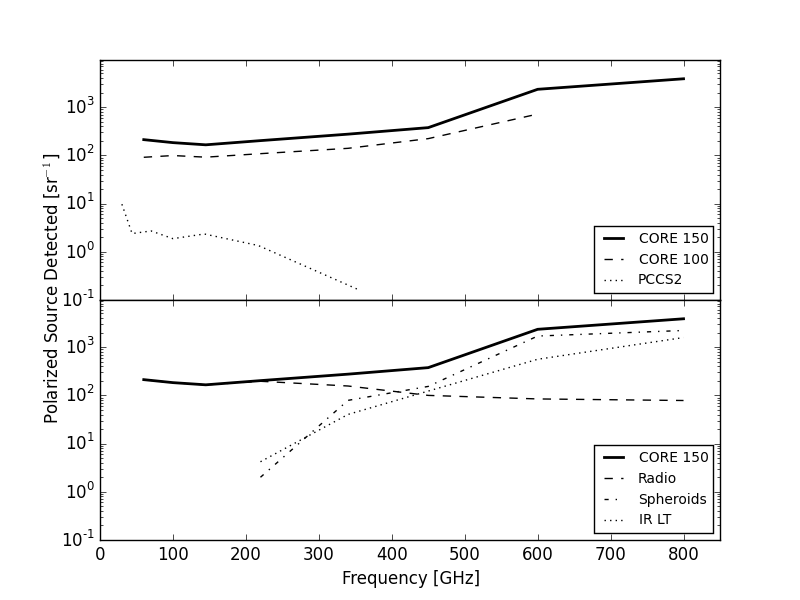}
\caption{\textbf{Upper panel:} predicted numbers of extragalactic sources detected in polarized flux density as a function of frequency for the two \mission configurations (solid and dashed lines) compared to the numbers of sources detected in polarization by \textit{Planck} (PCCS2, dotted line). \textbf{Lower panel:} contributions of the different source populations to the counts in polarized flux density as a function of frequency for the \mission\!\!150 configuration. The dashed, dot-dashed and dotted lines refer to radio sources (``radio''), proto-spheroidal galaxies (``spheroids'') and late-type galaxies ``IR LT''), respectively, while the thick solid line shows the total.  }
\label{fig:pol_det_lim}
\end{center}
\end{figure}

\section{Detecting sources in polarization}\label{sect:counts_pol}

The preliminary estimates \cite{DeZotti2015} suggested that \mission will allow a real breakthrough in the characterization of the polarization properties of extragalactic sources. However, those estimates were only tentative, being based on simplifying assumptions that allowed  analytical calculations.  We have now re-assessed them by means of realistic simulations in polarization (the simulations described in Ref.~\cite{DeZotti2015} were only in total intensity), carried out for the two options for the telescope size.

The relevant technical details of the simulations are summarized in Table~\ref{tab:simula}. The size of the simulated patch varies with frequency, on account of the varying angular resolution of the survey, quantified by the FWHM of the telescope beam. In the 1-m telescope case, the patch size decreases from $619\,\hbox{deg}^2$ at 60\,GHz to $129\,\hbox{deg}^2$ at 600\,GHz; in the 1.5-m case it decreases from $245\,\hbox{deg}^2$ at 60\,GHz to $33\,\hbox{deg}^2$ at 800\,GHz. Given the smallness of the patches, a flat-sky approximation was used.


Our simulations include both radio sources and star-forming galaxies. The latter comprise two sub-populations, late-type galaxies and what we are referring to as proto-spheroids, which have different clustering properties \cite[e.g.][]{Xia2012, Cai2013}. We started from source counts using models that accurately reproduce observational data, specifically the model in Ref.~\cite{Tucci2011} for radio sources and the model in Ref.~\cite{Cai2013} for star-forming galaxies.

We have selected the sets of frequencies, specified in Table~\ref{tab:simula}, in the range covered by \mission and for each frequency we populated sky patches with sources in the flux density range 0.01\,mJy--100\,Jy, assigning to each source a flux density drawn at random from the number counts. For each simulation we checked the consistency, within Poisson errors, of the counts of simulated sources with those given by the model. Then we associated with each source a polarization degree randomly extracted from probability distributions appropriate for each source population, as detailed below. The polarization angles were drawn at random from a uniform probability distribution.

The most extensive study of the polarization properties of extragalactic radio sources at high radio frequencies was carried out in Ref.~\cite{Massardi2013}. These authors obtained polarization data for 180 extragalactic sources extracted from the Australia Telescope 20-GHz (AT20G) survey catalogue and observed with the Australia Telescope Compact Array (ATCA) during a dedicated, high-sensitivity run ($\sigma_{\rm p}\simeq 1\,$mJy). Complementing their data with polarization information for seven extended sources from the 9--yr Wilkinson Microwave Anisotropy Probe (WMAP) co-added maps at 23\,GHz, they obtained a roughly $99\%$ complete sample of extragalactic sources brighter than $S_{20\rm GHz} = 500\,$mJy at the selection epoch.

The sample has a 91.4\% detection rate in polarization at 20\,GHz. The results are in general agreement with those of other polarization surveys at high radio frequencies, carefully reviewed in Ref.~\cite{TucciToffolatti2012} and, most recently, in Ref.~\cite{GalluzziMassardi2016}. The distribution of polarization degrees was found to be well described by a log-normal function with mean and dispersion of $2.14\%$ and $0.90\%$, respectively. We have assumed that this distribution holds at all the frequencies considered, consistent with the results of Ref.~\cite{Battye2011}.

Most recently, Ref.~\cite{Bonavera2017} applied the stacking technique to estimate the average fractional polarization from 30 to 353 GHz of a sample of 1560 compact sources - essentially all radio sources - detected in the 30 GHz Planck all-sky map. The average and median polarization fractions for the 881 sources, presumably extragalactic, outside the Galactic mask were found to be $\simeq 3\%$ and $\simeq 1.9\%$, approximately independent of frequency.

In the case of star-forming galaxies, the polarized emission in the \mission frequency range is dominated by dust, except perhaps at the lowest frequencies, where the polarization due to synchrotron emission may take over, at least for low-$z$ galaxies. But at low frequencies the polarized emission of these galaxies is undetectable by \mission in any case.

Polarization properties of dusty galaxies as a whole are almost completely unexplored. The only available information has come from  SCUPOL, the polarimeter for SCUBA on the James Clerk Maxwell Telescope, that has provided polarization measurements at $850\,\mu$m for only two galaxies, M\,82 \cite{GreavesHolland2002} and M\,87 \cite{Matthews2009}. However the global polarization degree has been published only for M\,82 and is $\Pi = 0.4\%$. This low value is in keeping with the notion that the polarized dust emission integrated over the whole galaxy is reduced because of the complex structure of galactic magnetic fields, with reversals along the line of sight, and also of the disordered alignment of dust grains. For our simulations we adopted a log-normal probability distribution of $\Pi$ with mean of 0.5\% and dispersion of $1.0\%$.

Integrating the \textit{Planck} dust polarization maps over a $20^\circ$ wide band centred on the Galactic plane we find an average value of the Stokes $Q$ parameter of about 2.7\%. We may then expect a similar value for spiral galaxies seen edge-on. For a galaxy seen with an inclination angle $\theta$ we expect that the polarization degree is reduced by a factor $\cos(\theta)$. If all galaxies are about as polarized as ours, the average polarization fraction for unresolved galaxies, averaged over all possible orientations, should be about half of 2.7\%, i.e.~around 1.4\%. If so our choice of the mean polarisation degree would be quite conservative and the number of detections in polarisation of dusty galaxies would be underestimated by a factor of around $4$.

The angular clustering of the radio sources is strongly diluted by the broadness of their luminosity functions, implying that objects distributed over distance ranges much larger than the clustering radius contribute to the counts at any flux density. This effect makes their clustering essentially irrelevant in the present context.

Analyses of clustering properties of star-forming galaxies \cite{Xia2012, Cai2013} have demonstrated a strong difference between late-types and proto-spheroids. The clustering of the former objects adds a minor contribution to the intensity fluctuations due to point sources, which, in the \mission frequency range, are dominated by the latter population. Ignoring the clustering properties of late-type galaxies is thus a good approximation, since we are only interested in characterizing the fluctuation field; these galaxies have therefore been distributed at random inside the simulated sky area.

In contrast, for proto-spheroidal galaxies we used the method elaborated  in Ref.~\cite{GonzalezNuevo2005}, which allowed us to distribute the sources consistently with their clustering properties, as described by the angular power spectrum, $P(k)$. At each frequency we adopted the $P(k\vert \nu)$ given by the  model of Ref.~\cite{Cai2013} that fits the available clustering data on dusty galaxies.

We have used the latest version of the Planck Sky Model \cite[PSM;][]{Delabrouille2013} to check that, in ``clean'' regions at high Galactic latitudes ($|b|> 30^\circ$), the contribution of Galactic emission to fluctuations at the \mission resolution is small compared to noise and, except at the highest \mission frequencies, to CMB fluctuations. The PSM currently incorporates all the available information on the various emission components, thus providing a realistic representation of the sky. We have therefore neglected the Galactic emissions. This implies that the detection limits derived below are lower limits for regions with substantial Galactic emission.

As for the CMB, we made use of the PSM software package to generate all-sky intensity and polarization maps in the HEALPix pixelization with high resolution, i.e. with $\hbox{N}_{\rm side}=8192$, corresponding to a pixel size of 25.65\,arcsec. Regions of the chosen size and angular resolution were projected onto the plane and added to the source maps (that were built on a planar surface). Each map was then filtered with a Gaussian beam function with the appropriated FWHM, and the instrumental noise was added; the adopted values for the FWHM and the noise are specified in Table~\ref{tab:simula}.

\begin{table*}
\centering
\begin{small}
\caption{Estimated detection limits in polarized flux density and surface densities of sources brighter than such limits for the \mission configuration with a 1-m telescope.} \vspace{7pt}
\begin{tabular}{lrrrrrrrrrr}
\hline\hline
 & & & \multicolumn{2}{c}{Radio} &\ &\multicolumn{2}{c}{Proto-sph} &\ & \multicolumn{2}{c}{Late-type} \\
 \cline{4-5}\cline{7-8}\cline{10-11} \vspace{-8pt}\\
Freq. & $P_{4\sigma}$ & $P_{5\sigma}$ & $N_{4\sigma}$ & $N_{5\sigma}$ & & $N_{4\sigma}$ & $N_{5\sigma}$ & & $N_{4\sigma}$ & $N_{5\sigma}$\\
 \hbox{[GHz]}  & [mJy] & [mJy] & [sr$^{-1}$] & [sr$^{-1}$] & & [sr$^{-1}$] & [sr$^{-1}$] & & [sr$^{-1}$] & [sr$^{-1}$] \\
\hline%
 60 & 11.0 &  13.9 &  91 &  69 & &  ...  &    ... & &   ... &   ... \\
100 &  9.0 &  13.2 &  99 &   65 & &  ...  &    ... &  &  ... &   ... \\
145 & 8.0 &  11.7 &  92 &  61 &  & ...  &    ... &   & ... &   ... \\
220 & 5.2 &   7.0 &  106 &  77 &  &  0.3 &    0.1 &   & 1.7 &    1.1 \\
340 & 4.5 &   5.8 &  103 &  78 &  & 17.0 &    6.1 &  & 20.3 &   13.3 \\
450 & 5.3 &   7.0 &  79 &  59 &  & 62.2 &   19.6 &  & 81.1 &   49.7 \\
600 & 5.7 &   8.7 &  55 &  34 &  & 403.4 &   69.7 & & 260.5 &  116.9\\
\hline \hline
\end{tabular}
\label{tab:light}
\end{small}
\end{table*}

\begin{table*}
\centering
\begin{small}
\caption{Estimated detection limits in polarized flux density and surface densities of sources brighter than such limits for the \mission configuration with a 1.5-m telescope.}
\vspace{7pt}
\begin{tabular}{rrrrrrrrrrr}
\hline\hline%
 & & & \multicolumn{2}{c}{Radio} &\ & \multicolumn{2}{c}{Proto-sph} &\ & \multicolumn{2}{c}{Late-type} \\
 \cline{4-5}\cline{7-8}\cline{10-11} \vspace{-8pt} \\
Freq. & $P_{4\sigma}$ & $P_{5\sigma}$ & $N_{4\sigma}$ & $N_{5\sigma}$ & & $N_{4\sigma}$ & $N_{5\sigma}$ & & $N_{4\sigma}$ & $N_{5\sigma}$ \\
\hbox{[GHz]} & [mJy] & [mJy] & [sr$^{-1}$] & [sr$^{-1}$] & & [sr$^{-1}$] & [sr$^{-1}$] & & [sr$^{-1}$] & [sr$^{-1}$] \\
\hline%
 60 & 5.2 & 7.7 & 212 & 137 & & ... &  ...  & & ... &  ...\\
100 & 5.2 & 6.9 & 184 & 134 &  & ... &  ...  &  & ... &  ...\\
145 & 4.6 & 5.6 & 165 & 134 &  & ... & ...  &  & ... &  ...\\
220 & 3.0 & 3.7 & 196 & 154 &  &  2 &   1  &  &  4 &   3 \\
340 & 3.1 & 3.9 & 156 & 122 &  & 79 &  32  & &  40 &  27 \\
450 & 4.2 & 5.3 & 100 &  78 &  & 153 &  60  & & 123 &  80 \\
600 & 3.9 & 5.2 &  84 &  61 & & 1699 & 596 & & 560 & 317 \\
800 & 4.7 & 6.2 &  78 &  58 & & 2215 & 817 & & 1581 & 891 \\
\hline\hline %
\end{tabular}
\label{tab:extended}
\end{small}
\end{table*}

We have applied the Mexican-hat wavelet\,2 as described in Ref.~\cite{GonzalezNuevo2006} and \cite{LopezCaniego2006}. This filtering approach allows an efficient point-source detection and extraction, since it effectively removes CMB structures or extended emission. This cleaning was carried out before attempting any detection. The rms of the final filtered patch was computed after masking all sources detected with signal-to-noise ratio $\hbox{S/N} > 4$.


In fact the size of patches over which we simulated the source distribution were too small to contain a statistically significant number of sources with polarized flux densities detected with $\hbox{S/N} > 4$. To determine the detection limits we then injected, at each frequency, fake sources with gradually increasing polarized flux densities. At least five sources per polarized flux density bin were introduced in the maps, distributed with a separation at least three times larger than the beam size. 


The source detection algorithm was then applied blindly to each of the simulated patches, producing a list of $4\sigma$ detections. Once the detection limits were determined at each frequency, the surface density of sources brighter than such limits were obtained from the model. The estimated detection limits and number of sources per steradian above them are given in Tables~\ref{tab:light} and \ref{tab:extended} and illustrated by Fig.~\ref{fig:both}; results are presented for a set of frequencies in the \mission range and for the 1-m and 1.5-m options.

Objects detected in polarization are essentially only radio sources for frequencies up to around $200\,$GHz. The number of detected dusty galaxies increases with increasing frequency. For the adopted distribution of polarization fraction, \mission will detect similar numbers of radio sources and of dusty galaxies at $\simeq 500\,$GHz. At 600\,GHz the latter population will dominate, reaching an integral count of $\simeq 660\,\hbox{sr}^{-1}$ for sources detected with $\hbox{S/N} > 4$. Of course these estimates are only guesses, since the distribution of the global polarization fraction of dusty galaxies is unknown. We will need the measurements of \mission to probe the polarization distribution.

The \mission\!\!150 configuration will do much better, reaching thousands of $\hbox{S/N} > 4$ detections per steradian in polarization of both proto-spheroidal and late-type galaxies at its highest frequency (800\,GHz). As illustrated by Fig.~\ref{fig:pol_det_lim} the progress with respect to \textit{Planck} polarization results is spectacular. In fact, in the ``extragalactic zone'', \textit{Planck} detected only a few tens of extragalactic objects in polarisation, all of them radio sources.\footnote{The PCCS2 lists hundreds of detections in polarization, mostly at low Galactic latitudes. The surface densities shown in Fig.~\ref{fig:pol_det_lim} are our own estimates based on detections in the ``extragalactic zone'' above the 90\% completeness limit. }

Polarization surveys of extragalactic sources are essential to control the contamination of CMB polarimetry experiments with arcmin pixel sizes, and carry crucial information on the source physics. Polarimetric properties of radio-loud AGN at millimetric and sub-millimetric wavelengths tell us about magnetic fields and about the plasma in the inner, unresolved regions of their relativistic jets. Additionally, polarimetry of dusty galaxies as a function of their inclination is informative on the structure and on the ordering of large-scale magnetic fields.

Finally, we note that, like in the total intensity case, CMB-S4 will primarily detect radio sources, reaching deeper flux density levels than \mission\!\!. In turn, \mission will extend the frequency coverage and the time-span for variability studies. For studies of polarization properties of dusty galaxies, \mission data will be unique.


\section{Conclusions}\label{sec:conclusions}

In spite of their sensitivities, now approaching fundamental limits, modern space-borne CMB experiments have provided only shallow surveys of extragalactic sources. The key limitation is source confusion, i.e. intensity peaks due to random positive fluctuations of faint sources that mimic real individual sources. The rms confusion noise scales approximately as the square of the {FWHM} (see figure~3 of \cite{DeZotti2015}). Hence the substantially better resolution of the diffraction-limited \mission telescope, compared to \textit{Planck}, especially at the highest frequencies, offers a large advantage in terms of detection limits. The advantage is further boosted in terms of the number of detected sources by the steepness of the source counts.

The power of all-sky surveys at mm and sub-mm wavelengths has already been vividly demonstrated by \textit{Planck}, which has detected thousands of dusty galaxies as well as many hundreds of extragalactic radio sources in this spectral range, which is difficult or impossible to explore from the ground and only lightly surveyed by other space missions. \textit{Planck} photometry proved to be crucially important to characterize the synchrotron peak of blazars. Surveys at mm wavelengths are the most effective way to select this class of sources, which, among other things, constitute the overwhelming majority of the identified extragalactic $\gamma$--ray sources detected by the \textit{Fermi}-LAT. \textit{Planck} data have also provided key information on the energy spectrum of relativistic electrons responsible for the synchrotron emission with interesting implications for their acceleration mechanisms.

\textit{Planck} detected several of the most extreme, strongly-lensed, high-$z$ galaxies, with estimated gravitational amplifications, $\mu$, of up to 50; \mission will detect thousands of strongly lensed galaxies. Strong lensing offers the opportunity of detailed follow-up studies of high-$z$ galaxies with otherwise unattainable sensitivity: the exposure time to reach a given flux density limit varies as $\mu^{-2}$ and, since lensing conserves surface brightness, it stretches the image, thus effectively increasing the angular resolution by a substantial factor (at least in one dimension).

Moreover, \textit{Planck} proved to be a powerful probe of the evolution of the large-scale structure of the Universe in the phase when early-type galaxies, the dominant population in rich clusters and groups of galaxies, were forming the bulk of their stars.

Beyond strongly expanding the samples of source populations detected by \textit{Planck}, \mission will open new windows. In particular, it will: (i) provide unbiased, flux limited samples of dense proto-cluster cores of star-forming galaxies, some examples of which were detected by \textit{Herschel}; (ii) allow a detailed investigation, via direct detections complemented with stacking analysis, of the evolution of the star-formation rate in virialized galaxy clusters detected by surveys of the SZ effect (including those carried out by \mission itself) or by X-ray surveys; (iii) provide the spectrum of the CIB dipole anisotropy, which contains important information on the average CIB intensity spectrum; and (iv) provide the first blind high frequency census of the polarization properties of radio sources and of star-forming galaxies.

\acknowledgments
Work supported in part by ASI/INAF agreement n.~2014-024-R.1 and by PRIN--INAF 2014 ``Probing the AGN/galaxy co-evolution through ultra-deep and ultra-high resolution radio surveys''. JGN acknowledges financial support from the Spanish MINECO for a ``Ramon y Cajal'' fellowship (RYC-2013-13256) and the I$+$D 2015 project AYA2015-65887-P (MINECO/FEDER). CH-M acknowledges the financial support of the Spanish MINECO through I$+$D project AYA-2015-66211-C2-2-P. MB is supported by the Netherlands Organization for Scientific Research, NWO, through grant number 614.001.451, and by the Polish National Science Centre under contract UMO-2012/07/D/ST9/02785. CJM is supported by an FCT Research Professorship, contract reference IF/00064/2012, funded by FCT/MCTES (Portugal) and POPH/FSE (EC). GR acknowledges support from the National Research Foundation of Korea (NRF) through NRF-SGER 2014055950 funded by the Korean Ministry of Education, Science and Technology (MoEST), and from the faculty research fund of Sejong University in 2016.

\bibliography{extragal_ref}

\providecommand{\href}[2]{#2}\begingroup\raggedright\begin{thebibliography}{100}

\bibitem{DeZotti2015}
G.~{De Zotti}, G.~{Castex}, J.~{Gonz{\'a}lez-Nuevo}, M.~{Lopez-Caniego},
  M.~{Negrello}, Z.-Y. {Cai}, M.~{Clemens}, J.~{Delabrouille}, D.~{Herranz},
  L.~{Bonavera}, J.-B. {Melin}, M.~{Tucci}, S.~{Serjeant}, M.~{Bilicki},
  P.~{Andreani}, D.~L. {Clements}, L.~{Toffolatti}, and B.~F. {Roukema}, {\it
  {Extragalactic sources in Cosmic Microwave Background maps}},  {\em \jcap}
  {\bf 6} (June, 2015) 018, [\href{http://arxiv.org/abs/1501.02170}{{\tt
  arXiv:1501.02170}}].

\bibitem{Planck_parameters2015}
{Planck Collaboration XIII}, {\it {Planck 2015 results. XIII. Cosmological
  parameters}},  {\em \aap} {\bf 594} (Sept., 2016) A13,
  [\href{http://arxiv.org/abs/1502.01589}{{\tt arXiv:1502.01589}}].

\bibitem{Delabrouille2017}
J.~{Delabrouille}, P.~{de Bernardis}, M.~{Bersanelli}, F.~{Bouchet},
  F.~{Boulanger}, S.~{Hanani}, B.~{Maffei}, D.~{McCarthy}, and {et al.},
  ``{Exploring Cosmic Origins with CORE: Survey requirements and mission
  design}.'' Feb., 2017.

\bibitem{deBernardis2017}
P.~{de Bernardis}, P.~A.~R. {Ade}, J.~{Baselmans}, E.~S. {Battistelli},
  A.~{Benoit}, M.~{Bersanelli}, A.~{Bideaud}, M.~{Calvo}, and {et al.},
  ``{Exploring Cosmic Origins with CORE: The Instrument}.'' Feb., 2017.

\bibitem{Ashdown2017}
M.~{Ashdown}, R.~{Banerji}, J.~{Borrill}, A.~{Buzzelli}, G.~{De Gasperis},
  J.~{Delabrouille}, E.~{Hivon}, H.~D. {Thuong}, and {et al.}, ``{Exploring
  Cosmic Origins with CORE: Mitigation of systematic effects}.'' Feb., 2017.

\bibitem{Remazeilles2017}
M.~{Remazeilles}, A.~J. {Banday}, C.~{Baccigalupi}, S.~{Basak}, A.~{Bonaldi},
  G.~{De Zotti}, J.~{Delabrouille}, C.~{Dickinson}, and {et al.}, ``{Exploring
  Cosmic Origins with CORE: $B$-mode Component Separation}.'' Feb., 2017.

\bibitem{DiValentino2016}
E.~{Di Valentino}, T.~{Brinckmann}, M.~{Gerbino}, V.~{Poulin}, F.~R. {Bouchet},
  J.~{Lesgourgues}, A.~{Melchiorri}, J.~{Chluba}, S.~{Clesse},
  J.~{Delabrouille}, C.~{Dvorkin}, F.~{Forastieri}, S.~{Galli}, D.~C. {Hooper},
  M.~{Lattanzi}, C.~J.~A.~P. {Martins}, L.~{Salvati}, G.~{Cabass}, A.~{Caputo},
  E.~{Giusarma}, E.~{Hivon}, P.~{Natoli}, L.~{Pagano}, S.~{Paradiso}, J.~A.
  {Rubino-Martin}, A.~{Achucarro}, M.~{Ballardini}, N.~{Bartolo}, D.~{Baumann},
  J.~G. {Bartlett}, P.~{de Bernardis}, A.~{Bonaldi}, M.~{Bucher}, Z.-Y. {Cai},
  G.~{De Zotti}, J.~M. {Diego}, J.~{Errard}, S.~{Ferraro}, F.~{Finelli}, R.~T.
  {Genova-Santos}, J.~{Gonzalez-Nuevo}, S.~{Grandis}, J.~{Greenslade},
  S.~{Hagstotz}, W.~{Handley}, M.~{Hindmarsh}, C.~{Hernandez-Monteagudo},
  K.~{Kiiveri}, M.~{Kunz}, A.~{Lasenby}, M.~{Liguori}, M.~{Lopez-Caniego},
  G.~{Luzzi}, J.-B. {Melin}, J.~J. {Mohr}, M.~{Negrello}, D.~{Paoletti},
  M.~{Remazeilles}, C.~{Ringeval}, J.~{Valiviita}, B.~{Van Tent}, V.~{Vennin},
  N.~{Vittorio}, and {the CORE collaboration}, {\it {Exploring Cosmic Origins
  with CORE: Cosmological Parameters}},  {\em ArXiv e-prints} (Nov., 2016)
  [\href{http://arxiv.org/abs/1612.00021}{{\tt arXiv:1612.00021}}].

\bibitem{Finelli2016}
{CORE Collaboration}, F.~{Finelli}, M.~{Bucher}, A.~{Ach{\'u}carro},
  M.~{Ballardini}, N.~{Bartolo}, D.~{Baumann}, S.~{Clesse}, J.~{Errard},
  W.~{Handley}, M.~{Hindmarsh}, K.~{Kiiveri}, M.~{Kunz}, A.~{Lasenby},
  M.~{Liguori}, D.~{Paoletti}, C.~{Ringeval}, J.~{V{\"a}liviita}, B.~{van
  Tent}, V.~{Vennin}, F.~{Arroja}, M.~{Ashdown}, A.~J. {Banday}, R.~{Banerji},
  J.~{Baselmans}, J.~G. {Bartlett}, P.~{de Bernardis}, M.~{Bersanelli},
  A.~{Bonaldi}, J.~{Borril}, F.~R. {Bouchet}, F.~{Boulanger}, T.~{Brinckmann},
  Z.-Y. {Cai}, M.~{Calvo}, A.~{Challinor}, J.~{Chluba}, G.~{D'Amico},
  J.~{Delabrouille}, J.~{Mar{\'{\i}}a Diego}, G.~{De Zotti}, V.~{Desjacques},
  E.~{Di Valentino}, S.~{Feeney}, J.~R. {Fergusson}, S.~{Ferraro},
  F.~{Forastieri}, S.~{Galli}, J.~{Garc{\'{\i}}a-Bellido}, R.~T.
  {G{\'e}nova-Santos}, M.~{Gerbino}, J.~{Gonz{\'a}lez-Nuevo}, S.~{Grandis},
  J.~{Greenslade}, S.~{Hagstotz}, S.~{Hanany}, D.~K. {Hazra},
  C.~{Hern{\'a}ndez-Monteagudo}, E.~{Hivon}, B.~{Hu}, E.~D. {Kovetz},
  H.~{Kurki-Suonio}, M.~{Lattanzi}, J.~{Lesgourgues}, J.~{Lizarraga},
  M.~{L{\'o}pez-Caniego}, G.~{Luzzi}, B.~{Maffei}, C.~J.~A.~P. {Martins},
  E.~{Mart{\'{\i}}nez-Gonz{\'a}lez}, D.~{McCarthy}, S.~{Matarrese},
  A.~{Melchiorri}, J.-B. {Melin}, A.~{Monfardini}, P.~{Natoli}, M.~{Negrello},
  F.~{Oppizzi}, E.~{Pajer}, S.~P. {Patil}, M.~{Piat}, G.~{Pisano}, V.~{Poulin},
  A.~{Ravenni}, M.~{Remazeilles}, A.~{Renzi}, D.~{Roest}, L.~{Salvati},
  A.~{Tartari}, G.~{Tasinato}, J.~{Torrado}, N.~{Trappe}, M.~{Tucci},
  J.~{Urrestilla}, P.~{Vielva}, and R.~{Van de Weygaert}, {\it {Exploring
  Cosmic Origins with CORE: Inflation}},  {\em ArXiv e-prints} (Dec., 2016)
  [\href{http://arxiv.org/abs/1612.08270}{{\tt arXiv:1612.08270}}].

\bibitem{Bartlett2017}
J.~G. {Bartlett}, J.-B. {Melin}, D.~{Tramonte}, J.~A. {Rubi\~no-Martin}, and
  {et al.}, ``{Exploring Cosmic Origins with CORE: Large-Scale Structure
  Science}.'' Feb., 2017.

\bibitem{Melin2017}
J.-B. {Melin}, A.~{Bonaldi}, M.~{Remazeilles}, S.~{Hagstotz}, J.~M. {Diego},
  C.~{Hern\'andez-Monteagudo}, R.-T. {G\'enova-Santos}, G.~{Luzzi}, and {et
  al.}, ``{Exploring Cosmic Origins with CORE: Cluster Science}.'' Feb., 2017.

\bibitem{Burigana2017}
C.~{Burigana}, C.~S. {Carvalho}, T.~{Trombetti}, A.~{Notari}, M.~{Quartin},
  G.~{De Gasperis}, A.~{Buzzelli}, N.~{Vittorio}, and {et al.}, ``{Exploring
  Cosmic Origins with CORE: implications of observer peculiar motion effects in
  the CMB}.'' Feb., 2017.

\bibitem{PCCS2_2015}
{Planck Collaboration XXVI}, {\it {Planck 2015 results. XXVI. The Second Planck
  Catalogue of Compact Sources}},  {\em \aap} {\bf 594} (Sept., 2016) A26,
  [\href{http://arxiv.org/abs/1507.02058}{{\tt arXiv:1507.02058}}].

\bibitem{Negrello2013}
M.~{Negrello}, M.~{Clemens}, J.~{Gonzalez-Nuevo}, G.~{De Zotti}, L.~{Bonavera},
  G.~{Cosco}, G.~{Guarese}, L.~{Boaretto}, S.~{Serjeant}, L.~{Toffolatti},
  A.~{Lapi}, M.~{Bethermin}, G.~{Castex}, D.~L. {Clements}, J.~{Delabrouille},
  H.~{Dole}, A.~{Franceschini}, N.~{Mandolesi}, L.~{Marchetti}, B.~{Partridge},
  and A.~{Sajina}, {\it {The local luminosity function of star-forming galaxies
  derived from the Planck Early Release Compact Source Catalogue}},  {\em
  \mnras} {\bf 429} (Feb., 2013) 1309--1323,
  [\href{http://arxiv.org/abs/1211.3832}{{\tt arXiv:1211.3832}}].

\bibitem{Marchetti2016}
L.~{Marchetti}, M.~{Vaccari}, A.~{Franceschini}, V.~{Arumugam}, H.~{Aussel},
  M.~{B{\'e}thermin}, J.~{Bock}, A.~{Boselli}, V.~{Buat}, D.~{Burgarella},
  D.~L. {Clements}, A.~{Conley}, L.~{Conversi}, A.~{Cooray}, C.~D. {Dowell},
  D.~{Farrah}, A.~{Feltre}, J.~{Glenn}, M.~{Griffin}, E.~{Hatziminaoglou},
  S.~{Heinis}, E.~{Ibar}, R.~J. {Ivison}, H.~T. {Nguyen}, B.~{O'Halloran},
  S.~J. {Oliver}, M.~J. {Page}, A.~{Papageorgiou}, C.~P. {Pearson},
  I.~{P{\'e}rez-Fournon}, M.~{Pohlen}, D.~{Rigopoulou}, I.~G. {Roseboom},
  M.~{Rowan-Robinson}, B.~{Schulz}, D.~{Scott}, N.~{Seymour}, D.~L. {Shupe},
  A.~J. {Smith}, M.~{Symeonidis}, I.~{Valtchanov}, M.~{Viero}, L.~{Wang},
  J.~{Wardlow}, C.~K. {Xu}, and M.~{Zemcov}, {\it {The HerMES submillimetre
  local and low-redshift luminosity functions}},  {\em \mnras} {\bf 456} (Feb.,
  2016) 1999--2023, [\href{http://arxiv.org/abs/1511.06167}{{\tt
  arXiv:1511.06167}}].

\bibitem{Kuehn2014}
K.~{Kuehn}, J.~{Lawrence}, D.~M. {Brown}, S.~{Case}, M.~{Colless},
  R.~{Content}, L.~{Gers}, J.~{Gilbert}, M.~{Goodwin}, A.~M. {Hopkins},
  M.~{Ireland}, N.~P.~F. {Lorente}, R.~{Muller}, V.~{Nichani}, A.~{Rakman},
  S.~N. {Richards}, W.~{Saunders}, N.~F. {Staszak}, J.~{Tims}, and L.~G.
  {Waller}, {\it {TAIPAN: optical spectroscopy with StarBugs}},  in {\em
  Ground-based and Airborne Instrumentation for Astronomy V}, vol.~9147 of {\em
  \procspie}, p.~914710, July, 2014.

\bibitem{Comparat2016}
J.~{Comparat}, C.-H. {Chuang}, S.~{Rodr{\'{\i}}guez-Torres},
  M.~{Pellejero-Ibanez}, F.~{Prada}, G.~{Yepes}, H.~M. {Courtois}, G.-B.
  {Zhao}, Y.~{Wang}, J.~{Sanchez}, C.~{Maraston}, R.~B. {Metcalf},
  J.~{Peiro-Perez}, F.~S. {Kitaura}, E.~{P{\'e}rez}, and R.~M. {Gonz{\'a}lez
  Delgado}, {\it {The Low Redshift survey at Calar Alto (LoRCA)}},  {\em
  \mnras} {\bf 458} (May, 2016) 2940--2952,
  [\href{http://arxiv.org/abs/1510.00147}{{\tt arXiv:1510.00147}}].

\bibitem{Dore2014}
O.~{Dor{\'e}}, J.~{Bock}, M.~{Ashby}, P.~{Capak}, A.~{Cooray}, R.~{de Putter},
  T.~{Eifler}, N.~{Flagey}, Y.~{Gong}, S.~{Habib}, K.~{Heitmann}, C.~{Hirata},
  W.-S. {Jeong}, R.~{Katti}, P.~{Korngut}, E.~{Krause}, D.-H. {Lee},
  D.~{Masters}, P.~{Mauskopf}, G.~{Melnick}, B.~{Mennesson}, H.~{Nguyen},
  K.~{{\"O}berg}, A.~{Pullen}, A.~{Raccanelli}, R.~{Smith}, Y.-S. {Song},
  V.~{Tolls}, S.~{Unwin}, T.~{Venumadhav}, M.~{Viero}, M.~{Werner}, and
  M.~{Zemcov}, {\it {Cosmology with the SPHEREX All-Sky Spectral Survey}},
  {\em ArXiv e-prints} (Dec., 2014) [\href{http://arxiv.org/abs/1412.4872}{{\tt
  arXiv:1412.4872}}].

\bibitem{Bilicki2014}
M.~{Bilicki}, T.~H. {Jarrett}, J.~A. {Peacock}, M.~E. {Cluver}, and
  L.~{Steward}, {\it {Two Micron All Sky Survey Photometric Redshift Catalog: A
  Comprehensive Three-dimensional Census of the Whole Sky}},  {\em \apjs} {\bf
  210} (Jan., 2014) 9, [\href{http://arxiv.org/abs/1311.5246}{{\tt
  arXiv:1311.5246}}].

\bibitem{Bilicki2016}
M.~{Bilicki}, J.~A. {Peacock}, T.~H. {Jarrett}, M.~E. {Cluver}, N.~{Maddox},
  M.~J.~I. {Brown}, E.~N. {Taylor}, N.~C. {Hambly}, A.~{Solarz}, B.~W.
  {Holwerda}, I.~{Baldry}, J.~{Loveday}, A.~{Moffett}, A.~M. {Hopkins}, S.~P.
  {Driver}, M.~{Alpaslan}, and J.~{Bland-Hawthorn}, {\it {WISE {$\times$}
  SuperCOSMOS Photometric Redshift Catalog: 20 Million Galaxies over $3\,\pi$
  Steradians}},  {\em \apjs} {\bf 225} (July, 2016) 5,
  [\href{http://arxiv.org/abs/1607.01182}{{\tt arXiv:1607.01182}}].

\bibitem{LopezCaniego2007}
M.~{L{\'o}pez-Caniego}, J.~{Gonz{\'a}lez-Nuevo}, D.~{Herranz}, M.~{Massardi},
  J.~L. {Sanz}, G.~{De Zotti}, L.~{Toffolatti}, and F.~{Arg{\"u}eso}, {\it
  {Nonblind Catalog of Extragalactic Point Sources from the Wilkinson Microwave
  Anisotropy Probe (WMAP) First 3 Year Survey Data}},  {\em \apjs} {\bf 170}
  (May, 2007) 108--125, [\href{http://arxiv.org/abs/astro-ph/0701473}{{\tt
  astro-ph/0701473}}].

\bibitem{Malek2014}
K.~{Ma{\l}ek}, A.~{Pollo}, T.~T. {Takeuchi}, V.~{Buat}, D.~{Burgarella},
  M.~{Malkan}, E.~{Giovannoli}, A.~{Kurek}, and S.~{Matsuura}, {\it {Properties
  of star forming galaxies in AKARI Deep Field-South}},  {\em \aap} {\bf 562}
  (Feb., 2014) A15, [\href{http://arxiv.org/abs/1312.0765}{{\tt
  arXiv:1312.0765}}].

\bibitem{Solarz2015}
A.~{Solarz}, A.~{Pollo}, T.~T. {Takeuchi}, K.~{Ma{\l}ek}, H.~{Matsuhara}, G.~J.
  {White}, A.~{P{\c e}piak}, T.~{Goto}, T.~{Wada}, S.~{Oyabu}, T.~{Takagi},
  Y.~{Ohyama}, C.~P. {Pearson}, H.~{Hanami}, T.~{Ishigaki}, and M.~{Malkan},
  {\it {Clustering of the AKARI NEP deep field 24 {$\mu$}m selected galaxies}},
   {\em \aap} {\bf 582} (Oct., 2015) A58,
  [\href{http://arxiv.org/abs/1509.00219}{{\tt arXiv:1509.00219}}].

\bibitem{Pollo2013}
A.~{Pollo}, T.~T. {Takeuchi}, A.~{Solarz}, P.~{Rybka}, T.~L. {Suzuki}, A.~{P{\c
  e}piak}, and S.~{Oyabu}, {\it {Clustering of far-infrared galaxies in the
  AKARI All-Sky Survey North}},  {\em Earth, Planets, and Space} {\bf 65}
  (Oct., 2013) 1109--1116.

\bibitem{Abazajian2016}
K.~N. {Abazajian}, P.~{Adshead}, Z.~{Ahmed}, S.~W. {Allen}, D.~{Alonso}, K.~S.
  {Arnold}, C.~{Baccigalupi}, J.~G. {Bartlett}, N.~{Battaglia}, B.~A. {Benson},
  C.~A. {Bischoff}, J.~{Borrill}, V.~{Buza}, E.~{Calabrese}, R.~{Caldwell},
  J.~E. {Carlstrom}, C.~L. {Chang}, T.~M. {Crawford}, F.-Y. {Cyr-Racine},
  F.~{De Bernardis}, T.~{de Haan}, S.~{di Serego Alighieri}, J.~{Dunkley},
  C.~{Dvorkin}, J.~{Errard}, G.~{Fabbian}, S.~{Feeney}, S.~{Ferraro}, J.~P.
  {Filippini}, R.~{Flauger}, G.~M. {Fuller}, V.~{Gluscevic}, D.~{Green},
  D.~{Grin}, E.~{Grohs}, J.~W. {Henning}, J.~C. {Hill}, R.~{Hlozek},
  G.~{Holder}, W.~{Holzapfel}, W.~{Hu}, K.~M. {Huffenberger}, R.~{Keskitalo},
  L.~{Knox}, A.~{Kosowsky}, J.~{Kovac}, E.~D. {Kovetz}, C.-L. {Kuo},
  A.~{Kusaka}, M.~{Le Jeune}, A.~T. {Lee}, M.~{Lilley}, M.~{Loverde}, M.~S.
  {Madhavacheril}, A.~{Mantz}, D.~J.~E. {Marsh}, J.~{McMahon}, P.~D.
  {Meerburg}, J.~{Meyers}, A.~D. {Miller}, J.~B. {Munoz}, H.~N. {Nguyen}, M.~D.
  {Niemack}, M.~{Peloso}, J.~{Peloton}, L.~{Pogosian}, C.~{Pryke}, M.~{Raveri},
  C.~L. {Reichardt}, G.~{Rocha}, A.~{Rotti}, E.~{Schaan}, M.~M. {Schmittfull},
  D.~{Scott}, N.~{Sehgal}, S.~{Shandera}, B.~D. {Sherwin}, T.~L. {Smith},
  L.~{Sorbo}, G.~D. {Starkman}, K.~T. {Story}, A.~{van Engelen}, J.~D.
  {Vieira}, S.~{Watson}, N.~{Whitehorn}, and W.~L. {Kimmy Wu}, {\it {CMB-S4
  Science Book, First Edition}},  {\em ArXiv e-prints} (Oct., 2016)
  [\href{http://arxiv.org/abs/1610.02743}{{\tt arXiv:1610.02743}}].

\bibitem{RowanRobinsonClements2015}
M.~{Rowan-Robinson} and D.~L. {Clements}, {\it {Cold galaxies}},  {\em \mnras}
  {\bf 453} (Oct., 2015) 2050--2057,
  [\href{http://arxiv.org/abs/1507.08778}{{\tt arXiv:1507.08778}}].

\bibitem{Negrello2016lensed}
M.~{Negrello}, S.~{Amber}, A.~{Amvrosiadis}, Z.-Y. {Cai}, A.~{Lapi},
  J.~{Gonzalez-Nuevo}, G.~{De Zotti}, C.~{Furlanetto}, S.~J. {Maddox},
  M.~{Allen}, T.~{Bakx}, R.~S. {Bussmann}, A.~{Cooray}, G.~{Covone},
  L.~{Danese}, H.~{Dannerbauer}, H.~{Fu}, J.~{Greenslade}, M.~{Gurwell},
  R.~{Hopwood}, L.~V.~E. {Koopmans}, N.~{Napolitano}, H.~{Nayyeri}, A.~{Omont},
  C.~E. {Petrillo}, D.~A. {Riechers}, S.~{Serjeant}, C.~{Tortora},
  E.~{Valiante}, G.~{Verdoes Kleijn}, G.~{Vernardos}, J.~L. {Wardlow},
  M.~{Baes}, A.~J. {Baker}, N.~{Bourne}, D.~{Clements}, S.~M. {Crawford},
  S.~{Dye}, L.~{Dunne}, S.~{Eales}, R.~J. {Ivison}, L.~{Marchetti}, M.~J.
  {Micha{\l}owski}, M.~W.~L. {Smith}, M.~{Vaccari}, and P.~{van der Werf}, {\it
  {The Herschel-ATLAS: a sample of 500 {$\mu$}m-selected lensed galaxies over
  600 deg$^{2}$}},  {\em \mnras} {\bf 465} (Mar., 2017) 3558--3580,
  [\href{http://arxiv.org/abs/1611.03922}{{\tt arXiv:1611.03922}}].

\bibitem{Glenn2010}
J.~{Glenn}, A.~{Conley}, M.~{B{\'e}thermin}, B.~{Altieri}, A.~{Amblard},
  V.~{Arumugam}, H.~{Aussel}, T.~{Babbedge}, A.~{Blain}, J.~{Bock},
  A.~{Boselli}, V.~{Buat}, N.~{Castro-Rodr{\'{\i}}guez}, A.~{Cava},
  P.~{Chanial}, D.~L. {Clements}, L.~{Conversi}, A.~{Cooray}, C.~D. {Dowell},
  E.~{Dwek}, S.~{Eales}, D.~{Elbaz}, T.~P. {Ellsworth-Bowers}, M.~{Fox},
  A.~{Franceschini}, W.~{Gear}, M.~{Griffin}, M.~{Halpern},
  E.~{Hatziminaoglou}, E.~{Ibar}, K.~{Isaak}, R.~J. {Ivison}, G.~{Lagache},
  G.~{Laurent}, L.~{Levenson}, N.~{Lu}, S.~{Madden}, B.~{Maffei},
  G.~{Mainetti}, L.~{Marchetti}, G.~{Marsden}, H.~T. {Nguyen}, B.~{O'Halloran},
  S.~J. {Oliver}, A.~{Omont}, M.~J. {Page}, P.~{Panuzzo}, A.~{Papageorgiou},
  C.~P. {Pearson}, I.~{P{\'e}rez-Fournon}, M.~{Pohlen}, D.~{Rigopoulou},
  D.~{Rizzo}, I.~G. {Roseboom}, M.~{Rowan-Robinson}, M.~S. {Portal},
  B.~{Schulz}, D.~{Scott}, N.~{Seymour}, D.~L. {Shupe}, A.~J. {Smith}, J.~A.
  {Stevens}, M.~{Symeonidis}, M.~{Trichas}, K.~E. {Tugwell}, M.~{Vaccari},
  I.~{Valtchanov}, J.~D. {Vieira}, L.~{Vigroux}, L.~{Wang}, R.~{Ward},
  G.~{Wright}, C.~K. {Xu}, and M.~{Zemcov}, {\it {HerMES: deep galaxy number
  counts from a P(D) fluctuation analysis of SPIRE Science Demonstration Phase
  observations}},  {\em \mnras} {\bf 409} (Nov., 2010) 109--121,
  [\href{http://arxiv.org/abs/1009.5675}{{\tt arXiv:1009.5675}}].

\bibitem{Cai2013}
Z.-Y. {Cai}, A.~{Lapi}, J.-Q. {Xia}, G.~{De Zotti}, M.~{Negrello},
  C.~{Gruppioni}, E.~{Rigby}, G.~{Castex}, J.~{Delabrouille}, and L.~{Danese},
  {\it {A Hybrid Model for the Evolution of Galaxies and Active Galactic Nuclei
  in the Infrared}},  {\em \apj} {\bf 768} (May, 2013) 21,
  [\href{http://arxiv.org/abs/1303.2335}{{\tt arXiv:1303.2335}}].

\bibitem{Lapi2012}
A.~{Lapi}, M.~{Negrello}, J.~{Gonz{\'a}lez-Nuevo}, Z.-Y. {Cai}, G.~{De Zotti},
  and L.~{Danese}, {\it {Effective Models for Statistical Studies of
  Galaxy-scale Gravitational Lensing}},  {\em \apj} {\bf 755} (Aug., 2012) 46,
  [\href{http://arxiv.org/abs/1206.1142}{{\tt arXiv:1206.1142}}].

\bibitem{Tucci2011}
M.~{Tucci}, L.~{Toffolatti}, G.~{de Zotti}, and
  E.~{Mart{\'{\i}}nez-Gonz{\'a}lez}, {\it {High-frequency predictions for
  number counts and spectral properties of extragalactic radio sources. New
  evidence of a break at mm wavelengths in spectra of bright blazar sources}},
  {\em \aap} {\bf 533} (Sept., 2011) A57,
  [\href{http://arxiv.org/abs/1103.5707}{{\tt arXiv:1103.5707}}].

\bibitem{Lapi2011}
A.~{Lapi}, J.~{Gonz{\'a}lez-Nuevo}, L.~{Fan}, A.~{Bressan}, G.~{De Zotti},
  L.~{Danese}, M.~{Negrello}, L.~{Dunne}, S.~{Eales}, S.~{Maddox}, R.~{Auld},
  M.~{Baes}, D.~G. {Bonfield}, S.~{Buttiglione}, A.~{Cava}, D.~L. {Clements},
  A.~{Cooray}, A.~{Dariush}, S.~{Dye}, J.~{Fritz}, D.~{Herranz}, R.~{Hopwood},
  E.~{Ibar}, R.~{Ivison}, M.~J. {Jarvis}, S.~{Kaviraj}, M.~{L{\'o}pez-Caniego},
  M.~{Massardi}, M.~J. {Micha{\l}owski}, E.~{Pascale}, M.~{Pohlen}, E.~{Rigby},
  G.~{Rodighiero}, S.~{Serjeant}, D.~J.~B. {Smith}, P.~{Temi}, J.~{Wardlow},
  and P.~{van der Werf}, {\it {Herschel-ATLAS Galaxy Counts and High-redshift
  Luminosity Functions: The Formation of Massive Early-type Galaxies}},  {\em
  \apj} {\bf 742} (Nov., 2011) 24, [\href{http://arxiv.org/abs/1108.3911}{{\tt
  arXiv:1108.3911}}].

\bibitem{Canameras2015}
R.~{Ca{\~n}ameras}, N.~P.~H. {Nesvadba}, D.~{Guery}, T.~{McKenzie},
  S.~{K{\"o}nig}, G.~{Petitpas}, H.~{Dole}, B.~{Frye}, I.~{Flores-Cacho},
  L.~{Montier}, M.~{Negrello}, A.~{Beelen}, F.~{Boone}, D.~{Dicken},
  G.~{Lagache}, E.~{Le Floc'h}, B.~{Altieri}, M.~{B{\'e}thermin}, R.~{Chary},
  G.~{de Zotti}, M.~{Giard}, R.~{Kneissl}, M.~{Krips}, S.~{Malhotra},
  C.~{Martinache}, A.~{Omont}, E.~{Pointecouteau}, J.-L. {Puget}, D.~{Scott},
  G.~{Soucail}, I.~{Valtchanov}, N.~{Welikala}, and L.~{Yan}, {\it {Planck's
  dusty GEMS: The brightest gravitationally lensed galaxies discovered with the
  Planck all-sky survey}},  {\em \aap} {\bf 581} (Sept., 2015) A105.

\bibitem{Nayyeri2016}
H.~{Nayyeri}, M.~{Keele}, A.~{Cooray}, D.~A. {Riechers}, R.~J. {Ivison}, A.~I.
  {Harris}, D.~T. {Frayer}, A.~J. {Baker}, S.~C. {Chapman}, S.~{Eales},
  D.~{Farrah}, H.~{Fu}, L.~{Marchetti}, R.~{Marques-Chaves}, P.~I.
  {Martinez-Navajas}, S.~J. {Oliver}, A.~{Omont}, I.~{Perez-Fournon},
  D.~{Scott}, M.~{Vaccari}, J.~{Vieira}, M.~{Viero}, L.~{Wang}, and
  J.~{Wardlow}, {\it {Candidate Gravitationally Lensed Dusty Star-forming
  Galaxies in the Herschel Wide Area Surveys}},  {\em \apj} {\bf 823} (May,
  2016) 17, [\href{http://arxiv.org/abs/1601.03401}{{\tt arXiv:1601.03401}}].

\bibitem{Negrello2007}
M.~{Negrello}, F.~{Perrotta}, J.~{Gonz{\'a}lez-Nuevo}, L.~{Silva}, G.~{de
  Zotti}, G.~L. {Granato}, C.~{Baccigalupi}, and L.~{Danese}, {\it
  {Astrophysical and cosmological information from large-scale submillimetre
  surveys of extragalactic sources}},  {\em \mnras} {\bf 377} (June, 2007)
  1557--1568, [\href{http://arxiv.org/abs/astro-ph/0703210}{{\tt
  astro-ph/0703210}}].

\bibitem{Negrello2010}
M.~{Negrello}, R.~{Hopwood}, G.~{De Zotti}, A.~{Cooray}, A.~{Verma}, J.~{Bock},
  D.~T. {Frayer}, M.~A. {Gurwell}, A.~{Omont}, R.~{Neri}, H.~{Dannerbauer},
  L.~L. {Leeuw}, E.~{Barton}, J.~{Cooke}, S.~{Kim}, E.~{da Cunha},
  G.~{Rodighiero}, P.~{Cox}, D.~G. {Bonfield}, M.~J. {Jarvis}, S.~{Serjeant},
  R.~J. {Ivison}, S.~{Dye}, I.~{Aretxaga}, D.~H. {Hughes}, E.~{Ibar},
  F.~{Bertoldi}, I.~{Valtchanov}, S.~{Eales}, L.~{Dunne}, S.~P. {Driver},
  R.~{Auld}, S.~{Buttiglione}, A.~{Cava}, C.~A. {Grady}, D.~L. {Clements},
  A.~{Dariush}, J.~{Fritz}, D.~{Hill}, J.~B. {Hornbeck}, L.~{Kelvin},
  G.~{Lagache}, M.~{Lopez-Caniego}, J.~{Gonzalez-Nuevo}, S.~{Maddox},
  E.~{Pascale}, M.~{Pohlen}, E.~E. {Rigby}, A.~{Robotham}, C.~{Simpson},
  D.~J.~B. {Smith}, P.~{Temi}, M.~A. {Thompson}, B.~E. {Woodgate}, D.~G.
  {York}, J.~E. {Aguirre}, A.~{Beelen}, A.~{Blain}, A.~J. {Baker},
  M.~{Birkinshaw}, R.~{Blundell}, C.~M. {Bradford}, D.~{Burgarella},
  L.~{Danese}, J.~S. {Dunlop}, S.~{Fleuren}, J.~{Glenn}, A.~I. {Harris},
  J.~{Kamenetzky}, R.~E. {Lupu}, R.~J. {Maddalena}, B.~F. {Madore}, P.~R.
  {Maloney}, H.~{Matsuhara}, M.~J. {Micha{\l}owski}, E.~J. {Murphy}, B.~J.
  {Naylor}, H.~{Nguyen}, C.~{Popescu}, S.~{Rawlings}, D.~{Rigopoulou},
  D.~{Scott}, K.~S. {Scott}, M.~{Seibert}, I.~{Smail}, R.~J. {Tuffs}, J.~D.
  {Vieira}, P.~P. {van der Werf}, and J.~{Zmuidzinas}, {\it {The Detection of a
  Population of Submillimeter-Bright, Strongly Lensed Galaxies}},  {\em
  Science} {\bf 330} (Nov., 2010) 800--804,
  [\href{http://arxiv.org/abs/1011.1255}{{\tt arXiv:1011.1255}}].

\bibitem{Peacock2016}
J.~A. {Peacock}, N.~C. {Hambly}, M.~{Bilicki}, H.~T. {MacGillivray},
  L.~{Miller}, M.~A. {Read}, and S.~B. {Tritton}, {\it {The SuperCOSMOS all-sky
  galaxy catalogue}},  {\em \mnras} {\bf 462} (Oct., 2016) 2085--2098,
  [\href{http://arxiv.org/abs/1607.01189}{{\tt arXiv:1607.01189}}].

\bibitem{Skrutskie2006}
M.~F. {Skrutskie}, R.~M. {Cutri}, R.~{Stiening}, M.~D. {Weinberg},
  S.~{Schneider}, J.~M. {Carpenter}, C.~{Beichman}, R.~{Capps}, T.~{Chester},
  J.~{Elias}, J.~{Huchra}, J.~{Liebert}, C.~{Lonsdale}, D.~G. {Monet},
  S.~{Price}, P.~{Seitzer}, T.~{Jarrett}, J.~D. {Kirkpatrick}, J.~E. {Gizis},
  E.~{Howard}, T.~{Evans}, J.~{Fowler}, L.~{Fullmer}, R.~{Hurt}, R.~{Light},
  E.~L. {Kopan}, K.~A. {Marsh}, H.~L. {McCallon}, R.~{Tam}, S.~{Van Dyk}, and
  S.~{Wheelock}, {\it {The Two Micron All Sky Survey (2MASS)}},  {\em \aj} {\bf
  131} (Feb., 2006) 1163--1183.

\bibitem{Wright2010}
E.~L. {Wright}, P.~R.~M. {Eisenhardt}, A.~K. {Mainzer}, M.~E. {Ressler}, R.~M.
  {Cutri}, T.~{Jarrett}, J.~D. {Kirkpatrick}, D.~{Padgett}, R.~S. {McMillan},
  M.~{Skrutskie}, S.~A. {Stanford}, M.~{Cohen}, R.~G. {Walker}, J.~C. {Mather},
  D.~{Leisawitz}, T.~N. {Gautier}, III, I.~{McLean}, D.~{Benford}, C.~J.
  {Lonsdale}, A.~{Blain}, B.~{Mendez}, W.~R. {Irace}, V.~{Duval}, F.~{Liu},
  D.~{Royer}, I.~{Heinrichsen}, J.~{Howard}, M.~{Shannon}, M.~{Kendall}, A.~L.
  {Walsh}, M.~{Larsen}, J.~G. {Cardon}, S.~{Schick}, M.~{Schwalm}, M.~{Abid},
  B.~{Fabinsky}, L.~{Naes}, and C.-W. {Tsai}, {\it {The Wide-field Infrared
  Survey Explorer (WISE): Mission Description and Initial On-orbit
  Performance}},  {\em \aj} {\bf 140} (Dec., 2010) 1868--1881,
  [\href{http://arxiv.org/abs/1008.0031}{{\tt arXiv:1008.0031}}].

\bibitem{Condon1998}
J.~J. {Condon}, W.~D. {Cotton}, E.~W. {Greisen}, Q.~F. {Yin}, R.~A. {Perley},
  G.~B. {Taylor}, and J.~J. {Broderick}, {\it {The NRAO VLA Sky Survey}},  {\em
  \aj} {\bf 115} (May, 1998) 1693--1716.

\bibitem{Mauch2003}
T.~{Mauch}, T.~{Murphy}, H.~J. {Buttery}, J.~{Curran}, R.~W. {Hunstead},
  B.~{Piestrzynski}, J.~G. {Robertson}, and E.~M. {Sadler}, {\it {SUMSS: a
  wide-field radio imaging survey of the southern sky - II. The source
  catalogue}},  {\em \mnras} {\bf 342} (July, 2003) 1117--1130,
  [\href{http://arxiv.org/abs/astro-ph/0303188}{{\tt astro-ph/0303188}}].

\bibitem{Eales2015}
S.~A. {Eales}, {\it {Practical cosmology with lenses}},  {\em \mnras} {\bf 446}
  (Jan., 2015) 3224--3234.

\bibitem{Meng2015}
X.-L. {Meng}, T.~{Treu}, A.~{Agnello}, M.~W. {Auger}, K.~{Liao}, and P.~J.
  {Marshall}, {\it {Precision cosmology with time delay lenses: high resolution
  imaging requirements}},  {\em \jcap} {\bf 9} (Sept., 2015) 059,
  [\href{http://arxiv.org/abs/1506.07640}{{\tt arXiv:1506.07640}}].

\bibitem{Serjeant2016}
S.~{Serjeant}, {\it {Strong Gravitational Lenses and Multi-Wavelength Galaxy
  Surveys with AKARI, Herschel, SPICA and Euclid}},  {\em ArXiv e-prints}
  (Apr., 2016) [\href{http://arxiv.org/abs/1604.00282}{{\tt
  arXiv:1604.00282}}].

\bibitem{vanderWerf2011}
P.~P. {van der Werf}, A.~{Berciano Alba}, M.~{Spaans}, A.~F. {Loenen},
  R.~{Meijerink}, D.~A. {Riechers}, P.~{Cox}, A.~{Wei{\ss}}, and F.~{Walter},
  {\it {Water Vapor Emission Reveals a Highly Obscured, Star-forming Nuclear
  Region in the QSO Host Galaxy APM 08279+5255 at z = 3.9}},  {\em \apjl} {\bf
  741} (Nov., 2011) L38, [\href{http://arxiv.org/abs/1106.4825}{{\tt
  arXiv:1106.4825}}].

\bibitem{Ivison2013}
R.~J. {Ivison}, A.~M. {Swinbank}, I.~{Smail}, A.~I. {Harris}, R.~S. {Bussmann},
  A.~{Cooray}, P.~{Cox}, H.~{Fu}, A.~{Kov{\'a}cs}, M.~{Krips}, D.~{Narayanan},
  M.~{Negrello}, R.~{Neri}, J.~{Pe{\~n}arrubia}, J.~{Richard}, D.~A.
  {Riechers}, K.~{Rowlands}, J.~G. {Staguhn}, T.~A. {Targett}, S.~{Amber},
  A.~J. {Baker}, N.~{Bourne}, F.~{Bertoldi}, M.~{Bremer}, J.~A. {Calanog},
  D.~L. {Clements}, H.~{Dannerbauer}, A.~{Dariush}, G.~{De Zotti}, L.~{Dunne},
  S.~A. {Eales}, D.~{Farrah}, S.~{Fleuren}, A.~{Franceschini}, J.~E. {Geach},
  R.~D. {George}, J.~C. {Helly}, R.~{Hopwood}, E.~{Ibar}, M.~J. {Jarvis}, J.-P.
  {Kneib}, S.~{Maddox}, A.~{Omont}, D.~{Scott}, S.~{Serjeant}, M.~W.~L.
  {Smith}, M.~A. {Thompson}, E.~{Valiante}, I.~{Valtchanov}, J.~{Vieira}, and
  P.~{van der Werf}, {\it {Herschel-ATLAS: A Binary HyLIRG Pinpointing a
  Cluster of Starbursting Protoellipticals}},  {\em \apj} {\bf 772} (Aug.,
  2013) 137, [\href{http://arxiv.org/abs/1302.4436}{{\tt arXiv:1302.4436}}].

\bibitem{Wang2016}
T.~{Wang}, D.~{Elbaz}, E.~{Daddi}, A.~{Finoguenov}, D.~{Liu}, C.~{Schreiber},
  S.~{Mart{\'{\i}}n}, V.~{Strazzullo}, F.~{Valentino}, R.~{van der Burg},
  A.~{Zanella}, L.~{Ciesla}, R.~{Gobat}, A.~{Le Brun}, M.~{Pannella},
  M.~{Sargent}, X.~{Shu}, Q.~{Tan}, N.~{Cappelluti}, and Y.~{Li}, {\it
  {Discovery of a Galaxy Cluster with a Violently Starbursting Core at z =
  2.506}},  {\em \apj} {\bf 828} (Sept., 2016) 56,
  [\href{http://arxiv.org/abs/1604.07404}{{\tt arXiv:1604.07404}}].

\bibitem{Clements2014}
D.~L. {Clements}, F.~G. {Braglia}, A.~K. {Hyde}, I.~{P{\'e}rez-Fournon},
  J.~{Bock}, A.~{Cava}, S.~{Chapman}, A.~{Conley}, A.~{Cooray}, D.~{Farrah},
  E.~A. {Gonz{\'a}lez Solares}, L.~{Marchetti}, G.~{Marsden}, S.~J. {Oliver},
  I.~G. {Roseboom}, B.~{Schulz}, A.~J. {Smith}, M.~{Vaccari}, J.~{Vieira},
  M.~{Viero}, L.~{Wang}, J.~{Wardlow}, M.~{Zemcov}, and G.~{de Zotti}, {\it
  {Herschel Multitiered Extragalactic Survey: clusters of dusty galaxies
  uncovered by Herschel and Planck}},  {\em \mnras} {\bf 439} (Apr., 2014)
  1193--1211, [\href{http://arxiv.org/abs/1311.5758}{{\tt arXiv:1311.5758}}].

\bibitem{Negrello2005}
M.~{Negrello}, J.~{Gonz{\'a}lez-Nuevo}, M.~{Magliocchetti}, L.~{Moscardini},
  G.~{De Zotti}, L.~{Toffolatti}, and L.~{Danese}, {\it {Effect of clustering
  on extragalactic source counts with low-resolution instruments}},  {\em
  \mnras} {\bf 358} (Apr., 2005) 869--874,
  [\href{http://arxiv.org/abs/astro-ph/0406388}{{\tt astro-ph/0406388}}].

\bibitem{Planck_high_z2015}
{Planck Collaboration XXXIX}, {\it {Planck intermediate results. XXXIX. The
  Planck list of high-redshift source candidates}},  {\em \aap} {\bf 596}
  (Dec., 2016) A100, [\href{http://arxiv.org/abs/1508.04171}{{\tt
  arXiv:1508.04171}}].

\bibitem{PlanckHerschel2015}
{Planck Collaboration XXVII}, {\it {Planck intermediate results. XXVII.
  High-redshift infrared galaxy overdensity candidates and lensed sources
  discovered by Planck and confirmed by Herschel-SPIRE}},  {\em \aap} {\bf 582}
  (Oct., 2015) A30, [\href{http://arxiv.org/abs/1503.08773}{{\tt
  arXiv:1503.08773}}].

\bibitem{KennicuttEvans2012}
R.~C. {Kennicutt} and N.~J. {Evans}, {\it {Star Formation in the Milky Way and
  Nearby Galaxies}},  {\em \araa} {\bf 50} (Sept., 2012) 531--608,
  [\href{http://arxiv.org/abs/1204.3552}{{\tt arXiv:1204.3552}}].

\bibitem{Aversa2015}
R.~{Aversa}, A.~{Lapi}, G.~{de Zotti}, F.~{Shankar}, and L.~{Danese}, {\it
  {Black Hole and Galaxy Coevolution from Continuity Equation and Abundance
  Matching}},  {\em \apj} {\bf 810} (Sept., 2015) 74,
  [\href{http://arxiv.org/abs/1507.07318}{{\tt arXiv:1507.07318}}].

\bibitem{FloresCacho2016}
I.~{Flores-Cacho}, D.~{Pierini}, G.~{Soucail}, L.~{Montier}, H.~{Dole},
  E.~{Pointecouteau}, R.~{Pell{\'o}}, E.~{Le Floc'h}, N.~{Nesvadba},
  G.~{Lagache}, D.~{Guery}, and R.~{Ca{\~n}ameras}, {\it {Multi-wavelength
  characterisation of z \~{} 2 clustered, dusty star-forming galaxies
  discovered by Planck}},  {\em \aap} {\bf 585} (Jan., 2016) A54,
  [\href{http://arxiv.org/abs/1510.01585}{{\tt arXiv:1510.01585}}].

\bibitem{Eales2010}
S.~{Eales}, L.~{Dunne}, D.~{Clements}, A.~{Cooray}, G.~{De Zotti}, S.~{Dye},
  R.~{Ivison}, M.~{Jarvis}, G.~{Lagache}, S.~{Maddox}, M.~{Negrello},
  S.~{Serjeant}, M.~A. {Thompson}, E.~{Van Kampen}, A.~{Amblard},
  P.~{Andreani}, M.~{Baes}, A.~{Beelen}, G.~J. {Bendo}, D.~{Benford},
  F.~{Bertoldi}, J.~{Bock}, D.~{Bonfield}, A.~{Boselli}, C.~{Bridge},
  V.~{Buat}, D.~{Burgarella}, R.~{Carlberg}, A.~{Cava}, P.~{Chanial},
  S.~{Charlot}, N.~{Christopher}, P.~{Coles}, L.~{Cortese}, A.~{Dariush},
  E.~{da Cunha}, G.~{Dalton}, L.~{Danese}, H.~{Dannerbauer}, S.~{Driver},
  J.~{Dunlop}, L.~{Fan}, D.~{Farrah}, D.~{Frayer}, C.~{Frenk}, J.~{Geach},
  J.~{Gardner}, H.~{Gomez}, J.~{Gonz{\'a}lez-Nuevo}, E.~{Gonz{\'a}lez-Solares},
  M.~{Griffin}, M.~{Hardcastle}, E.~{Hatziminaoglou}, D.~{Herranz},
  D.~{Hughes}, E.~{Ibar}, W.-S. {Jeong}, C.~{Lacey}, A.~{Lapi}, A.~{Lawrence},
  M.~{Lee}, L.~{Leeuw}, J.~{Liske}, M.~{L{\'o}pez-Caniego}, T.~{M{\"u}ller},
  K.~{Nandra}, P.~{Panuzzo}, A.~{Papageorgiou}, G.~{Patanchon}, J.~{Peacock},
  C.~{Pearson}, S.~{Phillipps}, M.~{Pohlen}, C.~{Popescu}, S.~{Rawlings},
  E.~{Rigby}, M.~{Rigopoulou}, A.~{Robotham}, G.~{Rodighiero}, A.~{Sansom},
  B.~{Schulz}, D.~{Scott}, D.~J.~B. {Smith}, B.~{Sibthorpe}, I.~{Smail},
  J.~{Stevens}, W.~{Sutherland}, T.~{Takeuchi}, J.~{Tedds}, P.~{Temi},
  R.~{Tuffs}, M.~{Trichas}, M.~{Vaccari}, I.~{Valtchanov}, P.~{van der Werf},
  A.~{Verma}, J.~{Vieria}, C.~{Vlahakis}, and G.~J. {White}, {\it {The Herschel
  ATLAS}},  {\em \pasp} {\bf 122} (May, 2010) 499--515,
  [\href{http://arxiv.org/abs/0910.4279}{{\tt arXiv:0910.4279}}].

\bibitem{Oliver2012}
S.~J. {Oliver}, J.~{Bock}, B.~{Altieri}, A.~{Amblard}, V.~{Arumugam},
  H.~{Aussel}, T.~{Babbedge}, A.~{Beelen}, M.~{B{\'e}thermin}, A.~{Blain},
  A.~{Boselli}, C.~{Bridge}, D.~{Brisbin}, V.~{Buat}, D.~{Burgarella},
  N.~{Castro-Rodr{\'{\i}}guez}, A.~{Cava}, P.~{Chanial}, M.~{Cirasuolo}, D.~L.
  {Clements}, A.~{Conley}, L.~{Conversi}, A.~{Cooray}, C.~D. {Dowell}, E.~N.
  {Dubois}, E.~{Dwek}, S.~{Dye}, S.~{Eales}, D.~{Elbaz}, D.~{Farrah},
  A.~{Feltre}, P.~{Ferrero}, N.~{Fiolet}, M.~{Fox}, A.~{Franceschini},
  W.~{Gear}, E.~{Giovannoli}, J.~{Glenn}, Y.~{Gong}, E.~A. {Gonz{\'a}lez
  Solares}, M.~{Griffin}, M.~{Halpern}, M.~{Harwit}, E.~{Hatziminaoglou},
  S.~{Heinis}, P.~{Hurley}, H.~S. {Hwang}, A.~{Hyde}, E.~{Ibar}, O.~{Ilbert},
  K.~{Isaak}, R.~J. {Ivison}, G.~{Lagache}, E.~{Le Floc'h}, L.~{Levenson},
  B.~L. {Faro}, N.~{Lu}, S.~{Madden}, B.~{Maffei}, G.~{Magdis}, G.~{Mainetti},
  L.~{Marchetti}, G.~{Marsden}, J.~{Marshall}, A.~M.~J. {Mortier}, H.~T.
  {Nguyen}, B.~{O'Halloran}, A.~{Omont}, M.~J. {Page}, P.~{Panuzzo},
  A.~{Papageorgiou}, H.~{Patel}, C.~P. {Pearson}, I.~{P{\'e}rez-Fournon},
  M.~{Pohlen}, J.~I. {Rawlings}, G.~{Raymond}, D.~{Rigopoulou}, L.~{Riguccini},
  D.~{Rizzo}, G.~{Rodighiero}, I.~G. {Roseboom}, M.~{Rowan-Robinson},
  M.~{S{\'a}nchez Portal}, B.~{Schulz}, D.~{Scott}, N.~{Seymour}, D.~L.
  {Shupe}, A.~J. {Smith}, J.~A. {Stevens}, M.~{Symeonidis}, M.~{Trichas}, K.~E.
  {Tugwell}, M.~{Vaccari}, I.~{Valtchanov}, J.~D. {Vieira}, M.~{Viero},
  L.~{Vigroux}, L.~{Wang}, R.~{Ward}, J.~{Wardlow}, G.~{Wright}, C.~K. {Xu},
  and M.~{Zemcov}, {\it {The Herschel Multi-tiered Extragalactic Survey:
  HerMES}},  {\em \mnras} {\bf 424} (Aug., 2012) 1614--1635,
  [\href{http://arxiv.org/abs/1203.2562}{{\tt arXiv:1203.2562}}].

\bibitem{Granato2015}
G.~L. {Granato}, C.~{Ragone-Figueroa}, R.~{Dom{\'{\i}}nguez-Tenreiro},
  A.~{Obreja}, S.~{Borgani}, G.~{De Lucia}, and G.~{Murante}, {\it {The early
  phases of galaxy clusters formation in IR: coupling hydrodynamical
  simulations with GRASIL-3D}},  {\em \mnras} {\bf 450} (June, 2015)
  1320--1332, [\href{http://arxiv.org/abs/1412.6105}{{\tt arXiv:1412.6105}}].

\bibitem{Alberts2014}
S.~{Alberts}, A.~{Pope}, M.~{Brodwin}, D.~W. {Atlee}, Y.-T. {Lin}, A.~{Dey},
  P.~R.~M. {Eisenhardt}, D.~P. {Gettings}, A.~H. {Gonzalez}, B.~T. {Jannuzi},
  C.~L. {Mancone}, J.~{Moustakas}, G.~F. {Snyder}, S.~A. {Stanford},
  D.~{Stern}, B.~J. {Weiner}, and G.~R. {Zeimann}, {\it {The evolution of
  dust-obscured star formation activity in galaxy clusters relative to the
  field over the last 9 billion years}},  {\em \mnras} {\bf 437} (Jan., 2014)
  437--457, [\href{http://arxiv.org/abs/1310.6040}{{\tt arXiv:1310.6040}}].

\bibitem{Alberts2016}
S.~{Alberts}, A.~{Pope}, M.~{Brodwin}, S.~M. {Chung}, R.~{Cybulski}, A.~{Dey},
  P.~R.~M. {Eisenhardt}, A.~{Galametz}, A.~H. {Gonzalez}, B.~T. {Jannuzi},
  S.~A. {Stanford}, G.~F. {Snyder}, D.~{Stern}, and G.~R. {Zeimann}, {\it {Star
  Formation and AGN Activity in Galaxy Clusters from $z=1-2$: a
  Multi-Wavelength Analysis Featuring \textit{Herschel}/PACS}},  {\em \apj}
  {\bf 825} (July, 2016) 72, [\href{http://arxiv.org/abs/1604.03564}{{\tt
  arXiv:1604.03564}}].

\bibitem{Wagner2016}
C.~R. {Wagner}, S.~{Courteau}, M.~{Brodwin}, S.~A. {Stanford}, G.~F. {Snyder},
  and D.~{Stern}, {\it {The Evolution of Star formation Activity in Cluster
  Galaxies over $0.15< z < 1.5$}},  {\em \apj} {\bf 834} (Jan., 2017) 53,
  [\href{http://arxiv.org/abs/1610.01498}{{\tt arXiv:1610.01498}}].

\bibitem{Mehrtens2012}
N.~{Mehrtens}, A.~K. {Romer}, M.~{Hilton}, E.~J. {Lloyd-Davies}, C.~J.
  {Miller}, S.~A. {Stanford}, M.~{Hosmer}, B.~{Hoyle}, C.~A. {Collins}, A.~R.
  {Liddle}, P.~T.~P. {Viana}, R.~C. {Nichol}, J.~P. {Stott}, E.~N. {Dubois},
  S.~T. {Kay}, M.~{Sahl{\'e}n}, O.~{Young}, C.~J. {Short}, L.~{Christodoulou},
  W.~A. {Watson}, M.~{Davidson}, C.~D. {Harrison}, L.~{Baruah}, M.~{Smith},
  C.~{Burke}, J.~A. {Mayers}, P.-J. {Deadman}, P.~J. {Rooney}, E.~M.
  {Edmondson}, M.~{West}, H.~C. {Campbell}, A.~C. {Edge}, R.~G. {Mann},
  K.~{Sabirli}, D.~{Wake}, C.~{Benoist}, L.~{da Costa}, M.~A.~G. {Maia}, and
  R.~{Ogando}, {\it {The XMM Cluster Survey: optical analysis methodology and
  the first data release}},  {\em \mnras} {\bf 423} (June, 2012) 1024--1052,
  [\href{http://arxiv.org/abs/1106.3056}{{\tt arXiv:1106.3056}}].

\bibitem{Merloni2012}
A.~{Merloni}, P.~{Predehl}, W.~{Becker}, H.~{B{\"o}hringer}, T.~{Boller},
  H.~{Brunner}, M.~{Brusa}, K.~{Dennerl}, M.~{Freyberg}, P.~{Friedrich},
  A.~{Georgakakis}, F.~{Haberl}, G.~{Hasinger}, N.~{Meidinger}, J.~{Mohr},
  K.~{Nandra}, A.~{Rau}, T.~H. {Reiprich}, J.~{Robrade}, M.~{Salvato},
  A.~{Santangelo}, M.~{Sasaki}, A.~{Schwope}, J.~{Wilms}, and t.~{German
  eROSITA Consortium}, {\it {eROSITA Science Book: Mapping the Structure of the
  Energetic Universe}},  {\em ArXiv e-prints} (Sept., 2012)
  [\href{http://arxiv.org/abs/1209.3114}{{\tt arXiv:1209.3114}}].

\bibitem{Bohringer2013}
H.~{B{\"o}hringer}, G.~{Chon}, C.~A. {Collins}, L.~{Guzzo}, N.~{Nowak}, and
  S.~{Bobrovskyi}, {\it {The extended ROSAT-ESO flux limited X-ray galaxy
  cluster survey (REFLEX II) II. Construction and properties of the survey}},
  {\em \aap} {\bf 555} (July, 2013) A30.

\bibitem{Hasselfield2013}
M.~{Hasselfield}, M.~{Hilton}, T.~A. {Marriage}, G.~E. {Addison}, L.~F.
  {Barrientos}, N.~{Battaglia}, E.~S. {Battistelli}, J.~R. {Bond},
  D.~{Crichton}, S.~{Das}, M.~J. {Devlin}, S.~R. {Dicker}, J.~{Dunkley},
  R.~{D{\"u}nner}, J.~W. {Fowler}, M.~B. {Gralla}, A.~{Hajian}, M.~{Halpern},
  A.~D. {Hincks}, R.~{Hlozek}, J.~P. {Hughes}, L.~{Infante}, K.~D. {Irwin},
  A.~{Kosowsky}, D.~{Marsden}, F.~{Menanteau}, K.~{Moodley}, M.~D. {Niemack},
  M.~R. {Nolta}, L.~A. {Page}, B.~{Partridge}, E.~D. {Reese}, B.~L. {Schmitt},
  N.~{Sehgal}, B.~D. {Sherwin}, J.~{Sievers}, C.~{Sif{\'o}n}, D.~N. {Spergel},
  S.~T. {Staggs}, D.~S. {Swetz}, E.~R. {Switzer}, R.~{Thornton}, H.~{Trac}, and
  E.~J. {Wollack}, {\it {The Atacama Cosmology Telescope: Sunyaev-Zel'dovich
  selected galaxy clusters at 148 GHz from three seasons of data}},  {\em
  \jcap} {\bf 7} (July, 2013) 008, [\href{http://arxiv.org/abs/1301.0816}{{\tt
  arXiv:1301.0816}}].

\bibitem{Bleem2015}
L.~E. {Bleem}, B.~{Stalder}, T.~{de Haan}, K.~A. {Aird}, S.~W. {Allen}, D.~E.
  {Applegate}, M.~L.~N. {Ashby}, M.~{Bautz}, M.~{Bayliss}, B.~A. {Benson},
  S.~{Bocquet}, M.~{Brodwin}, J.~E. {Carlstrom}, C.~L. {Chang}, and et~al.,
  {\it {Galaxy Clusters Discovered via the Sunyaev-Zel'dovich Effect in the
  2500-Square-Degree SPT-SZ Survey}},  {\em \apjs} {\bf 216} (Feb., 2015) 27,
  [\href{http://arxiv.org/abs/1409.0850}{{\tt arXiv:1409.0850}}].

\bibitem{PlanckCollaboration2015SZcatalog}
{Planck Collaboration XXXII}, {\it {Planck 2013 results. XXXII. The updated
  Planck catalogue of Sunyaev-Zeldovich sources}},  {\em \aap} {\bf 581}
  (Sept., 2015) A14, [\href{http://arxiv.org/abs/1502.00543}{{\tt
  arXiv:1502.00543}}].

\bibitem{PlanckCollaboration2016cluster_dust}
{Planck Collaboration XLIII}, {\it {Planck intermediate results. XLIII.
  Spectral energy distribution of dust in clusters of galaxies}},  {\em \aap}
  {\bf 596} (Dec., 2016) A104, [\href{http://arxiv.org/abs/1603.04919}{{\tt
  arXiv:1603.04919}}].

\bibitem{PlanckCollaboration2015SZ_CIB}
{Planck Collaboration XXIII}, {\it {Planck 2015 results. XXIII. The thermal
  Sunyaev-Zeldovich effect--cosmic infrared background correlation}},  {\em
  \aap} {\bf 594} (Aug., 2016) A23,
  [\href{http://arxiv.org/abs/1509.06555}{{\tt arXiv:1509.06555}}].

\bibitem{Popesso2015}
P.~{Popesso}, A.~{Biviano}, A.~{Finoguenov}, D.~{Wilman}, M.~{Salvato},
  B.~{Magnelli}, C.~{Gruppioni}, F.~{Pozzi}, G.~{Rodighiero}, F.~{Ziparo},
  S.~{Berta}, D.~{Elbaz}, M.~{Dickinson}, D.~{Lutz}, B.~{Altieri}, H.~{Aussel},
  A.~{Cimatti}, D.~{Fadda}, O.~{Ilbert}, E.~{Le Floch}, R.~{Nordon},
  A.~{Poglitsch}, and C.~K. {Xu}, {\it {The evolution of galaxy star formation
  activity in massive haloes}},  {\em \aap} {\bf 574} (Feb., 2015) A105,
  [\href{http://arxiv.org/abs/1407.8214}{{\tt arXiv:1407.8214}}].

\bibitem{Gruppioni2013}
C.~{Gruppioni}, F.~{Pozzi}, G.~{Rodighiero}, I.~{Delvecchio}, S.~{Berta},
  L.~{Pozzetti}, G.~{Zamorani}, P.~{Andreani}, A.~{Cimatti}, O.~{Ilbert},
  E.~{Le Floc'h}, D.~{Lutz}, and et~al., {\it {The Herschel PEP/HerMES
  luminosity function - I. Probing the evolution of PACS selected Galaxies to
  $z \simeq 4$}},  {\em \mnras} {\bf 432} (June, 2013) 23--52,
  [\href{http://arxiv.org/abs/1302.5209}{{\tt arXiv:1302.5209}}].

\bibitem{Brodwin2013}
M.~{Brodwin}, S.~A. {Stanford}, A.~H. {Gonzalez}, G.~R. {Zeimann}, G.~F.
  {Snyder}, C.~L. {Mancone}, A.~{Pope}, P.~R. {Eisenhardt}, D.~{Stern},
  S.~{Alberts}, M.~L.~N. {Ashby}, M.~J.~I. {Brown}, R.-R. {Chary}, A.~{Dey},
  A.~{Galametz}, D.~P. {Gettings}, B.~T. {Jannuzi}, E.~D. {Miller},
  J.~{Moustakas}, and L.~A. {Moustakas}, {\it {The Era of Star Formation in
  Galaxy Clusters}},  {\em \apj} {\bf 779} (Dec., 2013) 138,
  [\href{http://arxiv.org/abs/1310.6039}{{\tt arXiv:1310.6039}}].

\bibitem{EnsslinKaiser2000}
T.~A. {En{\ss}lin} and C.~R. {Kaiser}, {\it {Comptonization of the cosmic
  microwave background by relativistic plasma}},  {\em \aap} {\bf 360} (Aug.,
  2000) 417--430, [\href{http://arxiv.org/abs/astro-ph/0001429}{{\tt
  astro-ph/0001429}}].

\bibitem{Colafrancesco2011}
S.~{Colafrancesco}, P.~{Marchegiani}, and R.~{Buonanno}, {\it {Untangling the
  atmosphere of the Bullet cluster with Sunyaev-Zeldovich effect
  observations}},  {\em \aap} {\bf 527} (Mar., 2011) L1.

\bibitem{PlanckCollaboration2011CIB}
{Planck Collaboration XVIII}, {\it {Planck early results. XVIII. The power
  spectrum of cosmic infrared background anisotropies}},  {\em \aap} {\bf 536}
  (Dec., 2011) A18, [\href{http://arxiv.org/abs/1101.2028}{{\tt
  arXiv:1101.2028}}].

\bibitem{PlanckCollaborationXXX2014}
{Planck Collaboration XXX}, {\it {Planck 2013 results. XXX. Cosmic infrared
  background measurements and implications for star formation}},  {\em \aap}
  {\bf 571} (Nov., 2014) A30, [\href{http://arxiv.org/abs/1309.0382}{{\tt
  arXiv:1309.0382}}].

\bibitem{Mak2016}
D.~S.~Y. {Mak}, A.~{Challinor}, G.~{Efstathiou}, and G.~{Lagache}, {\it
  {Measurement of CIB power spectra over large sky areas from Planck HFI
  maps}},  {\em \mnras} {\bf 466} (Apr., 2017) 286--319,
  [\href{http://arxiv.org/abs/1609.08942}{{\tt arXiv:1609.08942}}].

\bibitem{PlanckCollaboration2016CIB_Gal_dust}
{Planck Collaboration XLVIII}, {\it {Planck intermediate results. XLVIII.
  Disentangling Galactic dust emission and cosmic infrared background
  anisotropies}},  {\em \aap} {\bf 596} (Dec., 2016) A109,
  [\href{http://arxiv.org/abs/1605.09387}{{\tt arXiv:1605.09387}}].

\bibitem{WuDore2016}
H.-Y. {Wu} and O.~{Dor{\'e}}, {\it {Optimizing future experiments of cosmic
  far-infrared background: a principal component approach}},  {\em ArXiv
  e-prints} (Dec., 2016) [\href{http://arxiv.org/abs/1612.02474}{{\tt
  arXiv:1612.02474}}].

\bibitem{Tucci2016}
M.~{Tucci}, V.~{Desjacques}, and M.~{Kunz}, {\it {Cosmic infrared background
  anisotropies as a window into primordial non-Gaussianity}},  {\em \mnras}
  {\bf 463} (Dec., 2016) 2046--2063,
  [\href{http://arxiv.org/abs/1606.02323}{{\tt arXiv:1606.02323}}].

\bibitem{DaneseDeZotti1981}
L.~{Danese} and G.~{de Zotti}, {\it {Dipole anisotropy and distortions of the
  spectrum of the cosmic microwave background}},  {\em \aap} {\bf 94} (Feb.,
  1981) L33.

\bibitem{Piat2002}
M.~{Piat}, G.~{Lagache}, J.~P. {Bernard}, M.~{Giard}, and J.~L. {Puget}, {\it
  {Cosmic background dipole measurements with the Planck-High Frequency
  Instrument}},  {\em \aap} {\bf 393} (Oct., 2002) 359--368,
  [\href{http://arxiv.org/abs/astro-ph/0110650}{{\tt astro-ph/0110650}}].

\bibitem{Balashev2015}
S.~A. {Balashev}, E.~E. {Kholupenko}, J.~{Chluba}, A.~V. {Ivanchik}, and D.~A.
  {Varshalovich}, {\it {Spectral Distortions of the CMB Dipole}},  {\em \apj}
  {\bf 810} (Sept., 2015) 131, [\href{http://arxiv.org/abs/1505.06028}{{\tt
  arXiv:1505.06028}}].

\bibitem{DeZotti2016}
G.~{De Zotti}, M.~{Negrello}, G.~{Castex}, A.~{Lapi}, and M.~{Bonato}, {\it
  {Another look at distortions of the Cosmic Microwave Background spectrum}},
  {\em \jcap} {\bf 3} (Mar., 2016) 047,
  [\href{http://arxiv.org/abs/1512.04816}{{\tt arXiv:1512.04816}}].

\bibitem{Fixsen1998}
D.~J. {Fixsen}, E.~{Dwek}, J.~C. {Mather}, C.~L. {Bennett}, and R.~A. {Shafer},
  {\it {The Spectrum of the Extragalactic Far-Infrared Background from the COBE
  FIRAS Observations}},  {\em \apj} {\bf 508} (Nov., 1998) 123--128,
  [\href{http://arxiv.org/abs/astro-ph/9803021}{{\tt astro-ph/9803021}}].

\bibitem{FixsenKashlinsky2011}
D.~J. {Fixsen} and A.~{Kashlinsky}, {\it {Probing the Universe's Tilt with the
  Cosmic Infrared Background Dipole}},  {\em \apj} {\bf 734} (June, 2011) 61,
  [\href{http://arxiv.org/abs/1104.0901}{{\tt arXiv:1104.0901}}].

\bibitem{Basu04}
K.~{Basu}, C.~{Hern{\'a}ndez-Monteagudo}, and R.~A. {Sunyaev}, {\it {CMB
  observations and the production of chemical elements at the end of the dark
  ages}},  {\em \aap} {\bf 416} (Mar., 2004) 447--466,
  [\href{http://arxiv.org/abs/astro-ph/0311620}{{\tt astro-ph/0311620}}].

\bibitem{CHMetal07}
C.~{Hern{\'a}ndez-Monteagudo}, J.~A. {Rubi{\~n}o-Mart{\'{\i}}n}, and R.~A.
  {Sunyaev}, {\it {On the influence of resonant scattering on cosmic microwave
  background polarization anisotropies}},  {\em \mnras} {\bf 380} (Oct., 2007)
  1656--1668.

\bibitem{CHMetal06}
C.~{Hern{\'a}ndez-Monteagudo}, L.~{Verde}, and R.~{Jimenez}, {\it {Tomography
  of the Reionization Epoch with Multifrequency CMB Observations}},  {\em \apj}
  {\bf 653} (Dec., 2006) 1--10,
  [\href{http://arxiv.org/abs/astro-ph/0604324}{{\tt astro-ph/0604324}}].

\bibitem{Kogut2016}
A.~{Kogut}, J.~{Chluba}, D.~J. {Fixsen}, S.~{Meyer}, and D.~{Spergel}, {\it
  {The Primordial Inflation Explorer (PIXIE)}},  in {\em Society of
  Photo-Optical Instrumentation Engineers (SPIE) Conference Series}, vol.~9904
  of {\em \procspie}, p.~99040W, July, 2016.

\bibitem{righi08}
M.~{Righi}, C.~{Hern{\'a}ndez-Monteagudo}, and R.~A. {Sunyaev}, {\it {Carbon
  monoxide line emission as a CMB foreground: tomography of the star-forming
  universe with different spectral resolutions}},  {\em \aap} {\bf 489} (Oct.,
  2008) 489--504, [\href{http://arxiv.org/abs/0805.2174}{{\tt
  arXiv:0805.2174}}].

\bibitem{Mashian2016}
N.~{Mashian}, A.~{Loeb}, and A.~{Sternberg}, {\it {Spectral distortion of the
  CMB by the cumulative CO emission from galaxies throughout cosmic history}},
  {\em \mnras} {\bf 458} (May, 2016) L99--L103,
  [\href{http://arxiv.org/abs/1601.02618}{{\tt arXiv:1601.02618}}].

\bibitem{chmOI07}
C.~{Hern{\'a}ndez-Monteagudo}, Z.~{Haiman}, R.~{Jimenez}, and L.~{Verde}, {\it
  {Oxygen Pumping: Probing Intergalactic Metals at the Epoch of Reionization}},
   {\em \apjl} {\bf 660} (May, 2007) L85--L88.

\bibitem{chmOI08}
C.~{Hern{\'a}ndez-Monteagudo}, Z.~{Haiman}, L.~{Verde}, and R.~{Jimenez}, {\it
  {Oxygen Pumping. II. Probing the Inhomogeneous Metal Enrichment at the Epoch
  of Reionization with High-Frequency CMB Observations}},  {\em \apj} {\bf 672}
  (Jan., 2008) 33--39, [\href{http://arxiv.org/abs/0709.3313}{{\tt
  arXiv:0709.3313}}].

\bibitem{Greve2014}
T.~R. {Greve}, I.~{Leonidaki}, E.~M. {Xilouris}, A.~{Wei{\ss}}, Z.-Y. {Zhang},
  P.~{van der Werf}, S.~{Aalto}, L.~{Armus}, T.~{D{\'{\i}}az-Santos}, A.~S.
  {Evans}, J.~{Fischer}, Y.~{Gao}, E.~{Gonz{\'a}lez-Alfonso}, A.~{Harris},
  C.~{Henkel}, R.~{Meijerink}, D.~A. {Naylor}, H.~A. {Smith}, M.~{Spaans},
  G.~J. {Stacey}, S.~{Veilleux}, and F.~{Walter}, {\it {Star Formation
  Relations and CO Spectral Line Energy Distributions across the J-ladder and
  Redshift}},  {\em \apj} {\bf 794} (Oct., 2014) 142,
  [\href{http://arxiv.org/abs/1407.4400}{{\tt arXiv:1407.4400}}].

\bibitem{Bonato2014}
M.~{Bonato}, M.~{Negrello}, Z.-Y. {Cai}, G.~{De Zotti}, A.~{Bressan},
  A.~{Lapi}, C.~{Gruppioni}, L.~{Spinoglio}, and L.~{Danese}, {\it {Exploring
  the early dust-obscured phase of galaxy formation with blind
  mid-/far-infrared spectroscopic surveys}},  {\em \mnras} {\bf 438} (Mar.,
  2014) 2547--2564, [\href{http://arxiv.org/abs/1312.1891}{{\tt
  arXiv:1312.1891}}].

\bibitem{UrryPadovani1995}
C.~M. {Urry} and P.~{Padovani}, {\it {Unified Schemes for Radio-Loud Active
  Galactic Nuclei}},  {\em \pasp} {\bf 107} (Sept., 1995) 803,
  [\href{http://arxiv.org/abs/astro-ph/9506063}{{\tt astro-ph/9506063}}].

\bibitem{Acero2015}
F.~{Acero}, M.~{Ackermann}, M.~{Ajello}, A.~{Albert}, W.~B. {Atwood},
  M.~{Axelsson}, L.~{Baldini}, J.~{Ballet}, , and {Fermi-LAT Collaboration},
  {\it {Fermi Large Area Telescope Third Source Catalog}},  {\em \apjs} {\bf
  218} (June, 2015) 23, [\href{http://arxiv.org/abs/1501.02003}{{\tt
  arXiv:1501.02003}}].

\bibitem{DeZotti2005}
G.~{de Zotti}, R.~{Ricci}, D.~{Mesa}, L.~{Silva}, P.~{Mazzotta},
  L.~{Toffolatti}, and J.~{Gonz{\'a}lez-Nuevo}, {\it {Predictions for
  high-frequency radio surveys of extragalactic sources}},  {\em \aap} {\bf
  431} (Mar., 2005) 893--903,
  [\href{http://arxiv.org/abs/astro-ph/0410709}{{\tt astro-ph/0410709}}].

\bibitem{Massardi2010}
M.~{Massardi}, A.~{Bonaldi}, M.~{Negrello}, S.~{Ricciardi}, A.~{Raccanelli},
  and G.~{de Zotti}, {\it {A model for the cosmological evolution of
  low-frequency radio sources}},  {\em \mnras} {\bf 404} (May, 2010) 532--544,
  [\href{http://arxiv.org/abs/1001.1069}{{\tt arXiv:1001.1069}}].

\bibitem{Cowie1996}
L.~L. {Cowie}, A.~{Songaila}, E.~M. {Hu}, and J.~G. {Cohen}, {\it {New Insight
  on Galaxy Formation and Evolution From Keck Spectroscopy of the Hawaii Deep
  Fields}},  {\em \aj} {\bf 112} (Sept., 1996) 839,
  [\href{http://arxiv.org/abs/astro-ph/9606079}{{\tt astro-ph/9606079}}].

\bibitem{Rigby2015}
E.~E. {Rigby}, J.~{Argyle}, P.~N. {Best}, D.~{Rosario}, and H.~J.~A.
  {R{\"o}ttgering}, {\it {Cosmic downsizing of powerful radio galaxies to low
  radio luminosities}},  {\em \aap} {\bf 581} (Sept., 2015) A96,
  [\href{http://arxiv.org/abs/1507.00341}{{\tt arXiv:1507.00341}}].

\bibitem{PlanckCollaboration2011radioSED}
{Planck Collaboration XV}, {\it {Planck early results. XV. Spectral energy
  distributions and radio continuum spectra of northern extragalactic radio
  sources}},  {\em \aap} {\bf 536} (Dec., 2011) A15,
  [\href{http://arxiv.org/abs/1101.2047}{{\tt arXiv:1101.2047}}].

\bibitem{PlanckCollaboration2016radio_spectra}
{Planck Collaboration XLV}, {\it {Planck intermediate results. XLV. Radio
  spectra of northern extragalactic radio sources}},  {\em \aap} {\bf 596}
  (Dec., 2016) A106, [\href{http://arxiv.org/abs/1606.05120}{{\tt
  arXiv:1606.05120}}].

\bibitem{Massardi2016}
M.~{Massardi}, A.~{Bonaldi}, L.~{Bonavera}, G.~{De Zotti}, M.~{Lopez-Caniego},
  and V.~{Galluzzi}, {\it {The Planck-ATCA Co-eval Observations project:
  analysis of radio source properties between 5 and 217 GHz}},  {\em \mnras}
  {\bf 455} (Jan., 2016) 3249--3262,
  [\href{http://arxiv.org/abs/1511.02605}{{\tt arXiv:1511.02605}}].

\bibitem{Cutini2014}
S.~{Cutini}, S.~{Ciprini}, M.~{Orienti}, A.~{Tramacere}, F.~{D'Ammando},
  F.~{Verrecchia}, G.~{Polenta}, L.~{Carrasco}, V.~{D'Elia}, P.~{Giommi},
  J.~{Gonz{\'a}lez-Nuevo}, P.~{Grandi}, D.~{Harrison}, E.~{Hays}, S.~{Larsson},
  A.~{L{\"a}hteenm{\"a}ki}, J.~{Le{\'o}n-Tavares}, M.~{L{\'o}pez-Caniego},
  P.~{Natoli}, R.~{Ojha}, B.~{Partridge}, A.~{Porras}, L.~{Reyes},
  E.~{Recillas}, and E.~{Torresi}, {\it {Radio-gamma-ray connection and
  spectral evolution in 4C +49.22 (S4 1150+49): the Fermi, Swift and Planck
  view}},  {\em \mnras} {\bf 445} (Dec., 2014) 4316--4334,
  [\href{http://arxiv.org/abs/1409.8101}{{\tt arXiv:1409.8101}}].

\bibitem{DeZotti2010}
G.~{de Zotti}, M.~{Massardi}, M.~{Negrello}, and J.~{Wall}, {\it {Radio and
  millimeter continuum surveys and their astrophysical implications}},  {\em
  \aapr} {\bf 18} (Feb., 2010) 1--65,
  [\href{http://arxiv.org/abs/0908.1896}{{\tt arXiv:0908.1896}}].

\bibitem{Mocanu2013}
L.~M. {Mocanu}, T.~M. {Crawford}, J.~D. {Vieira}, K.~A. {Aird}, M.~{Aravena},
  J.~E. {Austermann}, B.~A. {Benson}, M.~{B{\'e}thermin}, L.~E. {Bleem}, and
  et~al., {\it {Extragalactic Millimeter-wave Point-source Catalog, Number
  Counts and Statistics from 771 deg$^{2}$ of the SPT-SZ Survey}},  {\em \apj}
  {\bf 779} (Dec., 2013) 61, [\href{http://arxiv.org/abs/1306.3470}{{\tt
  arXiv:1306.3470}}].

\bibitem{Marsden2014}
D.~{Marsden}, M.~{Gralla}, T.~A. {Marriage}, E.~R. {Switzer}, B.~{Partridge},
  M.~{Massardi}, G.~{Morales}, G.~{Addison}, J.~R. {Bond}, D.~{Crichton},
  S.~{Das}, M.~{Devlin}, R.~{D{\"u}nner}, A.~{Hajian}, M.~{Hilton},
  A.~{Hincks}, J.~P. {Hughes}, K.~{Irwin}, A.~{Kosowsky}, F.~{Menanteau},
  K.~{Moodley}, M.~{Niemack}, L.~{Page}, E.~D. {Reese}, B.~{Schmitt},
  N.~{Sehgal}, J.~{Sievers}, S.~{Staggs}, D.~{Swetz}, R.~{Thornton}, and
  E.~{Wollack}, {\it {The Atacama Cosmology Telescope: dusty star-forming
  galaxies and active galactic nuclei in the Southern survey}},  {\em \mnras}
  {\bf 439} (Apr., 2014) 1556--1574,
  [\href{http://arxiv.org/abs/1306.2288}{{\tt arXiv:1306.2288}}].

\bibitem{PlanckCollaboration2011stat_prop}
{Planck Collaboration XIII}, {\it {Planck early results. XIII. Statistical
  properties of extragalactic radio sources in the Planck Early Release Compact
  Source Catalogue}},  {\em \aap} {\bf 536} (Dec., 2011) A13,
  [\href{http://arxiv.org/abs/1101.2044}{{\tt arXiv:1101.2044}}].

\bibitem{Giommi2012}
P.~{Giommi}, G.~{Polenta}, A.~{L{\"a}hteenm{\"a}ki}, D.~J. {Thompson},
  M.~{Capalbi}, S.~{Cutini}, D.~{Gasparrini}, J.~{Gonz{\'a}lez-Nuevo},
  J.~{Le{\'o}n-Tavares}, M.~{L{\'o}pez-Caniego}, M.~N. {Mazziotta}, and et~al.,
  {\it {Simultaneous Planck, Swift, and Fermi observations of X-ray and
  {$\gamma$}-ray selected blazars}},  {\em \aap} {\bf 541} (May, 2012) A160,
  [\href{http://arxiv.org/abs/1108.1114}{{\tt arXiv:1108.1114}}].

\bibitem{LeonTavares2012}
J.~{Le{\'o}n-Tavares}, E.~{Valtaoja}, P.~{Giommi}, G.~{Polenta},
  M.~{Tornikoski}, A.~{L{\"a}hteenm{\"a}ki}, D.~{Gasparrini}, and S.~{Cutini},
  {\it {Exploring the Relation between (Sub-)Millimeter Radiation and
  {$\gamma$}-Ray Emission in Blazars with Planck and Fermi}},  {\em \apj} {\bf
  754} (July, 2012) 23, [\href{http://arxiv.org/abs/1204.3589}{{\tt
  arXiv:1204.3589}}].

\bibitem{Rachen2016}
J.~P. {Rachen}, L.~{Fuhrmann}, T.~{Krichbaum}, E.~{Angelakis}, I.~{Nestoras},
  A.~{Zensus}, A.~{Sievers}, H.~{Ungerechts}, E.~{Keih{\"a}nen}, and
  M.~{Reinecke}, {\it {Coeval Observations of a Complete Sample of Blazars with
  Effelsberg, IRAM 30m, and Planck}},  {\em ArXiv e-prints} (Mar., 2016)
  [\href{http://arxiv.org/abs/1603.02144}{{\tt arXiv:1603.02144}}].

\bibitem{Xia2012}
J.-Q. {Xia}, M.~{Negrello}, A.~{Lapi}, G.~{De Zotti}, L.~{Danese}, and
  M.~{Viel}, {\it {Clustering of submillimetre galaxies in a self-regulated
  baryon collapse model}},  {\em \mnras} {\bf 422} (May, 2012) 1324--1331,
  [\href{http://arxiv.org/abs/1111.4212}{{\tt arXiv:1111.4212}}].

\bibitem{Massardi2013}
M.~{Massardi}, S.~G. {Burke-Spolaor}, T.~{Murphy}, R.~{Ricci},
  M.~{L{\'o}pez-Caniego}, M.~{Negrello}, R.~{Chhetri}, G.~{De Zotti}, R.~D.
  {Ekers}, R.~B. {Partridge}, and E.~M. {Sadler}, {\it {A polarization survey
  of bright extragalactic AT20G sources}},  {\em \mnras} {\bf 436} (Dec., 2013)
  2915--2928, [\href{http://arxiv.org/abs/1309.2527}{{\tt arXiv:1309.2527}}].

\bibitem{TucciToffolatti2012}
M.~{Tucci} and L.~{Toffolatti}, {\it {The Impact of Polarized Extragalactic
  Radio Sources on the Detection of CMB Anisotropies in Polarization}},  {\em
  Advances in Astronomy} {\bf 2012} (Dec., 2012) 624987,
  [\href{http://arxiv.org/abs/1204.0427}{{\tt arXiv:1204.0427}}].

\bibitem{GalluzziMassardi2016}
V.~{Galluzzi} and M.~{Massardi}, {\it {The polarimetric multi-frequency radio
  sources properties}},  {\em International Journal of Modern Physics D} {\bf
  25} (Mar., 2016) 1640005, [\href{http://arxiv.org/abs/1611.08159}{{\tt
  arXiv:1611.08159}}].

\bibitem{Battye2011}
R.~A. {Battye}, I.~W.~A. {Browne}, M.~W. {Peel}, N.~J. {Jackson}, and
  C.~{Dickinson}, {\it {Statistical properties of polarized radio sources at
  high frequency and their impact on cosmic microwave background polarization
  measurements}},  {\em \mnras} {\bf 413} (May, 2011) 132--148,
  [\href{http://arxiv.org/abs/1003.5846}{{\tt arXiv:1003.5846}}].

\bibitem{Bonavera2017}
L.~{Bonavera}, J.~{Gonz{\'a}lez-Nuevo}, F.~{Arg{\"u}eso}, and L.~{Toffolatti},
  {\it {Statistics of the fractional polarisation of compact radio sources in
  Planck maps}},  {\em ArXiv e-prints} (Mar., 2017)
  [\href{http://arxiv.org/abs/1703.09952}{{\tt arXiv:1703.09952}}].

\bibitem{GreavesHolland2002}
J.~S. {Greaves} and W.~S. {Holland}, {\it {Submillimetre polarization of M82
  and the Galactic Center: Implications for CMB polarimetry}},  in {\em
  Astrophysical Polarized Backgrounds} (S.~{Cecchini}, S.~{Cortiglioni},
  R.~{Sault}, and C.~{Sbarra}, eds.), vol.~609 of {\em American Institute of
  Physics Conference Series}, pp.~267--270, Mar., 2002.

\bibitem{Matthews2009}
B.~C. {Matthews}, C.~A. {McPhee}, L.~M. {Fissel}, and R.~L. {Curran}, {\it {The
  Legacy of SCUPOL: 850 {$\mu$}m Imaging Polarimetry from 1997 to 2005}},  {\em
  \apjs} {\bf 182} (May, 2009) 143--204.

\bibitem{GonzalezNuevo2005}
J.~{Gonz{\'a}lez-Nuevo}, L.~{Toffolatti}, and F.~{Arg{\"u}eso}, {\it
  {Predictions of the Angular Power Spectrum of Clustered Extragalactic Point
  Sources at Cosmic Microwave Background Frequencies from Flat and All-Sky
  Two-dimensional Simulations}},  {\em \apj} {\bf 621} (Mar., 2005) 1--14,
  [\href{http://arxiv.org/abs/astro-ph/0405553}{{\tt astro-ph/0405553}}].

\bibitem{Delabrouille2013}
J.~{Delabrouille}, M.~{Betoule}, J.-B. {Melin}, M.-A. {Miville-Desch{\^e}nes},
  J.~{Gonzalez-Nuevo}, M.~{Le Jeune}, G.~{Castex}, G.~{de Zotti}, S.~{Basak},
  M.~{Ashdown}, J.~{Aumont}, C.~{Baccigalupi}, A.~J. {Banday}, J.-P. {Bernard},
  F.~R. {Bouchet}, D.~L. {Clements}, A.~{da Silva}, C.~{Dickinson}, F.~{Dodu},
  K.~{Dolag}, F.~{Elsner}, L.~{Fauvet}, G.~{Fa{\"y}}, G.~{Giardino},
  S.~{Leach}, J.~{Lesgourgues}, M.~{Liguori}, J.~F. {Mac{\'{\i}}as-P{\'e}rez},
  M.~{Massardi}, S.~{Matarrese}, P.~{Mazzotta}, L.~{Montier}, S.~{Mottet},
  R.~{Paladini}, B.~{Partridge}, R.~{Piffaretti}, G.~{Prezeau}, S.~{Prunet},
  S.~{Ricciardi}, M.~{Roman}, B.~{Schaefer}, and L.~{Toffolatti}, {\it {The
  pre-launch Planck Sky Model: a model of sky emission at submillimetre to
  centimetre wavelengths}},  {\em \aap} {\bf 553} (May, 2013) A96,
  [\href{http://arxiv.org/abs/1207.3675}{{\tt arXiv:1207.3675}}].

\bibitem{GonzalezNuevo2006}
J.~{Gonz{\'a}lez-Nuevo}, F.~{Arg{\"u}eso}, M.~{L{\'o}pez-Caniego},
  L.~{Toffolatti}, J.~L. {Sanz}, P.~{Vielva}, and D.~{Herranz}, {\it {The
  Mexican hat wavelet family: application to point-source detection in cosmic
  microwave background maps}},  {\em \mnras} {\bf 369} (July, 2006) 1603--1610,
  [\href{http://arxiv.org/abs/astro-ph/0604376}{{\tt astro-ph/0604376}}].

\bibitem{LopezCaniego2006}
M.~{L{\'o}pez-Caniego}, D.~{Herranz}, J.~{Gonz{\'a}lez-Nuevo}, J.~L. {Sanz},
  R.~B. {Barreiro}, P.~{Vielva}, F.~{Arg{\"u}eso}, and L.~{Toffolatti}, {\it
  {Comparison of filters for the detection of point sources in Planck
  simulations}},  {\em \mnras} {\bf 370} (Aug., 2006) 2047--2063,
  [\href{http://arxiv.org/abs/astro-ph/0606199}{{\tt astro-ph/0606199}}].

\end{thebibliography}\endgroup
\bibliographystyle{JHEP}

\end{document}